\documentclass[
reprint,onecolumn,
superscriptaddress, 12pt, 
nofootinbib,
 amsmath,amssymb,
floatfix
]{revtex4-2}

\usepackage{graphicx}% Include figure files
\usepackage{dcolumn}% Align table columns on decimal point
\usepackage{bm}% bold math
\usepackage{cancel}% to cancel terms in equations
\usepackage{amsfonts}%bold math
\usepackage{xcolor}%to color some texts
\usepackage{tcolorbox}% for boxing texts

\newcommand{\beqa}{\begin{eqnarray}}
\newcommand{\eeqa}{\end{eqnarray}}
\newcommand{\nn}{\nonumber}
\newcommand{\commutator}[3][1.2em]{[ \makebox[#1]{$#2$} , \makebox[#1]{$#3$} ]}%commuator in equations

\begin{document}

\title{Dynamics of Force Dipoles in Curved Biological Membranes }

\author{Sarthak Bagaria}
\affiliation {Goldman Sachs, London, United Kingdom}
\author{Rickmoy Samanta}
\affiliation{Indian Statistical Institute, 203 B. T. Road, Kolkata 700108, India\\
Email for correspondence: rickmoysamanta@gmail.com}
\date{\today}% It is always \today, today,

\begin{abstract}  
We construct a model to explore the hydrodynamic interactions of active inclusions in curved biological membranes. The curved membrane is modelled as a two dimensional layer of highly viscous fluid, surrounded by external solvents of different viscosities. The active inclusions are modelled as point force dipoles. The point dipole limit is taken along a geodesic of the curved geometry, incorporating the change in orientation of the forces due to curvature. We demonstrate this explicitly for the case of a spherical membrane, leading to an  analytic solution for the flow  generated by a single inclusion. We further show that the flow field features an additional defect of negative index, arising from the membrane topology, which is not present in the planar version of the model. We observe that a mutually perpendicular dipole pair moves along geodesics on the sphere and thus act as \textit{curvature checkers}, analogous to vortex dipoles. We finally explore the hydrodynamic interactions of a pair of inclusions in regimes of low and high curvature, as well as situations where the external fluid outside the membrane is confined. Our study suggests aggregation of dipoles  in curved biological membranes of both low and high curvatures, under strong confinement. However, very high curvatures tend to destroy dipole aggregation, even under strong confinement.
\end{abstract}

\maketitle
%\tableofcontents
\section {Introduction}
The collective dynamics of active particles has been a recurring theme in the study of living machines arising in a wide class of biological systems. Such many body systems, made up of living matter, provide an exciting platform to explore non-equilibrium physics. Moreover, many interesting phenomena that emerge at such length scales can be captured by mathematical models, often built out of relatively simple ingredients.\\\\
One such example, arising in biology and the focus of this work, is that of living machines like proteins and other inclusions embedded in  biological membranes. Typically such membranes are well approximated by a thin layer of highly viscous fluid. The membrane flows are thus described by Low Reynolds Hydrodynamics \cite{purcell}, where inertia effects are suppressed compared to the viscous terms in the Navier Stokes Equation. Moreover, the 2D membrane fluid also exchanges momentum with the external fluids surrounding the membrane and thus they are essentially \emph{quasi}-2D in nature.\\\\
In the context of flat membranes, the early works  Ref.\cite{saff1,saff2,hughes,evans} showed that the  quasi-2D nature of the membranes gives rise to a new length scale, Saffman Length, given by the ratio of the viscosity of the 2D membrane fluid and the viscosity of the external solvents. The Saffman length acts as a regulator for the long distance logarithmic divergence one usually encounters in such 2D flows. Beyond this length scale, the traction stress from the 3D external fluid  starts dominating  over the 2D membrane stress, thereby regulating the divergence. \\\\
Biological membranes arising in nature always have some curvature and typically form closed, compact surfaces. Several recent works have performed a detailed study of fluid flows in curved biological membranes  Ref.\cite{henlev2008,henlev2010,wg2012, wg2013, atzbergershape,atz2016, atz2018,atz2019}. In particular  Ref.\cite{henlev2008,henlev2010} extended the pioneering works of Saffman and Delbr\"uck  Ref.\cite{saff1,saff2} to account for membrane curvature. These works are a starting point for our analysis here.\\\\
Biological machines such as membrane proteins are examples of active matter. They can perform mechanical work using chemical energy, without any external force. The flows generated by such inclusions are essentially similar to that of force dipoles (stresslets)\footnote{We use the terms ''force dipoles" and ''stresslets" interchangeably in the paper.} to leading order. Force dipoles have been widely studied in the context of swimming in  Stokes fluids  Ref.\cite{lauga2009}. Recently, aggregation  of force dipoles has been studied in  planar membranes  Ref.\cite{mnk}. However, similar studies are yet to be performed in curved biological membranes.\\\\
The key results of this paper are listed below:
\begin{itemize}
\item  We illustrate the construction of flows  sourced by a point force dipole embedded in a viscous membrane of non-zero curvature and coupled to external fluids. While taking the point dipole limit, we keep track of how the curvature affects the orientation of point forces. We demonstrate this procedure in detail for a spherical membrane. The formula for the resulting flow Eq.(\ref{vdp_f}) is the central result of this paper. 
\item We also provide generalizations of the formula Eq.(\ref{vdp_f})  to situations where the external fluid outside the membrane is confined to a finite radius Eq.(\ref{conf_vdp_f}), as well as membranes of arbitrary but fixed geometry Eq.(\ref{rot_gen}).
\item We also examine the streamline topology of the resulting flow sourced by such a point dipole. The topology of the membrane gives rise to a new defect of negative index, not seen in the planar version of the model and is consistent with the topology of the membrane. 
\item 
 Using the formulas Eq.(\ref{vdp_f}) and Eq.(\ref{conf_vdp_f}), we explore the hydrodynamic interactions of a pair of point force dipoles in the curved membrane, in low and high curvature regimes, as well as situations where the external fluid outside the membrane is strongly confined. Our study suggests aggregation of dipoles in curved biological membranes, of both low and high curvature, under strong confinement. We also observe that for a fixed confinement depth, extreme high curvatures tend to destroy the dipole aggregation.
\end{itemize}
The work presented here builds upon the results of Ref.\cite{henlev2008,henlev2010} and Ref.\cite{mnk} and expands them in several important directions.   Ref.\cite{henlev2010} constructs Stokeslet flows in curved membranes, which we extend to force dipole flows using a novel and intuitive approach (Sec.\ref{fdp_cons}).  A close inspection of the resulting dynamical formulas (Eq.\ref{dynmeq}) reveals that curvature effects modify the hydrodynamic interactions as well as the rotation rate of the dipoles. Since fluidic membranes arising in nature typically form closed, curved surfaces, our results show that \textit{moderate} curvature effects do not destroy the physically interesting scenario of dipole aggregation. However, dipoles fail to aggregate in \textit{very} high curvature situations, even under strong confinement. Moreover, membrane topology creates new defects (Fig.\ref{fdp_flow}), while curvature modifies dipole dynamics (Fig.\ref{curv}) compared to flat membranes (Ref.\cite{mnk}). Thus the present work may be considered as a building block to understand the dynamics of biological motor proteins in curved membranes and other microfluidic environments arising in nature, paving the way for an improved understanding of cellular transport Ref.\cite{mik}. The model we study here is also  relevant to many recent experiments, particularly Ref.\cite{henlev2008,wg2013}.\\\\
%We believe  these results will be interesting in the context of dynamics of active machines like proteins and other inclusions in a biological membrane as well as designing artificial swimmers in thin biofilms as well as other bioengineering and microfluidic devices.\\\\
The paper is organized as follows: In Sec.\ref{hydrosummary}  we briefly review the generalization of Stokes equations to curved surfaces and discuss the flows sourced by a point force ie. a stokeslet. In Sec.\ref{fdp_cons} we illustrate the procedure to construct flows sourced by point dipoles in a curved membrane and apply it to obtain an analytic solution for the case of  spherical membranes. The streamline topology of the resulting flows is discussed in Sec.\ref{strm}. Next we analyze the dynamics of a pair of such inclusions in regimes of low and high curvatures, with and without confinement, in Sec.\ref{pair_dyn}. We present a summary of the simulations performed in the paper in Sec.\ref{sumsim}. Finally, we conclude in Sec.\ref{cncl} with possible directions for the future.\\\\
The main text is supplemented by Appendix \ref{plane_summary} which briefly reviews  the corresponding force dipole dynamics in flat membranes, and Appendix \ref{chyd_summary} presents a detailed derivation of the formulation of hydrodynamics in curved biological membranes and associated Stokeslet flows. \\

\section { Summary of viscous hydrodynamics in curved membranes }
\label{hydrosummary}
In this section we briefly review the formulation of hydrodynamics in viscous fluid membranes of generic curvature and then specialize to spherical membranes along the lines of  Ref.\cite{henlev2008,henlev2010}. The formulation of hydrodynamics presented here rests on several assumptions. First, the biological membrane is approximated as a 2D viscous Newtonian fluid surrounded above and below by 3D external fluids. This approximation is valid as long as  gradients of the velocity field components  are negligible along the normal to the membrane. The membrane fluid is incompressible and impermeable to the external fluids.  We also restrict to only tangential flows in the membrane such that the geometry of the membrane stays unaltered, allowing only in-plane shear modes. In such situations, the appropriate generalization of the Stokes equations describing viscous fluid flows for the 2D curved membrane, coupled to external 3D fluids, are given by
\beqa
&D^\alpha v_\alpha =0,\nn\\ 
&\sigma^{ext}_{\alpha} = \underbrace{-\eta_{2D} \left( K(\vec{x}) v_\alpha  + D^\mu D_\mu v_\alpha \right) + D_\alpha p}_{Membrane~contribution}
+ \underbrace{ \left( \sigma^{3D}_{\alpha z}|_{z\rightarrow 0^-} - \sigma^{3D}_{\alpha z}|_{z\rightarrow 0^+}\right)}_{Traction ~from~ 3D~ external ~fluids} 
\label{curvedstokes_main}
\eeqa
where x denotes generic surface coordinates for the membrane geometry, $v_\alpha$ represents the 2D velocity of the membrane fluid where $\alpha$ represents the surface coordinates. $\sigma^{ext}_{\alpha} $ represents the local stress exerted by the inclusions embedded in the membrane. $\eta_{2D}$ is the viscosity of the 2D membrane fluid, $ D$ is the covariant derivative compatible with the metric, 
$K(\vec{x})$ is the local Gaussian curvature, 
$p$ is the local 2D membrane pressure,
$\sigma^{3D}$ is the stress tensor of the external fluids (see Eq.(\ref{trc})in Appendix \ref{chyd_summary}) and 
$z$ is a generalized coordinate normal to the membrane surface. The first  equation in Eq.(\ref{curvedstokes_main}) is that of an incompressible fluid while the second equation is the hydrodynamic stress balance at the membrane surface. The inclusions embedded in the membrane exert  a local stress $\sigma^{ext}$.  This is balanced by the 2D membrane stress  and  the external traction from the fluids above and below the membrane. The membrane stress includes a curvature term $K(\vec{x})$. For a concise review of the formulation of viscous hydrodynamics adapted to curved membrane geometries in the current context, please see Appendix \ref{chyd_summary}. We refer to Ref.\cite{henlev2010} for a detailed and insightful discussion. The external fluid flows are governed by the usual 3D Stokes equations:
 \beqa 
 \eta _{\pm}\nabla^2 {\bf v}_{\pm} = \nabla_{\pm} p^{\pm},  \nabla \cdot {\bf v}_{\pm} = 0,
  \label{stokes3d}
  \eeqa
  where ${\bf v}^{+}$ (${\bf v}^{-}$) is the fluid velocity outside (inside) the membrane, similarly we define external pressures $p_{\pm}$ and viscosities $\eta_{\pm}$. We also have no slip boundary condition at the membrane surface. The 3D viscosities of the external fluids $\eta_{\pm}$ can be combined with the membrane 2D viscosity $\eta_{2D}$ to  construct two Saffman lengths $\lambda_{\pm}=\frac{\eta_{2D}}{\eta_{\pm}}$. For simplicity, we will set $\eta_+ =\eta_- \equiv \eta$ in the rest of the paper and denote the unique Saffman length simply by $\lambda$ \beqa
\lambda =\frac{\eta_{2D}}{\eta}
 \label{saffl}
\eeqa
The other length scales in the model are set by the local membrane curvature.   For membranes of arbitrary shape, these coupled system of equations require advanced numerical approaches, see for example Ref.\cite{atz2018}. However, for simpler situations where the curvature is constant, say a spherical membrane, analytic solutions can be constructed with relative ease.\\\\
Specializing to the spherical membrane of radius R, the velocity field $\bm{v}$ at ($\theta,\phi)$ due to a point force $\bm{f}$ located at $(\theta_0,\phi_0)$ can be expressed  compactly\footnote{We adopt the convention where gradient of a function is treated as a column vector and the force vector $\bm{f}$ is treated as a row vector} as (written in the usual $(\hat{\theta}, \hat{\phi})$ basis)
\beqa
\bm{v}_{Stokeslet} =  \frac{1}{4 \pi \eta_{2D}} \tilde{\bm{\nabla}}_{\theta,\phi}\left( \bm{f} \cdot \tilde{\bm{\nabla}}_{\theta_0, \phi_0}~ S[\gamma] \right)
\label{ptf_flow}
\eeqa
where 
$\tilde{\nabla}_{\theta,\phi}$ is the twisted gradient operator $ \left(\csc \theta ~ \partial_\phi, -\partial_ \theta\right)$ at the response location $(\theta,\phi)$ and $\tilde{\nabla}_{\theta_0,\phi_0}$ is a similar gradient at the source location and
\beqa
S[\gamma]:=\sum_l \frac{2l+1}{s_l l (l+1)} P_l[\cos \gamma]
\label{sdef_dp} 
\eeqa
where $\cos \gamma = \sin \theta \sin \theta_0 \cos(\phi -\phi_0) + \cos \theta \cos \theta_0$ is the cosine of the geodesic angle between source and response locations.  Further, we have
\beqa
s_l =l(l+1) -2 +\frac{R}{\lambda_-}(l-1)+\frac{R}{\lambda_+}(l+2)
\label{sldef_dp}
\eeqa
\begin{figure}[h]
\begin{tabular}{cc}
\includegraphics[width=4cm]{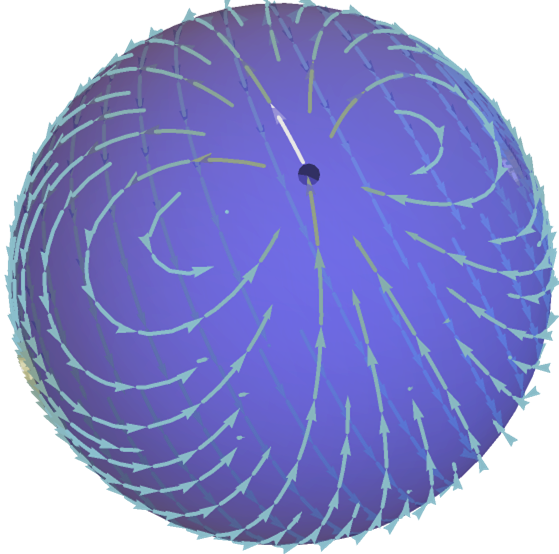}&~~~~
\includegraphics[width=5cm]{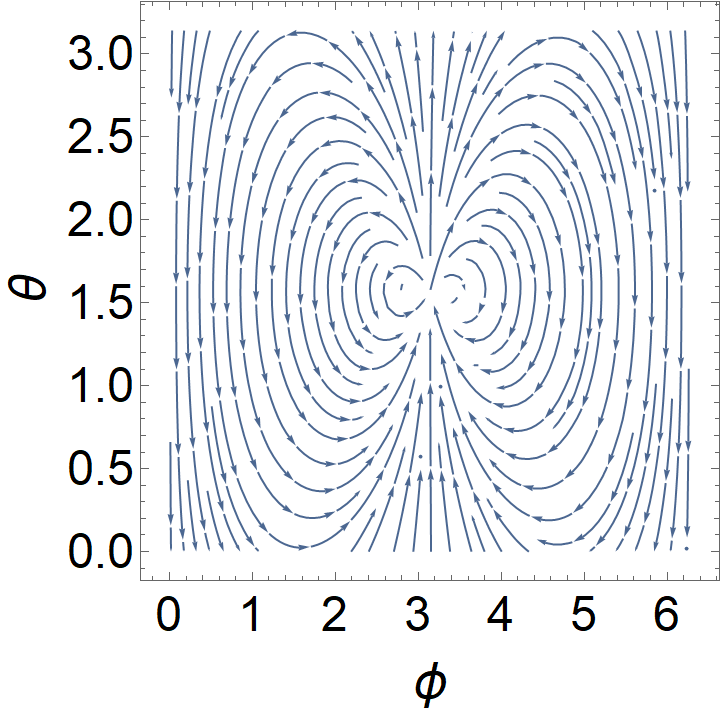}\\
\includegraphics[width=4cm]{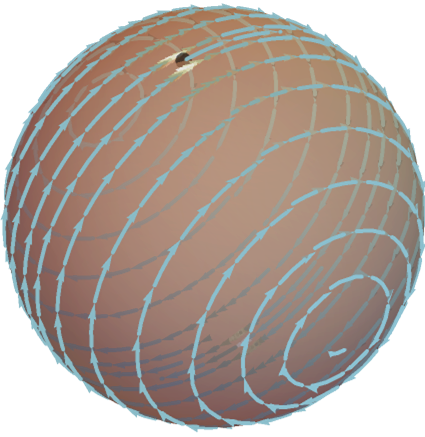}&~~~~
\includegraphics[width=5cm]{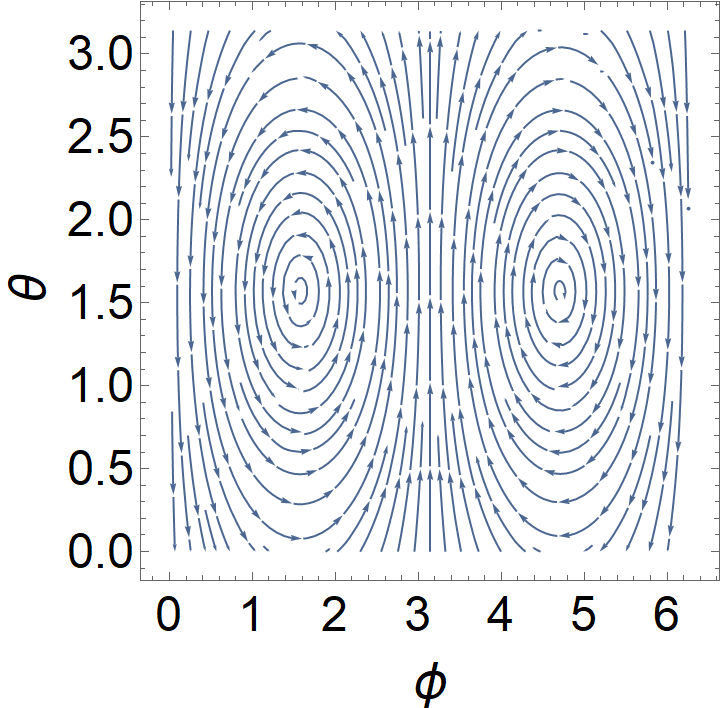}
    \end{tabular}\\
\caption{(Color online) Flow field around a point force  at low curvature (first row) and high  curvature (second row). In both situations radius R is held fixed and the fluid viscosity $\eta_{2D}$ is tuned, such that $\lambda/R=0.1$ in the first row (low curvature) and $\lambda/R=100$ in the second row (high curvature). The brown color in the high curvature regime corresponds to the more viscous membrane fluid.}
     \label{st}
\end{figure}
A brief derivation of Eq.(\ref{ptf_flow}) is presented in Appendix \ref{chyd_summary}.
Using Appell Hypergeometric functions, one can perform the sum in Eq.(\ref{sdef_dp}) analytically in the full parameter space of the model. The expressions are listed in Appendix \ref{chyd_summary}. We use these functions to simulate force dipole flows and interactions in this paper.\\\\
Let us add some comments on the streamline topology of the resulting flows. The hairy ball theorem dictates that the flow fields on a sphere must have singularities. Moreover, the sum of the index of these singularities must add up to the Euler Characteristic of the surface, via the Poincare Index Theorem. For a spherical membrane, the Euler characteristic is two.  We note that the Stokeslet flow field features two vortical defects around the point of application of the point force, see Fig.\ref{st}. Vortical defects have index $+1$, leading to a total index of two for the Stokeslet flow. In the low curvature regime, the velocity field  exhibits two vortical defects around the point of application of the force. With increasing curvature, the vortical defects migrate away to diametrically opposite points, as shown in Fig.\ref{st} and first reported in  Ref.\cite{henlev2008,henlev2010,atz2016}.

\section{Flows sourced by a force dipole}
\label{fdp_cons}
\begin{figure}[h]
\includegraphics[width=8cm]{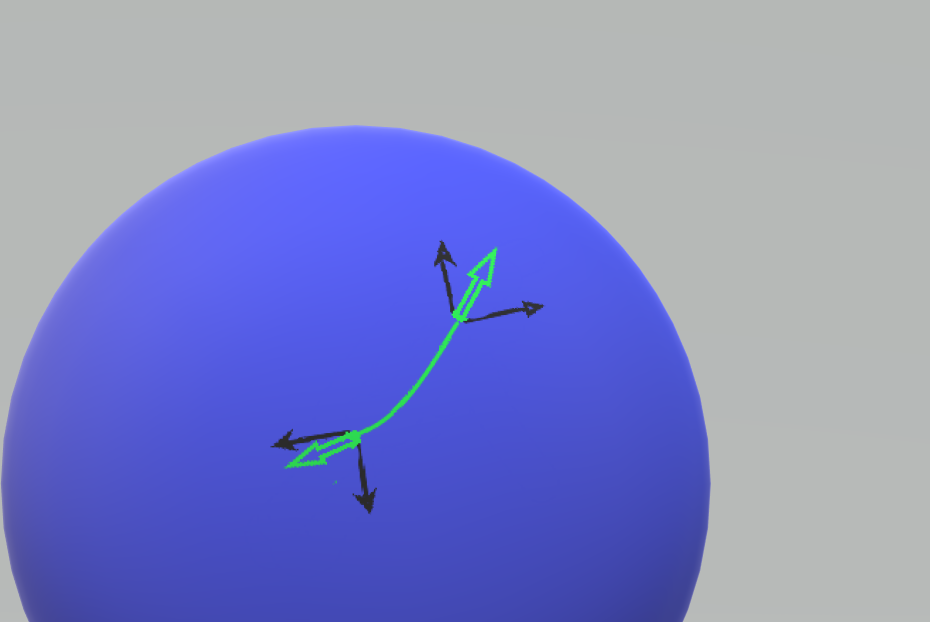}
   \caption{(Color online) \textbf{Construction of the point force dipole in a generic curved geometry}: The two point forces are shown as green arrows, initially at a finite separation and tangent to the geodesic (green curve) connecting them. Note that the point force at the right end of the geodesic makes a different angle wrt the background frame in the local tangent space (shown with black arrows), compared to the point force at the left end of the geodesic. This is due to the curvature of the surface. The point dipole limit arises in the limit of vanishing geodesic separation, when the two point forces are dragged closer to each other along the geodesic. }
     \label{limit_procedure}
\end{figure}
Now that we have understood the flow field generated by a Stokeslet in a spherical membrane, we move to the construction of the flow due to a point force dipole (Stresslet). Let us first recall how this is done for flat membrane flows. A simple and intuitive way is to consider two equal and oppositely oriented point forces (stokeslets) initially at a finite separation and then taking the point dipole limit by transporting the forces closer and closer, along the line of separation. In the limit of vanishing inclusion size, this amounts to keeping only the leading term ie. the dipole. It is clear that the resulting flow due to a stresslet is given by the directional derivative of the Stokeslet flow field along the orientation of the force, see Eq.(\ref{vdp_free}) in Appendix \ref{plane_summary} where we have presented a summary of force dipole solutions in flat membranes. We adopt a similar strategy for curved membranes. We first consider two equal and opposite point forces separated along a geodesic, the analogue of a straight line in a curved geometry, see Fig.\ref{limit_procedure}. The point forces (denoted by green arrows) are tangential to the geodesic (shown as a green curve), situated  at the two endpoints of the geodesic. The  point dipole arises in the limit of vanishing geodesic separation. However, the orientation of the forces can change due to curvature as they are dragged along the geodesic, as illustrated in Fig.\ref{limit_procedure} wrt to a background frame (shown in black) for a generic curved surface. As we will see, this will give rise to an extra term compared to the formula for flat membranes.\\\\
We now demonstrate this approach explicitly for the spherical membrane. Let us consider two  oppositely oriented point forces  of equal  magnitudes $|\bm{f}|$, one at $(\theta_0, \phi_0)$  and the other at $(\theta_0 + d\theta_0, \phi_0 + d \phi_0)$. The two point forces are separated along a geodesic of infinitesimal length $dL$. Each of these point forces generates flows in the spherical membrane $\bm{v}_{Stokeslet} $ given by Eq.(\ref{ptf_flow}). The point dipole limit is obtained by taking the limit $dL \rightarrow 0 ,~ |\bm{f}| \rightarrow \infty$ along the geodesic such that $ |\bm{f}| dL :=\kappa$ is constant.  
\beqa
& \bm{v}_{dipole} = \frac{1}{4 \pi \eta_{2D} } \tilde{\bm{\nabla}}_{\theta,\phi} \left( \bm{f}_{\substack{\theta_0 +d \theta_0 \\\phi_0 +d \phi_0 }} \cdot \tilde{\bm{\nabla}}_{\substack{\theta_0 +d \theta_0 \\\phi_0 +d \phi_0 }} S[\gamma {\scriptstyle ( \theta, \phi, \theta_0 + d\theta_0,\phi_0 +d\phi_0)}]  - \bm{f}_{\substack{\theta_0, \phi_0 }}  \cdot \tilde{\bm{\nabla}}_{\substack{\theta_0, \phi_0 }} S[\gamma {\scriptstyle ( \theta, \phi, \theta_0,\phi_0 )}] \right)\nn\\
&= \frac{1}{4 \pi \eta_{2D} }\tilde{\bm{\nabla}}_{\theta,\phi} \left(  [d\theta_0 \partial_{\theta_0}+ d\phi_0 \partial_{\phi_0}]~ \bm{f}_{\substack{\theta_0, \phi_0 }} \cdot \tilde{\bm{\nabla}}_{\theta_0, \phi_0} S[\gamma {\scriptstyle ( \theta, \phi, \theta_0,\phi_0 )}]+ \Delta \bm{f} \cdot \tilde{\bm{\nabla}}_{\theta_0, \phi_0}S[\gamma {\scriptstyle ( \theta, \phi, \theta_0,\phi_0 )}]+..\right)\nn\\
\label{fdpint}
\eeqa
where  $\bm{f}_{\substack{\theta_0, \phi_0 }}$   and  $\bm{f}_{\substack{\theta_0 +d \theta_0 \\\phi_0 +d \phi_0 }}$  represent the two point forces at locations $(\theta_0, \phi_0)$ and $(\theta_0 + d\theta_0, \phi_0 + d \phi_0)$ respectively. $\Delta  \bm{f}$ denotes the first order change in $\bm{f}$ due to parallel transport as one moves from $(\theta_0, \phi_0)$ to $(\theta_0 + d\theta_0, \phi_0 + d \phi_0)$ in the curved geometry. Also ``..." represents higher order terms. Henceforth, we will suppress the arguments of the geodesic distance $\gamma( \theta, \phi, \theta_0,\phi_0 )$ and denote it simply by $\gamma$.\\\\
Let us now write the point force 
$\bm{f}_{\substack{\theta_0, \phi_0 }}$ in the last line of Eq.(\ref{fdpint}) in terms of the vector $\hat{\bm{T}}$, which is tangent to the geodesic at the location $(\theta_0,\phi_0)$.  In the usual $(\hat{\theta}, \hat{\phi})$ basis on the sphere, the components of $\hat{\bm{T}}$ are given by
\beqa
 \hat{\bm{T}} := \left(\sin \alpha, \cos \alpha\right)\hspace{1cm} \text{where}~ \bm{f}_{\substack{\theta_0, \phi_0 }} \equiv |\bm{f}| \hat{\bm{T}}
 \label{Tdef}
 \eeqa
 where $\alpha$ represents the orientation of the force vector in the local tangent space at  $(\theta_0,\phi_0)$ and measured wrt $\hat{\phi}$ (this will also be the orientation of the point dipole  that we construct below). The infinitesimal geodesic of length $dL$ connects the two point forces, one at $(\theta_0, \phi_0)$ and the other at $(\theta_0 +d \theta_0, \phi_0 +d \phi_0)$. Thus it follows that $ R~ d \theta_0 = dL \sin \alpha, ~ R \sin \theta_0~ d \phi_0 = dL \cos \alpha$. Keeping the leading order terms in Eq.(\ref{fdpint}) for small sized inclusions, the force dipole velocity field at $(\theta, \phi)$ is given by
\beqa
\bm{v}_{dipole} =\frac{1}{4 \pi \eta_{2D}}\tilde{\bm{\nabla}}_{\theta,\phi} \left(  \frac{1}{R} |\bm{f}|~ dL ~\left[\hat{\bm{T}} \cdot \bm{\nabla}_{\theta_0, \phi_0}\right] \left(\hat{\bm{T}}\cdot \tilde{\bm{\nabla}}_{\theta_0, \phi_0} S[\gamma ]\right) +\Delta \bm{f} \cdot  \tilde{ \bm{\nabla}}_{\theta_0, \phi_0}S[\gamma ] \right) \label{vdpint}
\eeqa
The curvature induced $\Delta \bm{f}$ for transport along the infinitesimal geodesic of length $dL$  is 
\beqa
 \Delta {\bm{f}}=  |\bm{f}|~ dL~  (\hat{\bm{T}} \cdot \bm{M})
 \label{rotf}
\eeqa 
where the matrix $\bm{M}$ generates the  rotation of the tangent vector along an infinitesimal geodesic of length $dL$ on the sphere, computed in the $(\hat{\theta}, \hat{\phi})$ basis as
\beqa
 \bm{M}= \frac{1}{R}\begin{pmatrix} 
 0    & - \cot \theta_0 \cos \alpha\\ 
  \cot \theta_0 \cos \alpha & 0
\end{pmatrix}
\label{rot}
\eeqa
Plugging Eq.(\ref{rotf}) into Eq.(\ref{vdpint}) we get
\beqa
\bm{v}_{dipole} = \frac{|\bm{f}|~ dL}{4 \pi \eta_{2D} R}\tilde{\bm{\nabla}}_{\theta,\phi} \left(  \left[\hat{\bm{T}} \cdot \bm{\nabla}_{\theta_0, \phi_0}\right] \left(\hat{\bm{T}}\cdot \tilde{\bm{\nabla}}_{\theta_0, \phi_0} S[\gamma ]\right)+ (\hat{\bm{T}} \cdot \bm{M}) \cdot \tilde{ \bm{\nabla}}_{\theta_0, \phi_0}S[\gamma ] \right)
\eeqa
Taking the point dipole limit $dL \rightarrow 0 ,~ |\bm{f}| \rightarrow \infty$ along the geodesic such that $ |\bm{f}| dL :=\kappa$ is held fixed, we get

\beqa
\bm{v}_{dipole} = \frac{\kappa}{4 \pi \eta_{2D} R}\tilde{\bm{\nabla}}_{\theta,\phi}  \left(  \left[\hat{\bm{T}} \cdot \bm{\nabla}_{\theta_0, \phi_0}\right] \left(\hat{\bm{T}}\cdot \tilde{\bm{\nabla}}_{\theta_0, \phi_0} S[\gamma ]\right)+ (\hat{\bm{T}} \cdot \bm{M}) \cdot \tilde{ \bm{\nabla}}_{\theta_0, \phi_0}S[\gamma ] \right)
\label{vdp_f}
\eeqa
where the function $S[\gamma]$ is defined in Eq.(\ref{sdef_dp}). \\\\
The stresslet flow in the curved membrane given by Eq.(\ref{vdp_f}) is the central result of this paper. It gives the velocity field at the location $(\theta,\phi)$ in response to a point force dipole situated at $(\theta_0,\phi_0)$. Let us note that the second term in Eq.(\ref{ptf_flow}) takes into account the change in orientation of the point forces  as they are dragged along the infinitesimal geodesic while taking the point dipole limit. This arises purely due to curvature of the surface and is absent in the plane. \\\\
\textbf{ Dipole flows under confinement}: If additionally the external fluid outside the membrane is confined to a finite radial distance $R+H$, with spherically symmetric boundary conditions\footnote{Note that in our terminology, ``confinement" refers to confining the external fluid outside the  membrane ($r>R$) with viscosity $\eta_+$ so that it no longer extends to infinity. For any closed membrane surface, the fluid inside the membrane ($r<R$) with viscosity  $\eta_-$ is always confined to a compact region.},  then carrying out the steps sketched in Appendix \ref{chyd_summary} (allowing both rising and falling modes for the region $R<r<R+H$), we can derive the flow in the confined situation. The structure of the solution, denoted by $\bm{v^c}_{dipole}$, is essentially the same as Eq.(\ref{vdp_f}), but now with a modified $S^c[\gamma]$ ie.

\beqa
\bm{v^c}_{dipole} = \frac{\kappa}{4 \pi \eta_{2D} R}\tilde{\bm{\nabla}}_{\theta,\phi}  \left(  \left[\hat{\bm{T}} \cdot \bm{\nabla}_{\theta_0, \phi_0}\right] \left(\hat{\bm{T}}\cdot \tilde{\bm{\nabla}}_{\theta_0, \phi_0} S^c[\gamma ]\right)+ (\hat{\bm{T}} \cdot \bm{M}) \cdot \tilde{ \bm{\nabla}}_{\theta_0, \phi_0}S^c[\gamma ] \right)\nn\\
\label{conf_vdp_f}
\eeqa 
where 
\beqa
S^c[\gamma]:=\sum_l \frac{2l+1}{s_l^{c} l (l+1)} P_l[\cos \gamma]
\label{conf_sdef_dp} 
\eeqa
 with $s_l^{c}$ now given by
\beqa
&s_l^c = l(l+1)-2 +\left(\frac{R}{\lambda_-}(l-1) - \frac{R}{\lambda_+}\left(\frac{R h_l^\prime(R)}{h_l(R)} -1\right)\right)\nn\\
&\text{where}~h_l(r) =\frac{1}{r^{l+1}}-\frac{r^l}{(R+H)^{2l+1}}
%&h_l^\prime(R) =\frac{-(l+1)}{R^{l+2}}-\frac{l R^{l-1}}{(R+H)^{2l+1}}\nn\\
%&\tilde{C}_l =h_l(R) =\frac{1}{R^{l+1}}-\frac{R^l}{(R+H)^{2l+1}}\nn
\label{slc_conf}
\eeqa
In the limit $H \rightarrow \infty$ we recover the solution for the unconfined case Eq.(\ref{vdp_f}).\\\\
\textbf{Comments on generic curved membranes}: For a generic curved 2D membrane with orthogonal coordinate basis $(e_1,e_2)$, the general structure of the formula Eq.(\ref{vdp_f}) still holds\footnote{With the twisted gradient $\tilde{\bm{\nabla}}$ suitably generalized by the antisymmetric covariant derivative $\epsilon_{\alpha \beta} D^\beta$ for the given surface} but one needs the appropriate function $S[\gamma]$ for the given geometry of the membrane. Typically this requires a numerical evaluation of the spectrum of the Laplace Beltrami operator on the curved manifold, along the lines of Ref.\cite{atz2018}. On the other hand, the  matrix $\bm{M}$ appearing in the second term in Eq.(\ref{vdp_f}) can be evaluated for generic geometries with relative ease. This has the following form (denoted by $\bm{M}^{gen}$) in terms of Christoffel symbols $\Gamma$ (repeated coordinate index $c$ indicates summation and $A$ denotes anti-symmetrization)
 \beqa
 \bm{M}^{gen}=\begin{pmatrix} 
 - \Gamma^1_{1c} T^c  &  -\Gamma^2_{1c} T^c \frac{\left\lVert e^1\right\rVert}{\left\lVert e^2\right\rVert} \\ 
   -\Gamma^1_{2c} T^c \frac{\left\lVert e^2\right\rVert}{\left\lVert e^1\right\rVert}  &  - \Gamma^2_{2c} T^c 
\end{pmatrix}^{A}
\label{rot_gen}
\eeqa
Here the norms are denoted by $\left\lVert~\right\rVert$. The Christoffel symbols $\Gamma$ can be derived from the surface metric $g_{ab}$
\beqa
\Gamma^{i}_{kl}= \frac{1}{2} g^{im}\left(\partial_l g_{mk} +\partial_k g_{ml} -\partial _m g_{kl}\right).
\eeqa
The vector $T$ is the unit vector representing the orientation of the dipole ie. in our notation $ \sin \alpha =T^1 \left\lVert e_1\right\rVert, \cos \alpha =T^2 \left\lVert e_2\right\rVert$. 
 For the spherical basis $(e_\theta, e_\phi)$ the matrix $\bm{M}^{gen}$ reduces to that in Eq.(\ref{rot}).

 \begin{figure}[h]
\begin{tabular}{lcc}
\includegraphics[width=3.5cm]{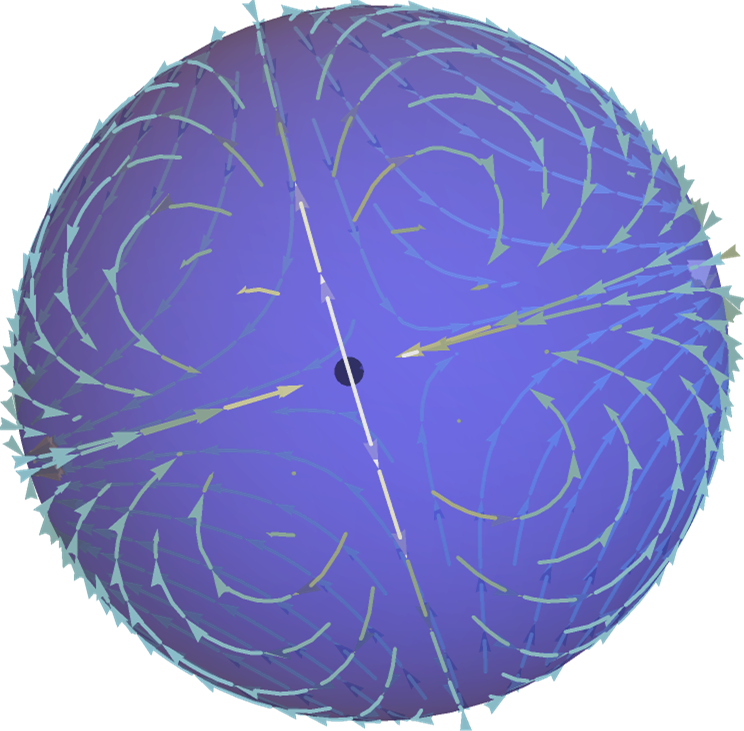}~~~~&
\includegraphics[width=3.5cm]{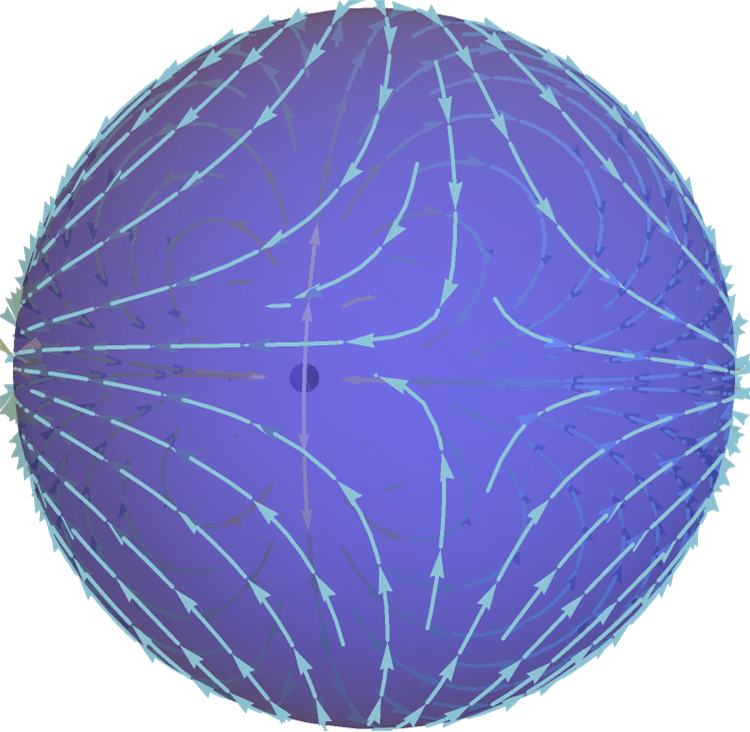}~~~~&
\includegraphics[width=3.5cm]{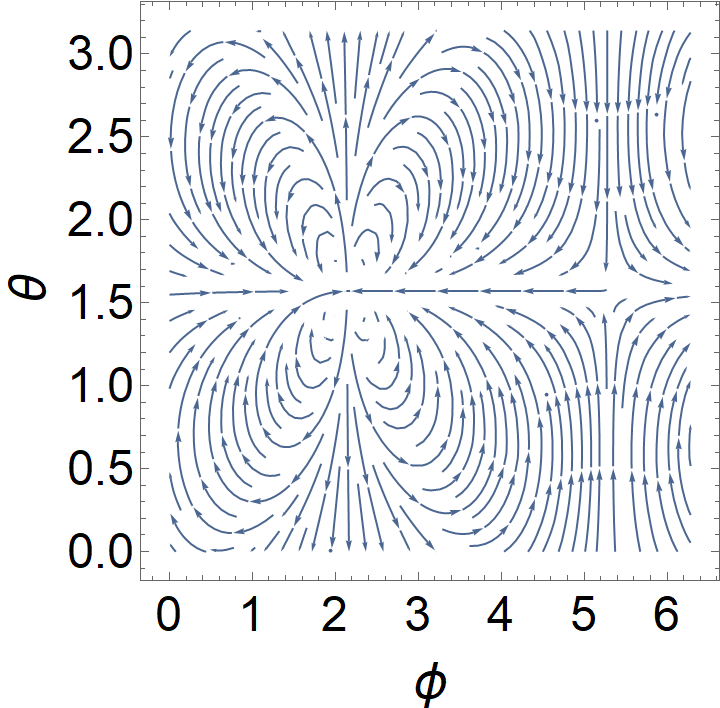}\\ \\
\includegraphics[width=3.5cm]{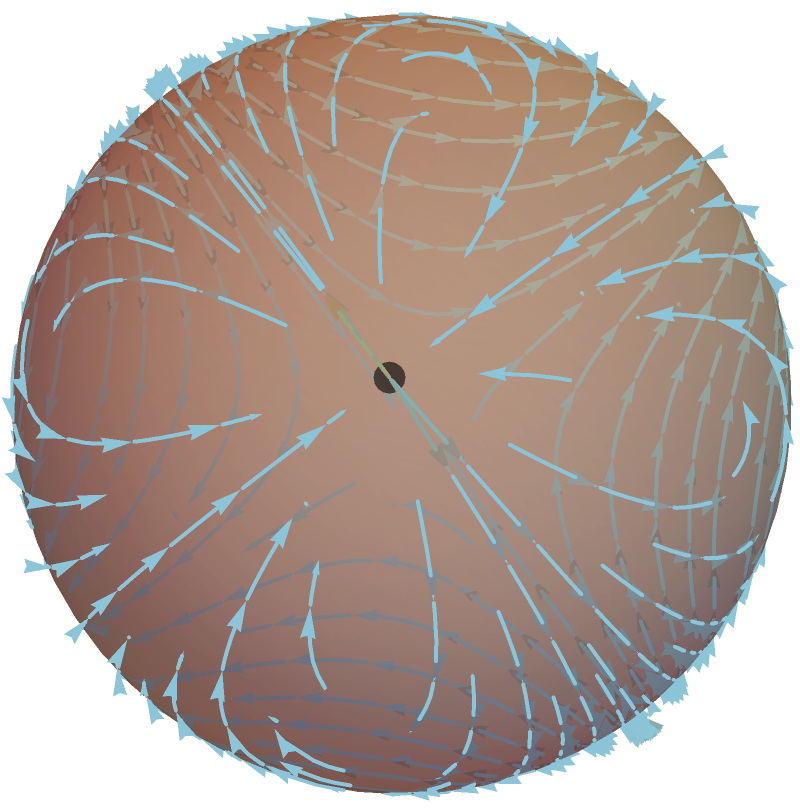}~~~~&
\includegraphics[width=3.5cm]{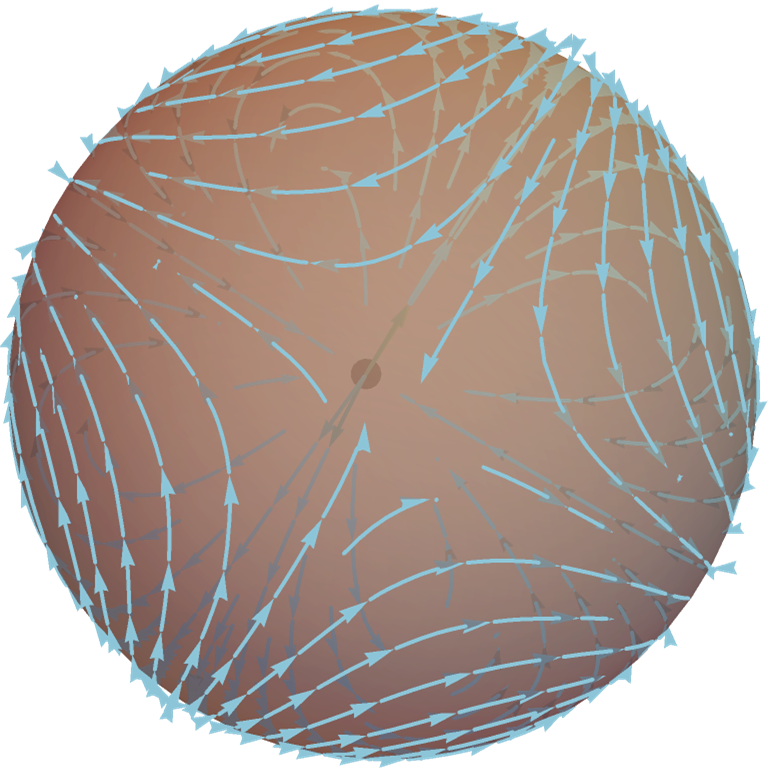}~~~~&
\includegraphics[width=3.5cm]{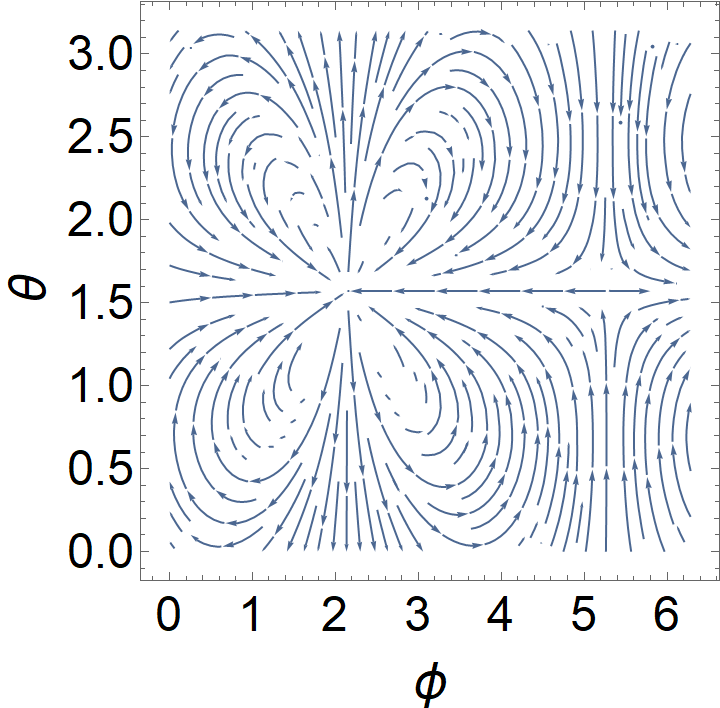}
    \end{tabular}\\
\caption{(Color online) Flow field around a force dipole oriented along $\hat{\theta}$ ie. $\alpha = \pi/2 $ and situated at $(\theta=\pi/2,~\phi=2.2)$, obtained by plotting Eq.(\ref{vdp_f}) in the low curvature (first row) and high curvature (second row) regimes. In each row, the figure on the left shows the flow field around the dipole center (represented by a black dot). The middle plot shows the additional defect of negative index which forms on the other side of the membrane, opposite to the location of the dipole.The plot on the right shows the flow lines in a $\theta,\phi$ chart, showing both the dipole flow along with the additional defect that arises due to spherical topology.  This defect is absent in the corresponding flows in flat membranes (see main text). Also note that the near field flow is more radial in character at high curvature.}
    \label{fdp_flow}
\end{figure}
\begin{figure}[h]
\begin{tabular}{lcccccccc}
\includegraphics[width=3.2cm]{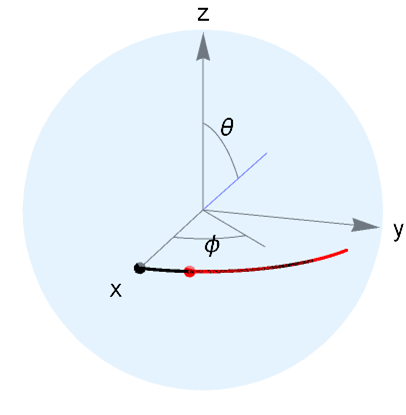}&&
\includegraphics[width=3cm]{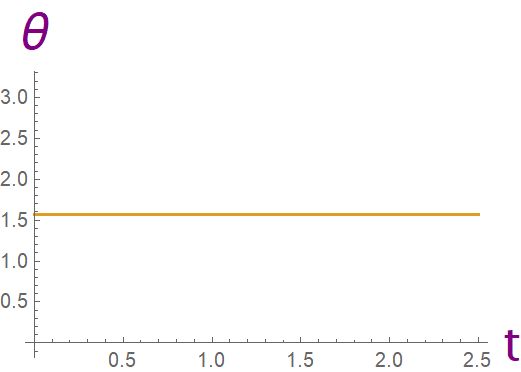}&&
\includegraphics[width=3cm]{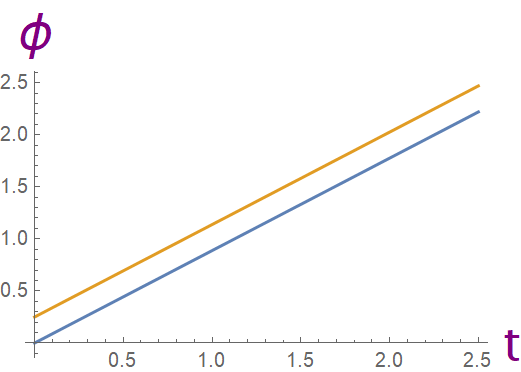}&&
\includegraphics[width=3cm]{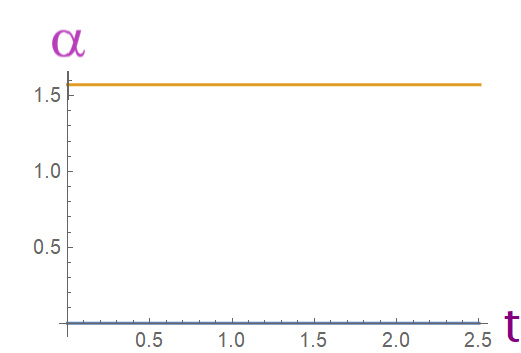}\\
\includegraphics[width=3cm]{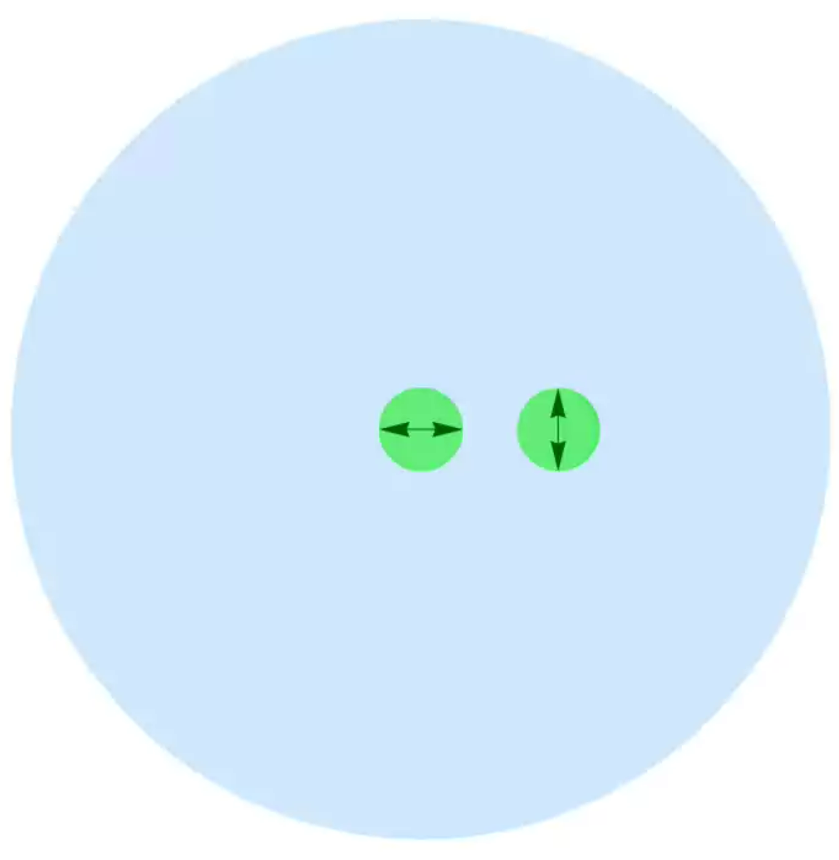}$\pmb{\rightarrow}$&&
\includegraphics[width=3cm]{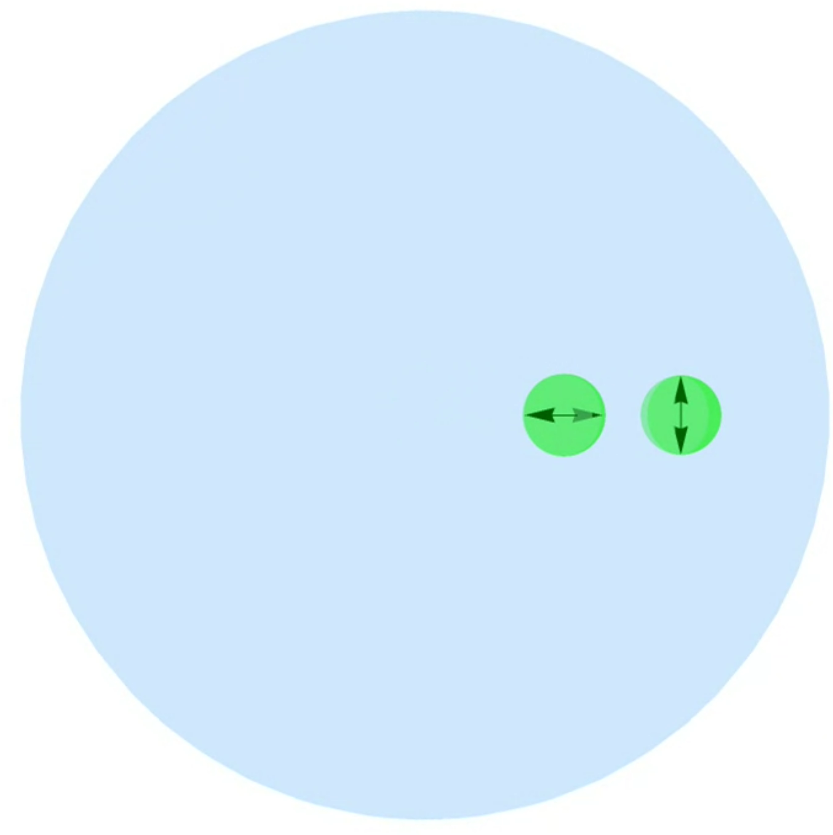}$\pmb{\rightarrow}$&&
\includegraphics[width=3cm]{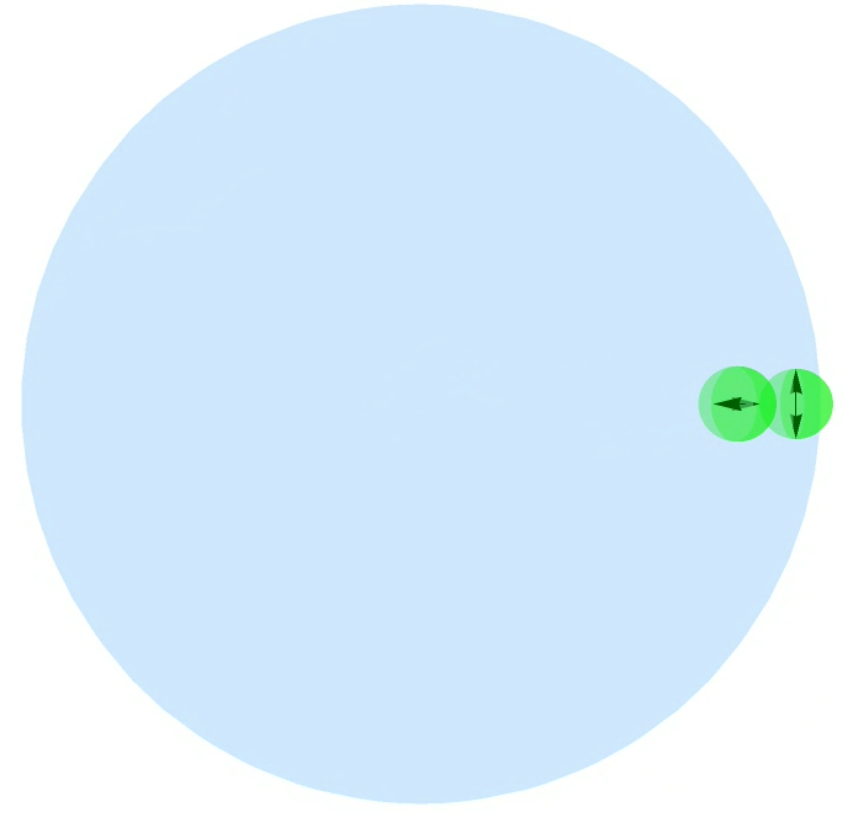}$\pmb{\rightarrow}$&&
\includegraphics[width=3cm]{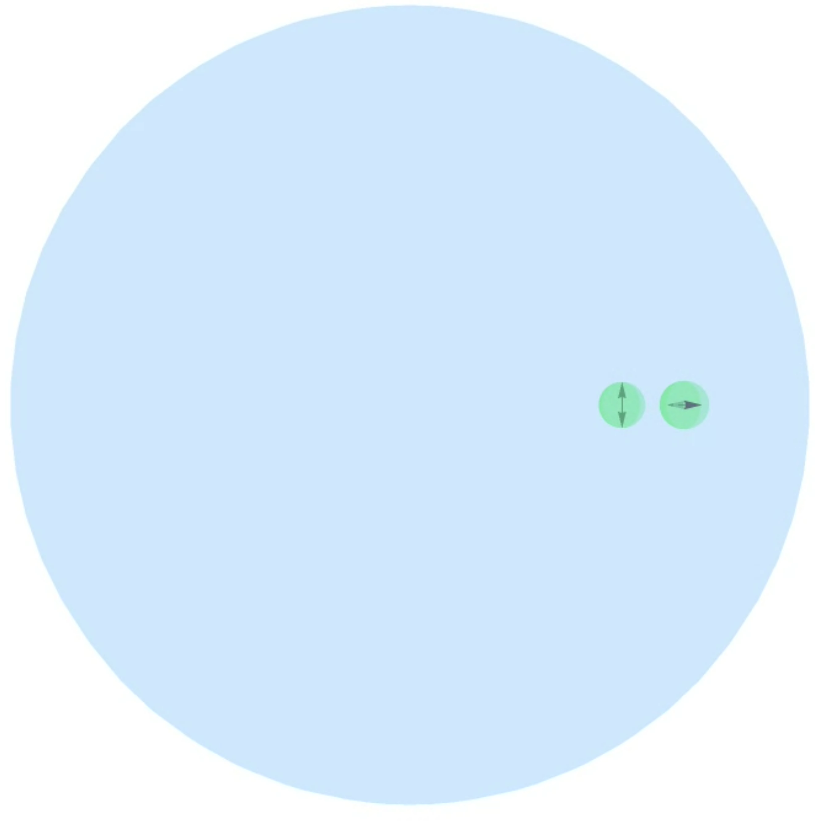}\\
\includegraphics[width=3.2cm]{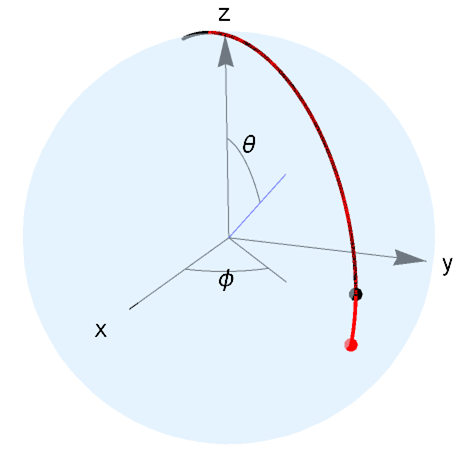}&&
\includegraphics[width=3cm]{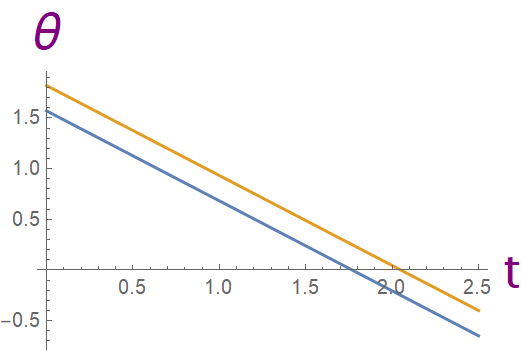}&&
\includegraphics[width=3cm]{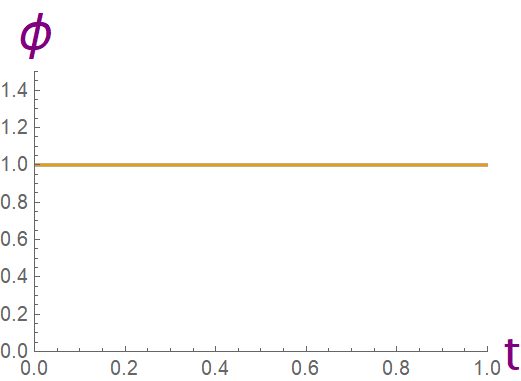}&&
\includegraphics[width=3cm]{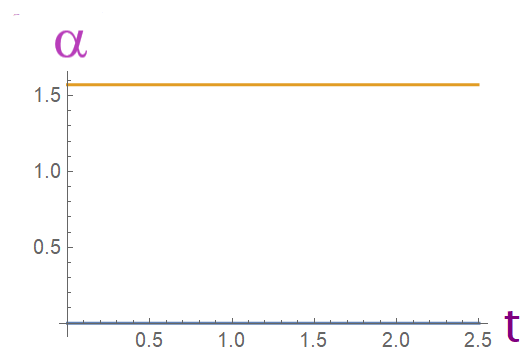}\\
\includegraphics[width=3cm]{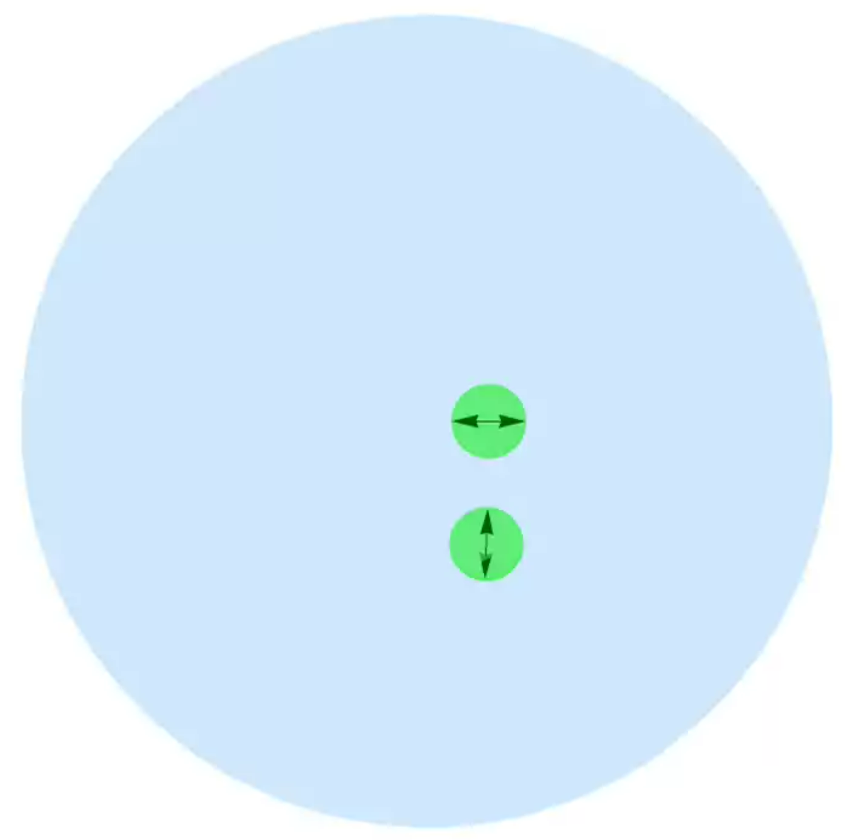}$\pmb{\rightarrow}$&&
\includegraphics[width=3cm]{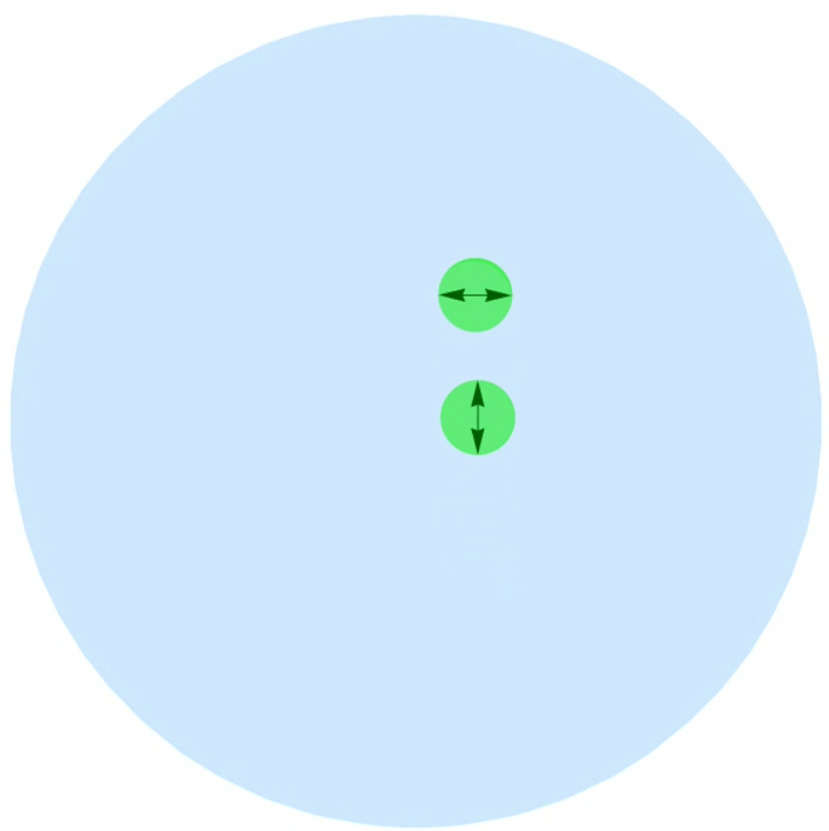}$\pmb{\rightarrow}$&&
\includegraphics[width=3cm]{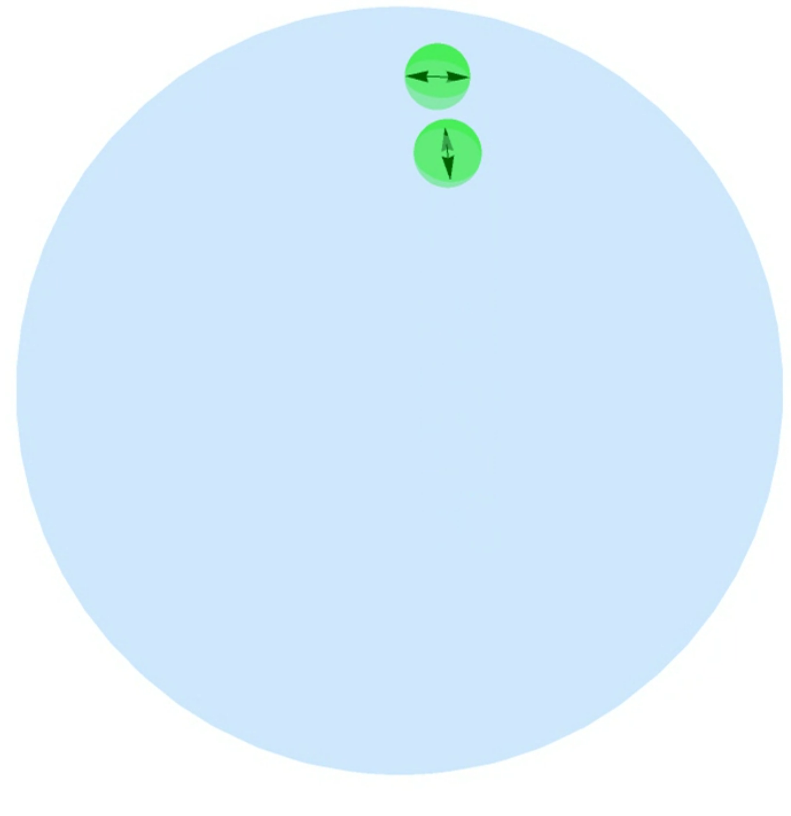}$\pmb{\rightarrow}$&&
\includegraphics[width=3cm]{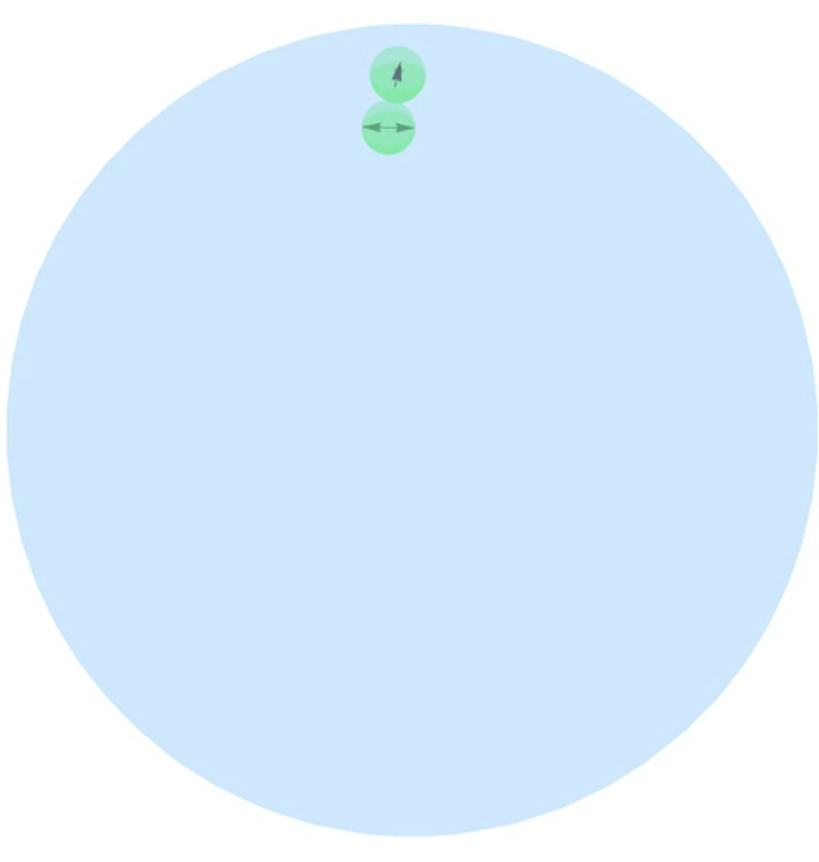}
\end{tabular}
 \caption{(Color online) Dipole Pair dynamics for mutually perpendicular orientations along the equatorial geodesic (first and second rows) and longitudinal geodesic $\phi=1$ (third amd fourth rows). For each case, in the first row we show the trace of the trajectories and the evolution of position $(\theta,\phi)$ and orientation $\alpha$ (measured wrt to $\hat{\phi}$) of the dipole pair wrt time. In the next row we display snapshots of the dynamics at different instants of time, starting with $t=0$ from the left. Time is measured in units of $\frac{\eta R^3}{\kappa}$.}
 \label{figsym} 
\end{figure}
\begin{figure}[h]
\begin{tabular}{lcccccccc}
\includegraphics[width=3cm]{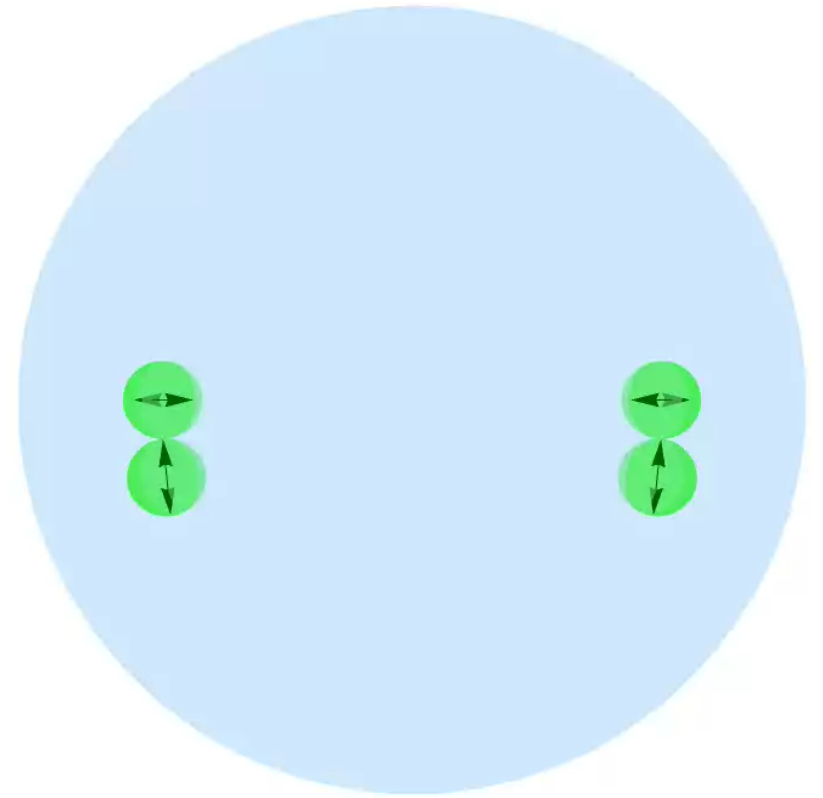}$\pmb{\rightarrow}$&&
\includegraphics[width=3cm]{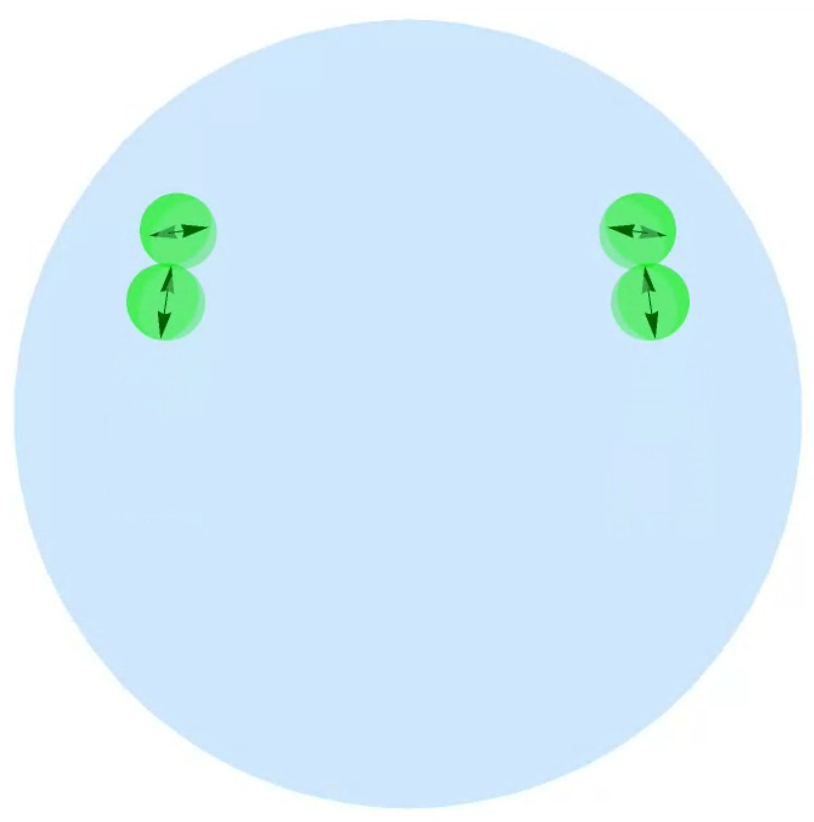}$\pmb{\rightarrow}$&&
\includegraphics[width=2.95cm]{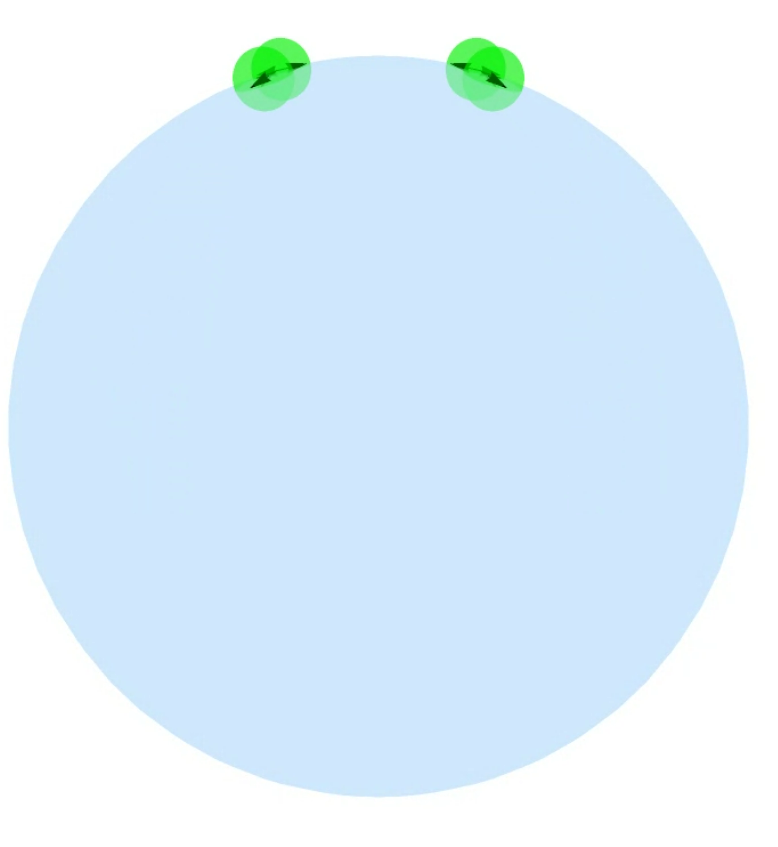}&&
\includegraphics[width=2.9cm]{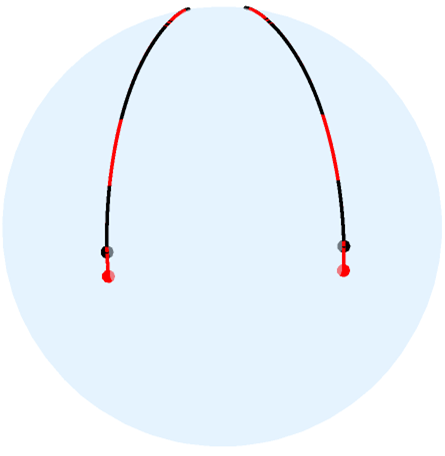}&&
\includegraphics[width=2.9cm]{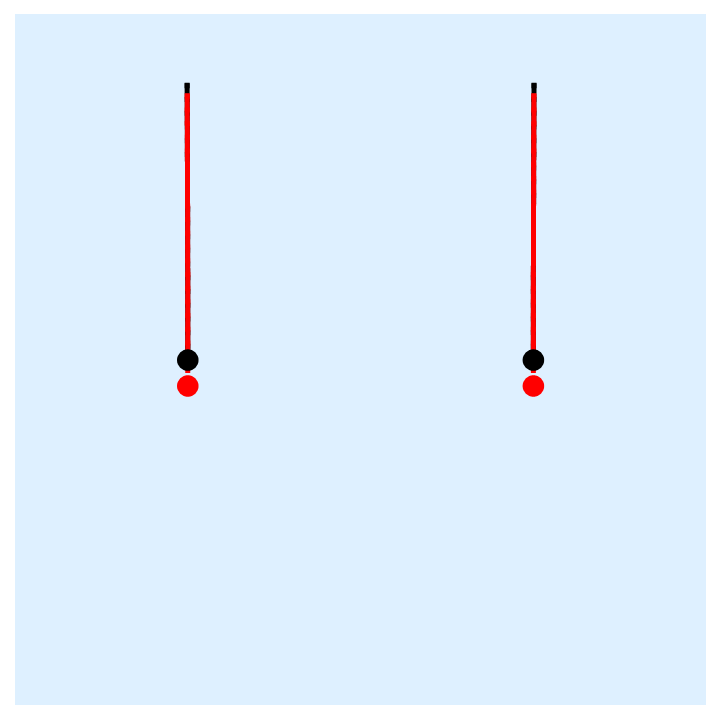}
\end{tabular}
 \caption{(Color online) Focusing action of curvature on dipole motion: Starting with an initial configuration of two dipole pairs (each pair being mutually perpendicular) and finitely separated at the equator, all the dipoles self propel towards the north pole. During this motion, curvature effects reduce the geodesic separation between the dipole pairs. This effect is absent in the flat membrane due to absence of curvature. From left to right, we have 3 snapshots of the motion, followed by a trace of the trajectories on the sphere (which shows the focusing action of curvature), followed by the corresponding  trajectories in the flat membrane. }
 \label{curv} 
\end{figure}
\begin{figure}[h]
\begin{tabular}{lcccccccc}
\includegraphics[width=3cm]{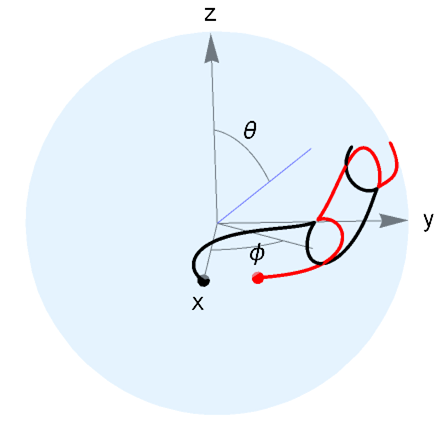}&&
\includegraphics[width=3cm]{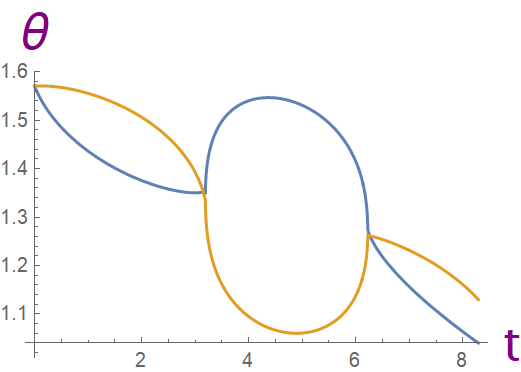}&&
\includegraphics[width=3cm]{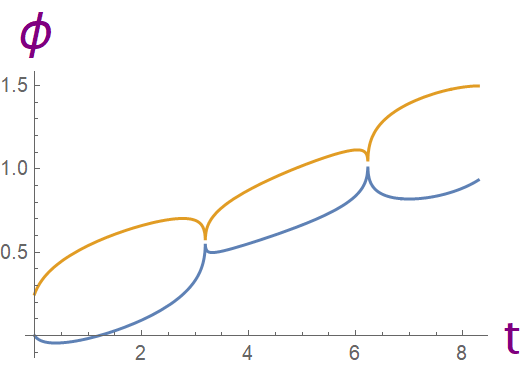}&&
\includegraphics[width=3cm]{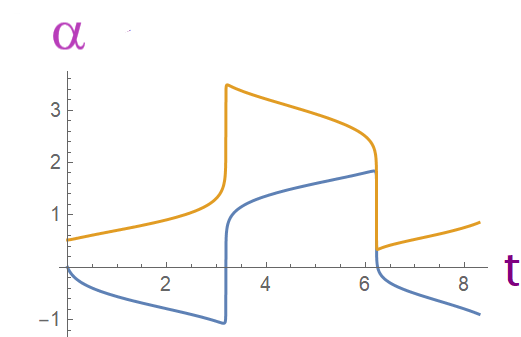}\\
\includegraphics[width=3cm]{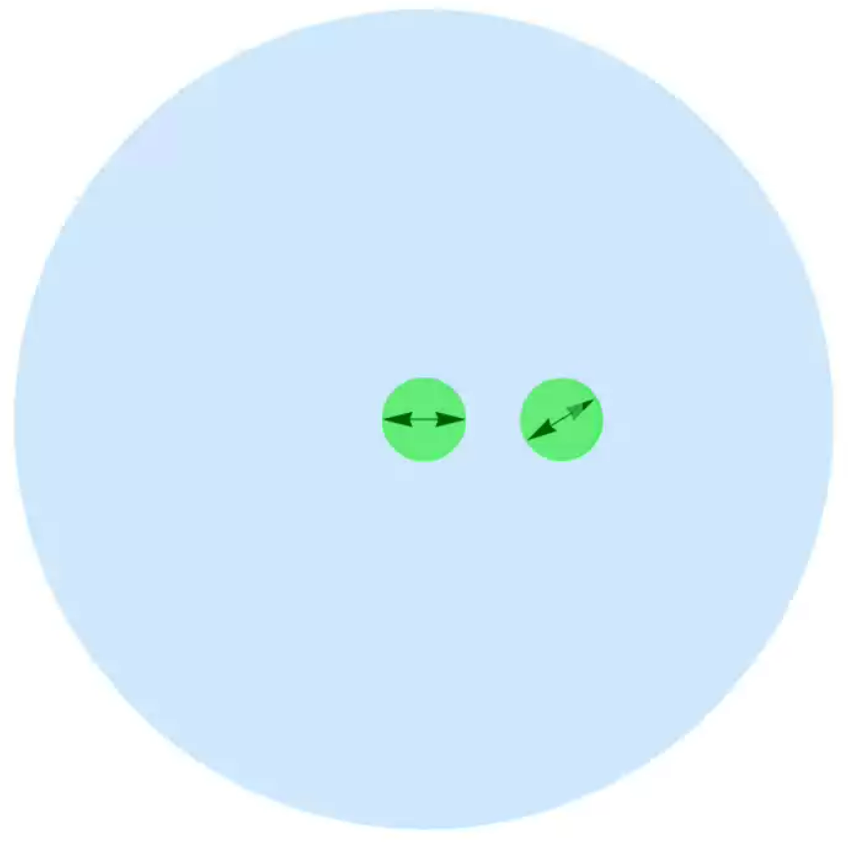}$\pmb{\rightarrow}$&&
\includegraphics[width=3cm]{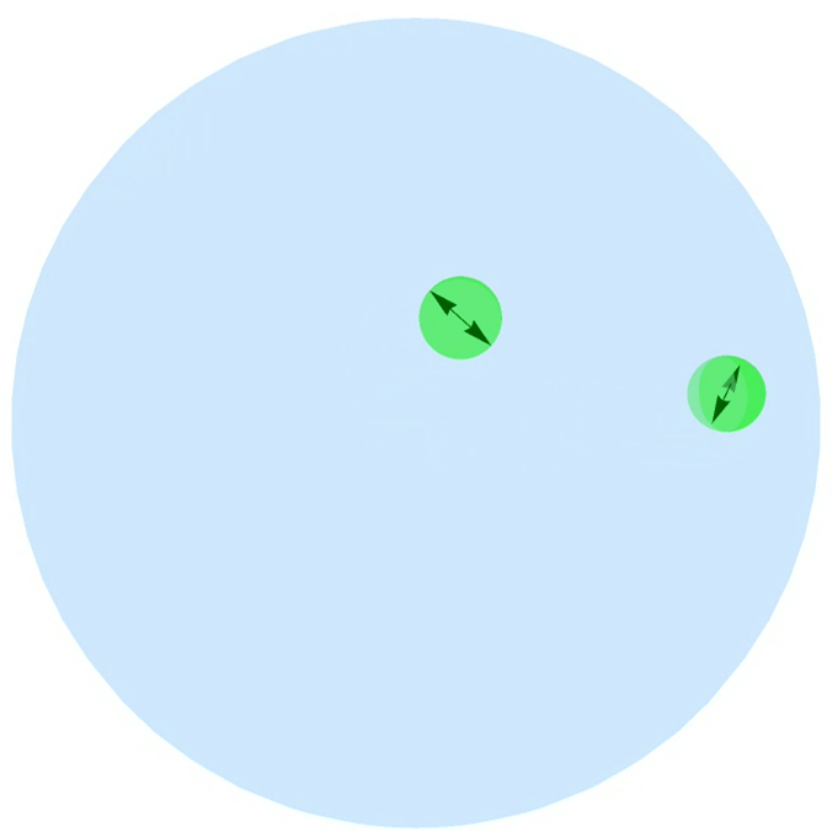}$\pmb{\rightarrow}$&&
\includegraphics[width=3cm]{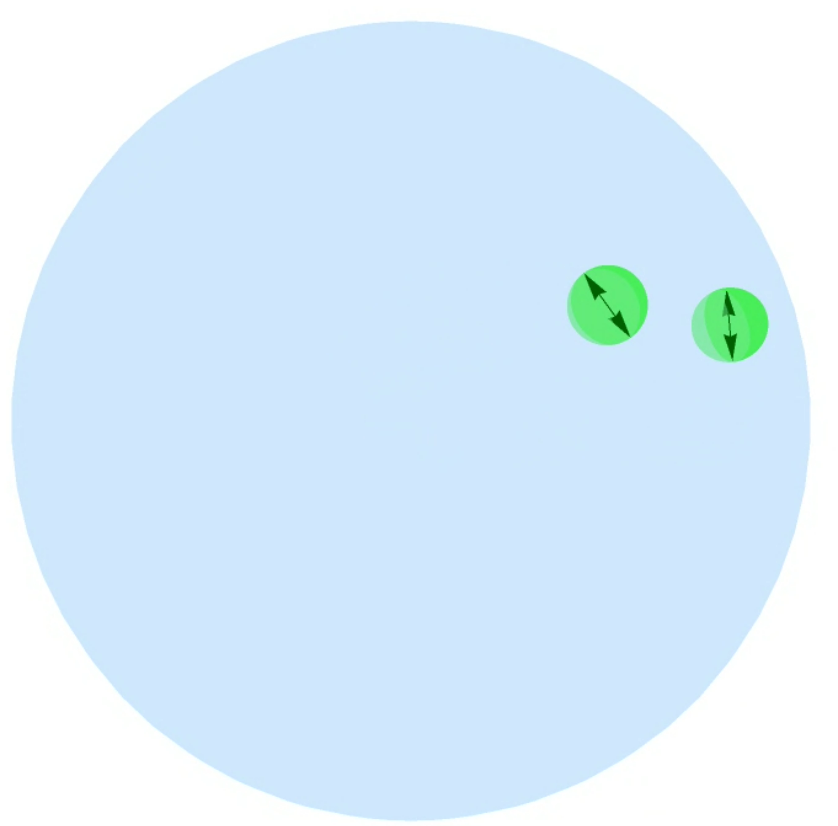}$\pmb{\rightarrow}$&&
\includegraphics[width=3cm]{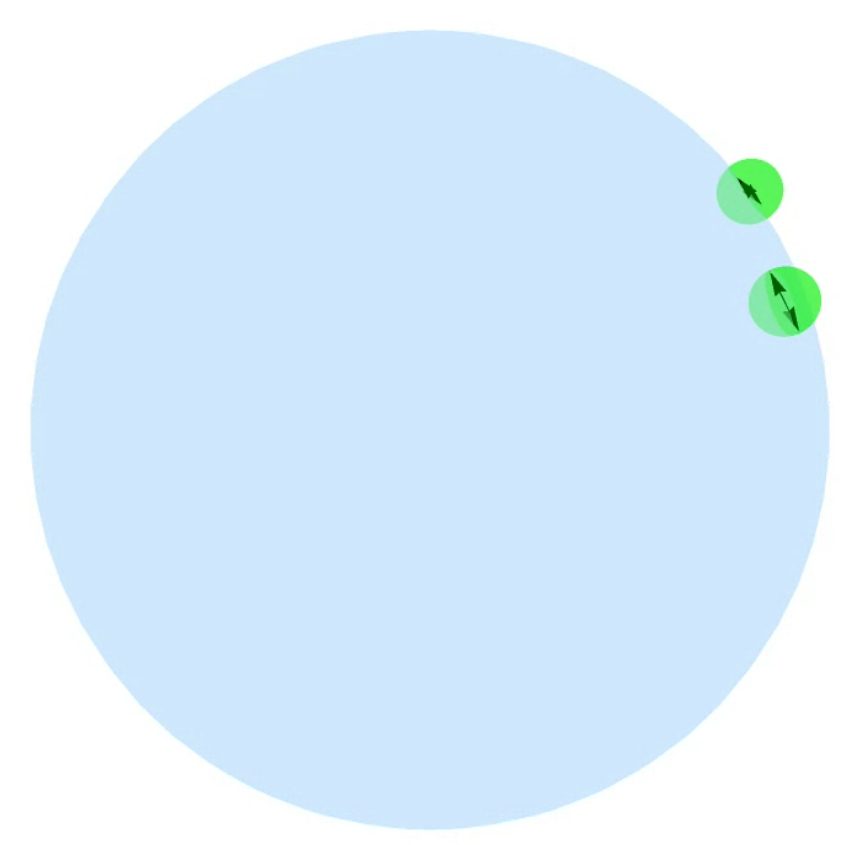}\\\\\\
\includegraphics[width=3cm]{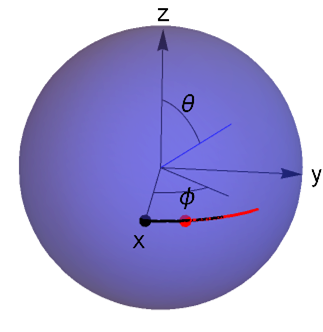}&&
\includegraphics[width=3cm]{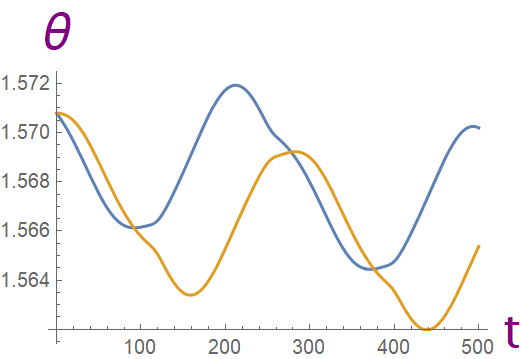}&&
\includegraphics[width=3cm]{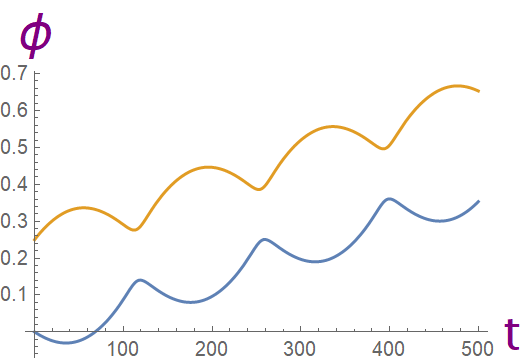}&&
\includegraphics[width=3cm]{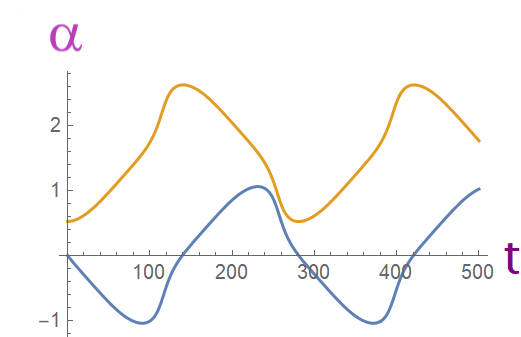}\\
\includegraphics[width=3cm]{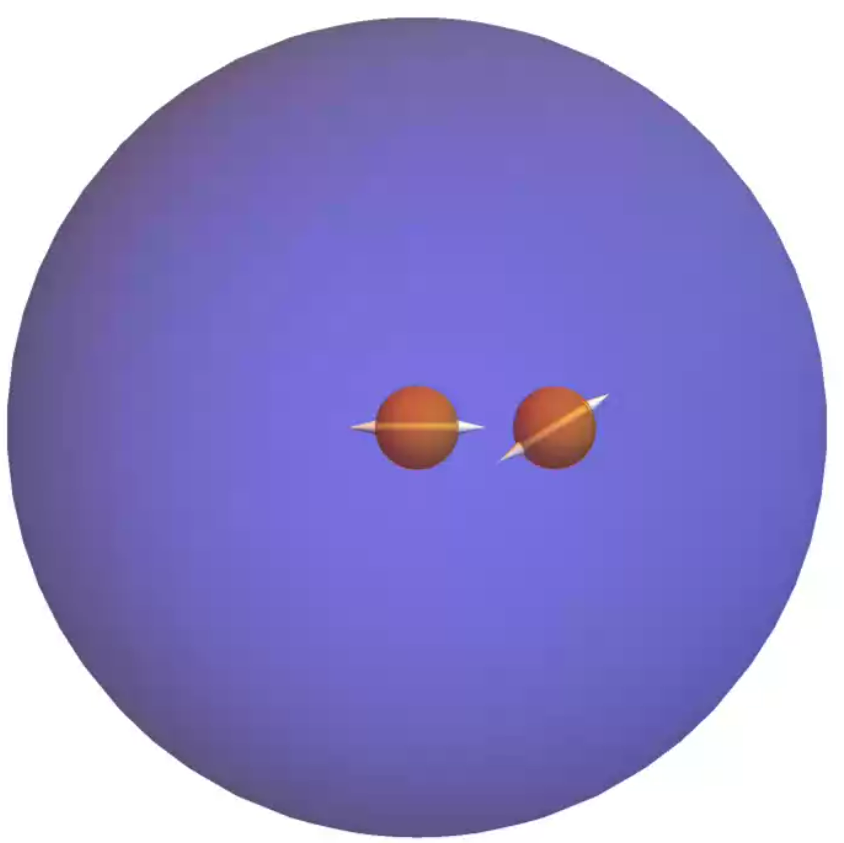}$\pmb{\rightarrow}$&&
\includegraphics[width=3cm]{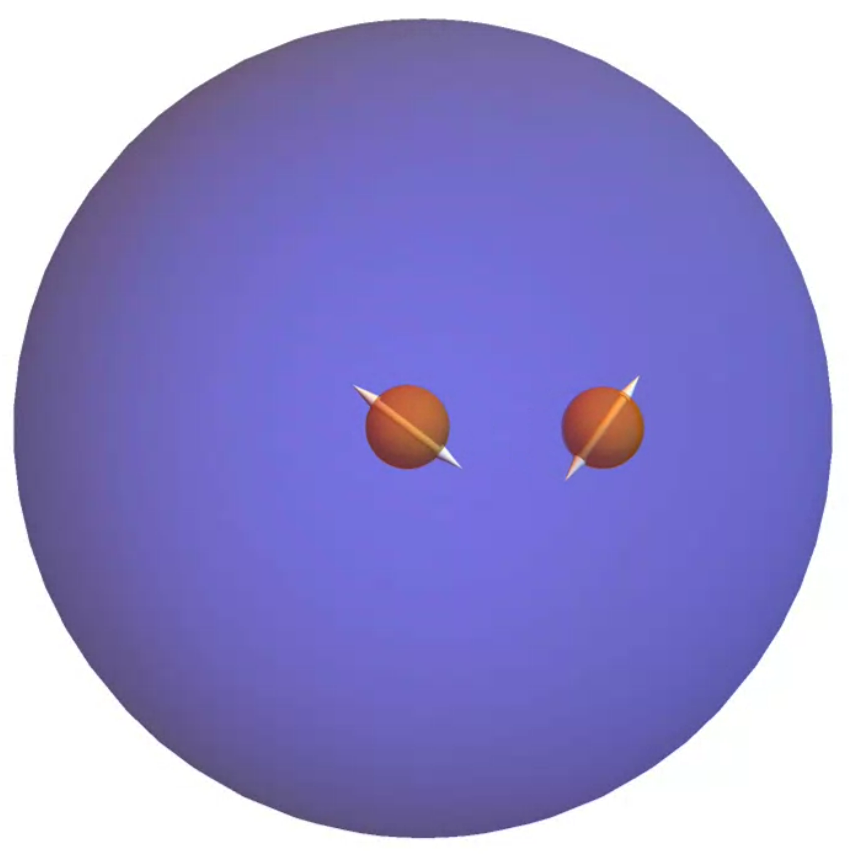}$\pmb{\rightarrow}$&&
\includegraphics[width=3cm]{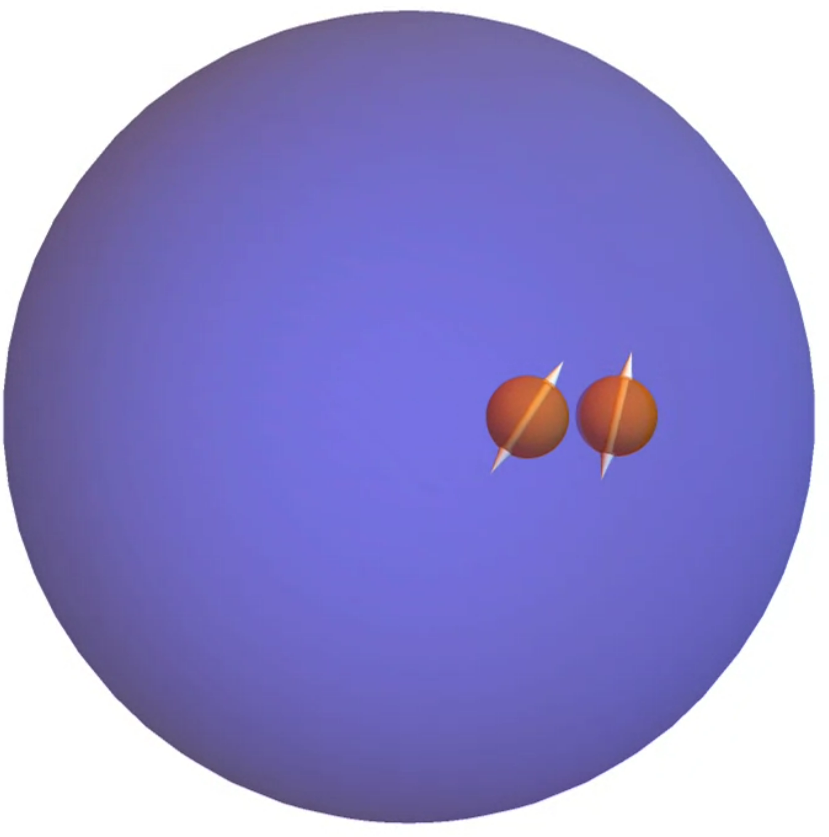}$\pmb{\rightarrow}$&&
\includegraphics[width=3cm]{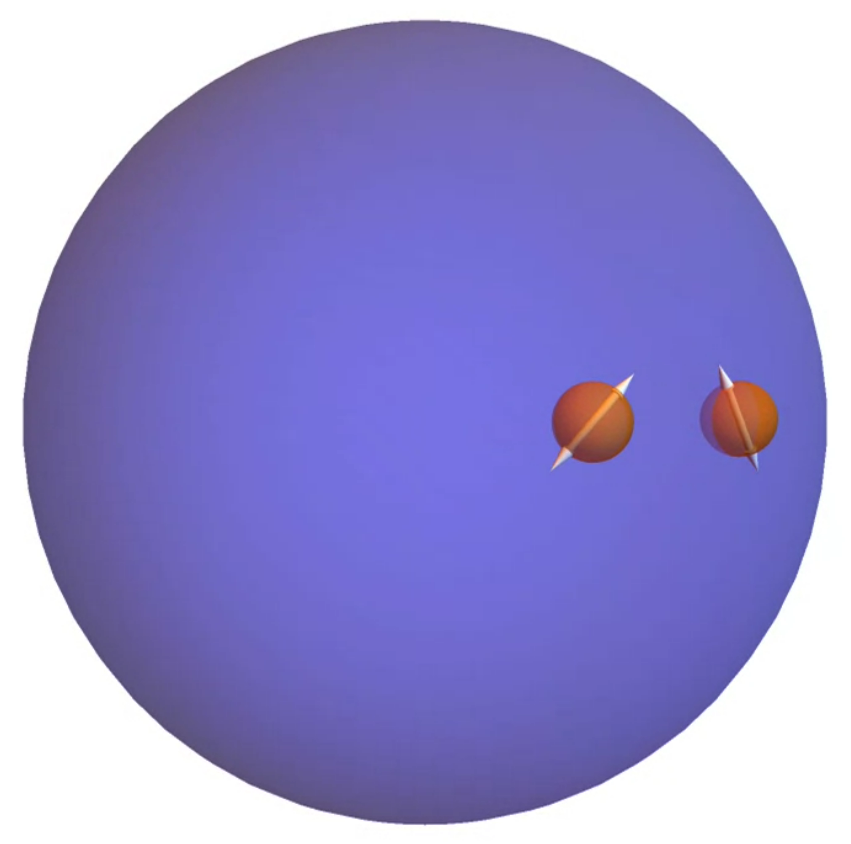}
\end{tabular}
 \caption{(Color online) Pair dynamics at low  curvature (first and second row) and high curvature (third and fourth row) without external confinement. In both regimes radius R is held fixed and membrane fluid viscosity $\eta_{2D}$ is tuned such that for low curvature $\lambda/R =0.1$, denoted by light blue color and for high curvature  $\lambda/R =100$, denoted by deep blue color. Time is measured in units of $\frac{\eta R^3}{\kappa}$. Dipole pairs in both regimes start from identical positions and relative orientation. For each regime, in the first row we show the trace of the trajectories and the evolution of position $(\theta,\phi)$ and orientation (measured wrt to $\hat{\phi}$) of the dipole pair wrt time. In the next row we display snapshots of the dynamics at different instants of time, starting with $t=0$ from the left. Both regimes feature non-linear oscillatory dynamics.}
 \label{figpairdyn} 
\end{figure}
\begin{figure}[h]
\begin{tabular}{lcccccccc}
\includegraphics[width=2cm]{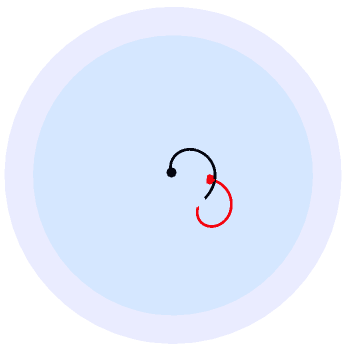}\llap{\raisebox{2cm}{\includegraphics[height=2cm]{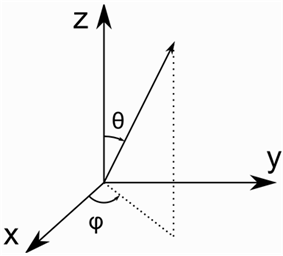}}}&&
\includegraphics[width=3cm]{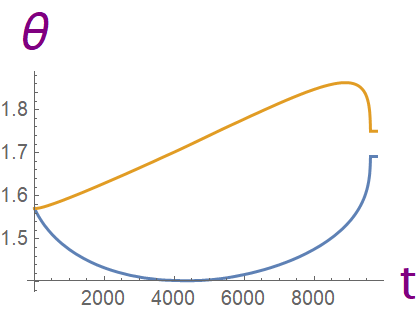}&&
\includegraphics[width=3cm]{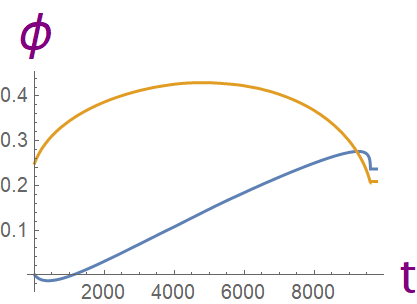}&&
\includegraphics[width=3cm]{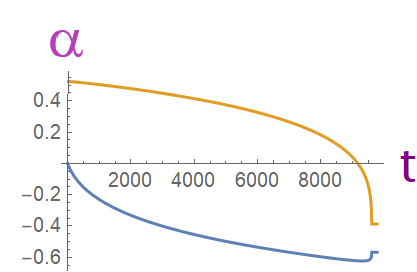}\\
\includegraphics[width=3cm]{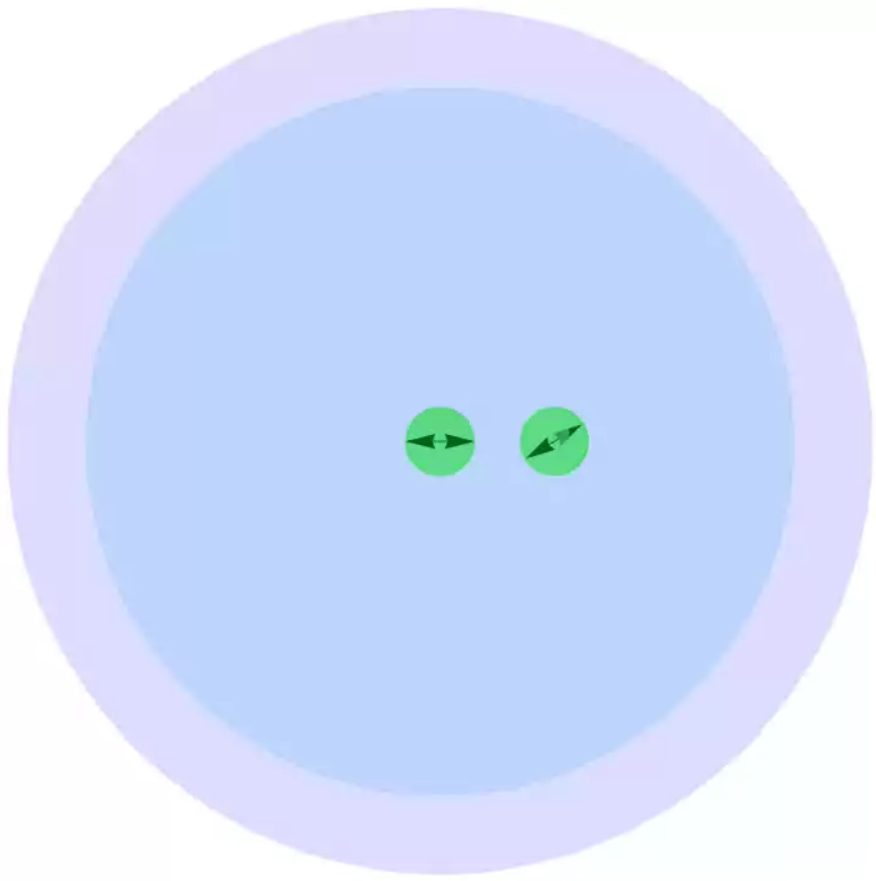}$\pmb{\rightarrow}$&&
\includegraphics[width=3cm]{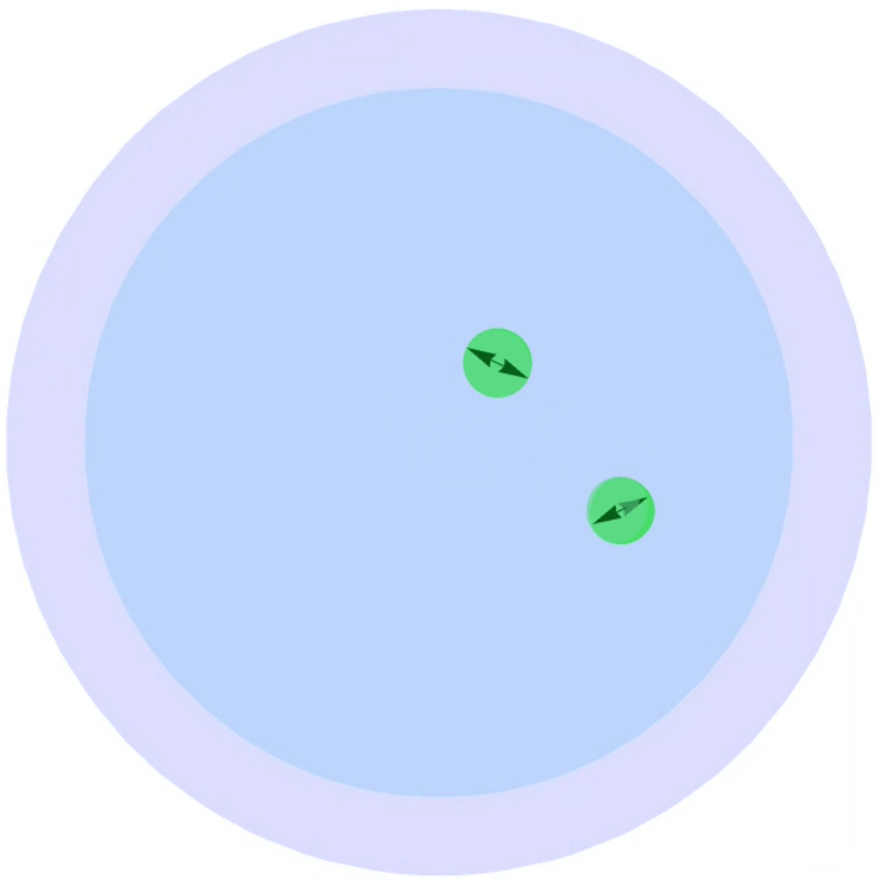}$\pmb{\rightarrow}$&&
\includegraphics[width=3cm]{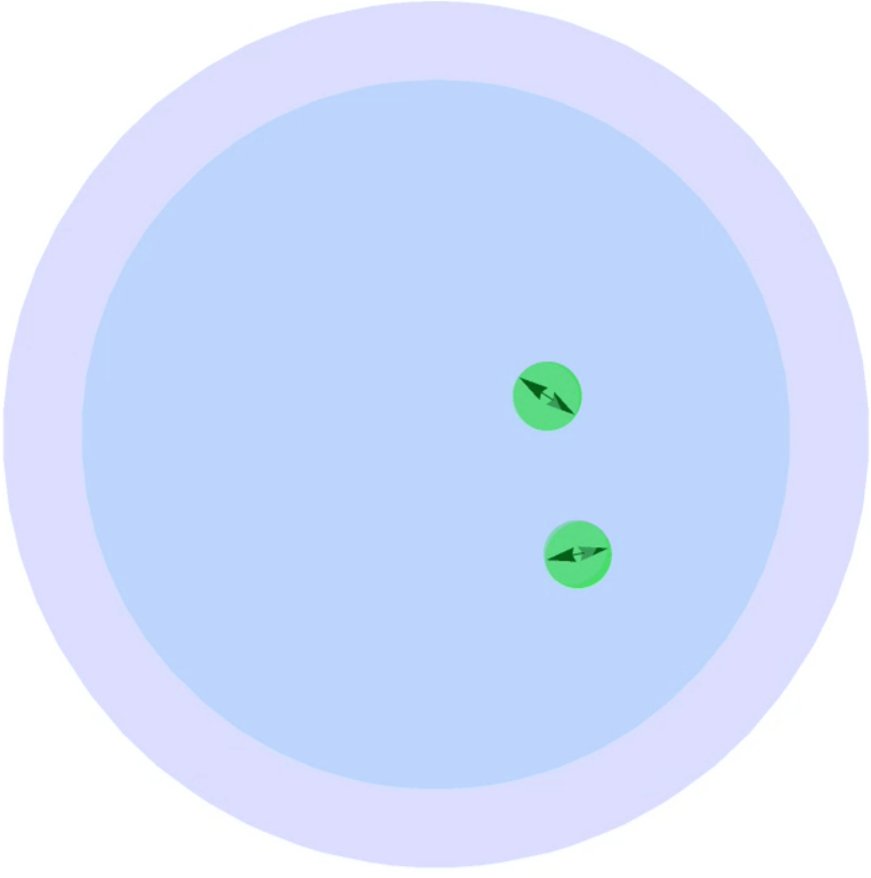}$\pmb{\rightarrow}$&&
\includegraphics[width=3cm]{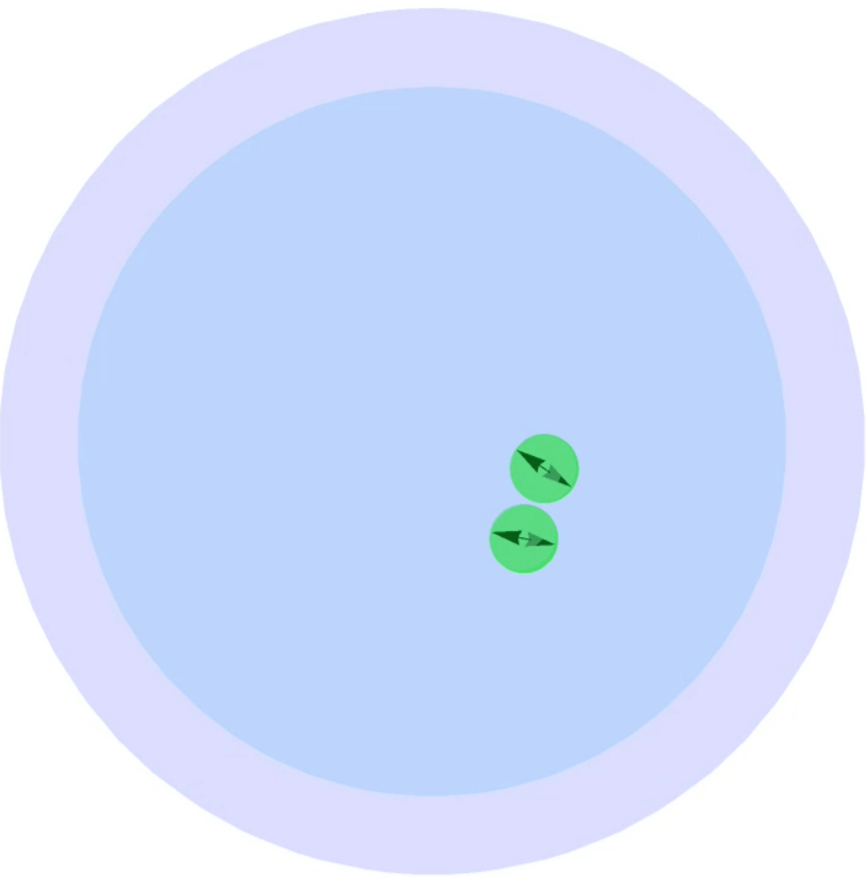}\\
\includegraphics[width=2cm]{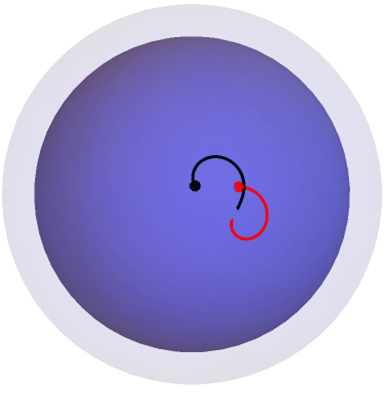}&&
\includegraphics[width=3cm]{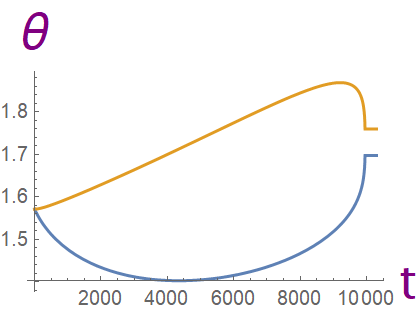}&&
\includegraphics[width=3cm]{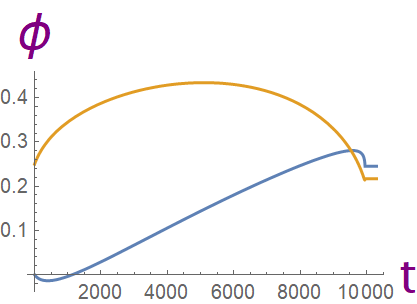}&&
\includegraphics[width=3cm]{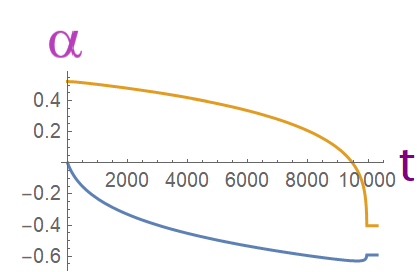}\\
\includegraphics[width=3cm]{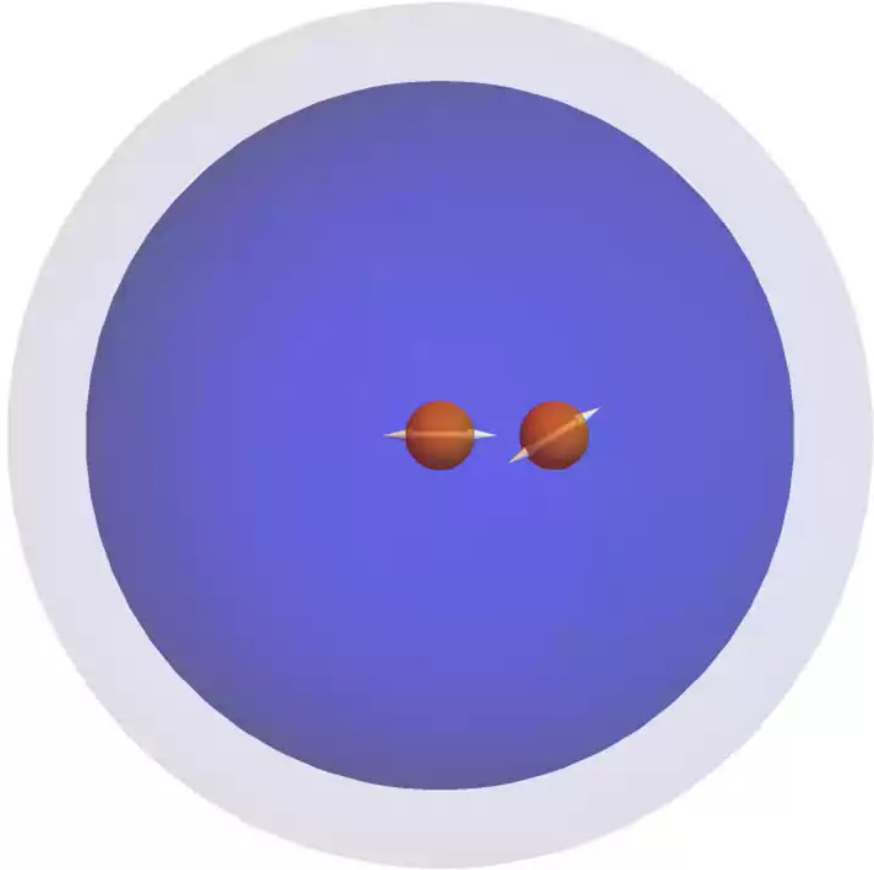}$\pmb{\rightarrow}$&&
\includegraphics[width=3cm]{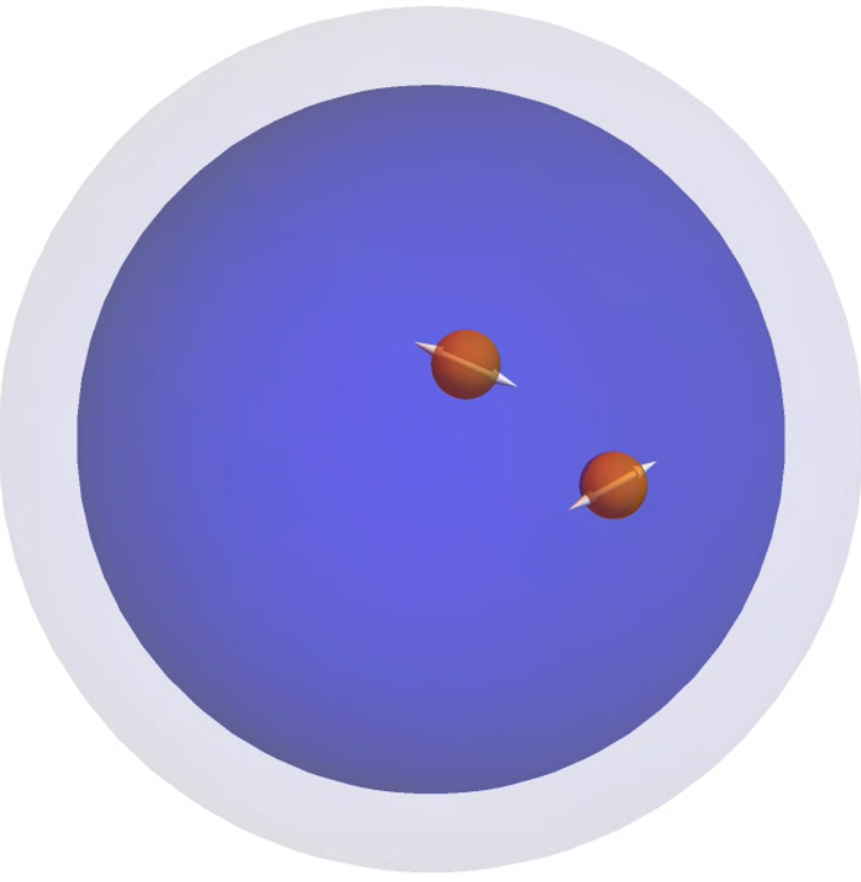}$\pmb{\rightarrow}$&&
\includegraphics[width=3cm]{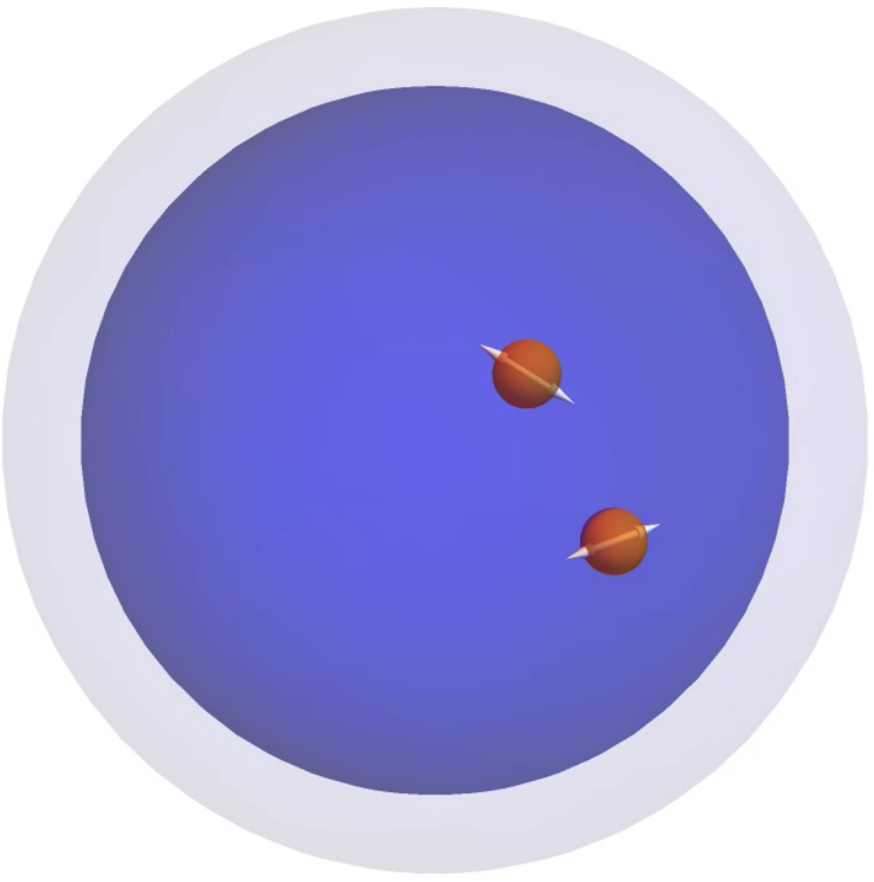}$\pmb{\rightarrow}$&&
\includegraphics[width=3cm]{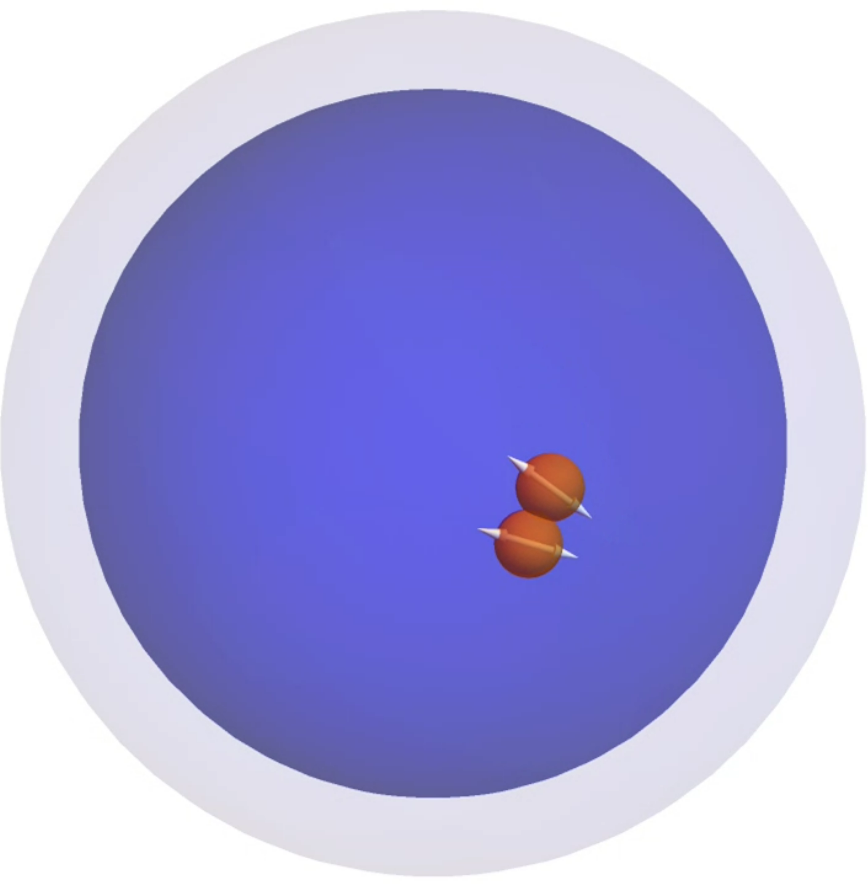}\\
\includegraphics[width=2cm]{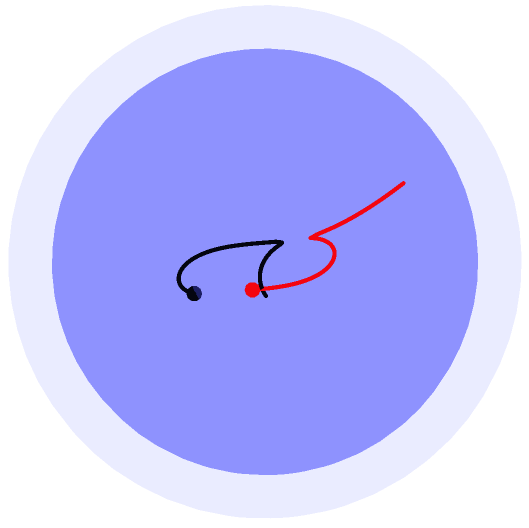}&&
\includegraphics[width=3cm]{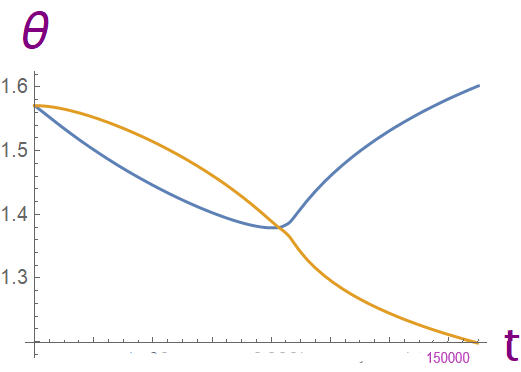}&&
\includegraphics[width=3cm]{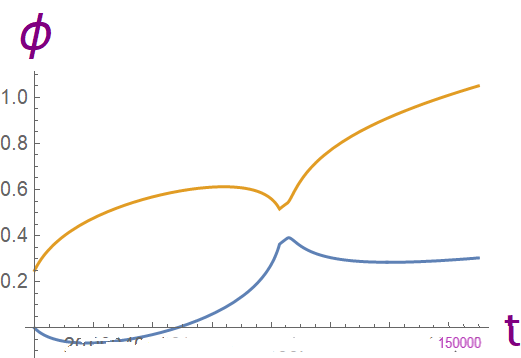}&&
\includegraphics[width=3cm]{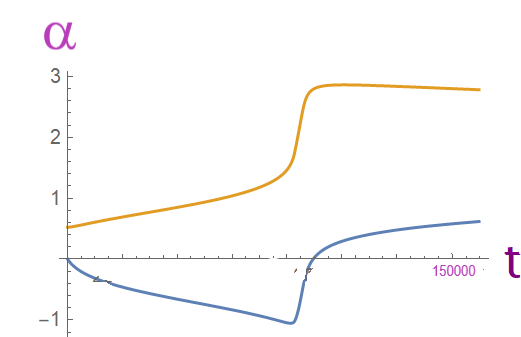}\\
\includegraphics[width=3cm]{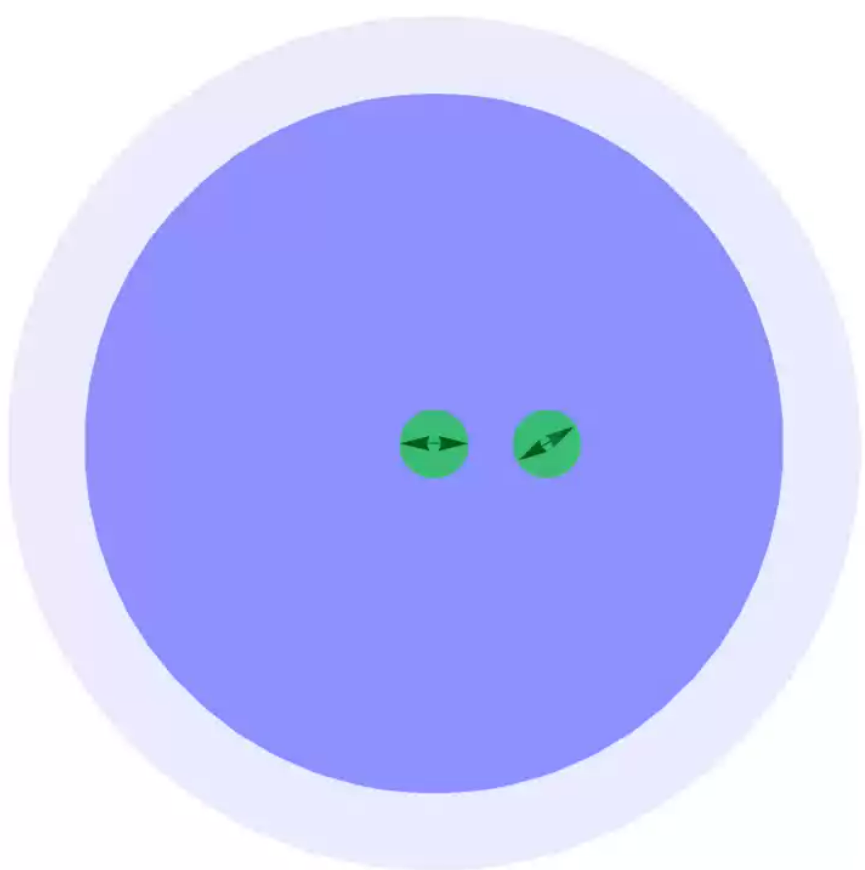}$\pmb{\rightarrow}$&&
\includegraphics[width=3cm]{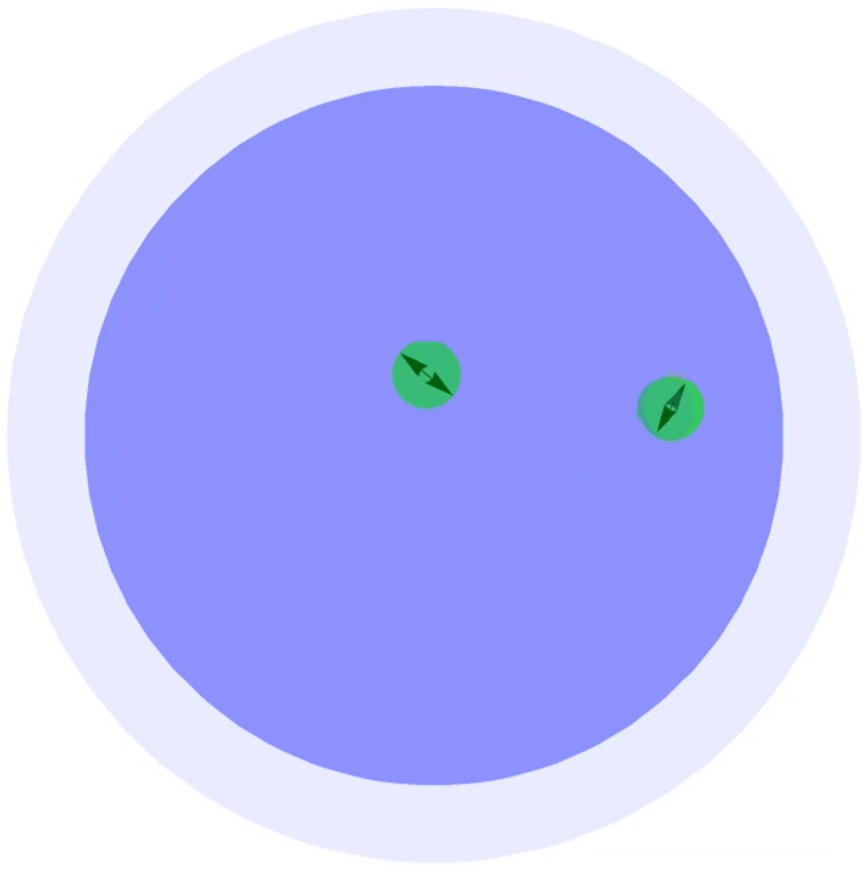}$\pmb{\rightarrow}$&&
\includegraphics[width=3cm]{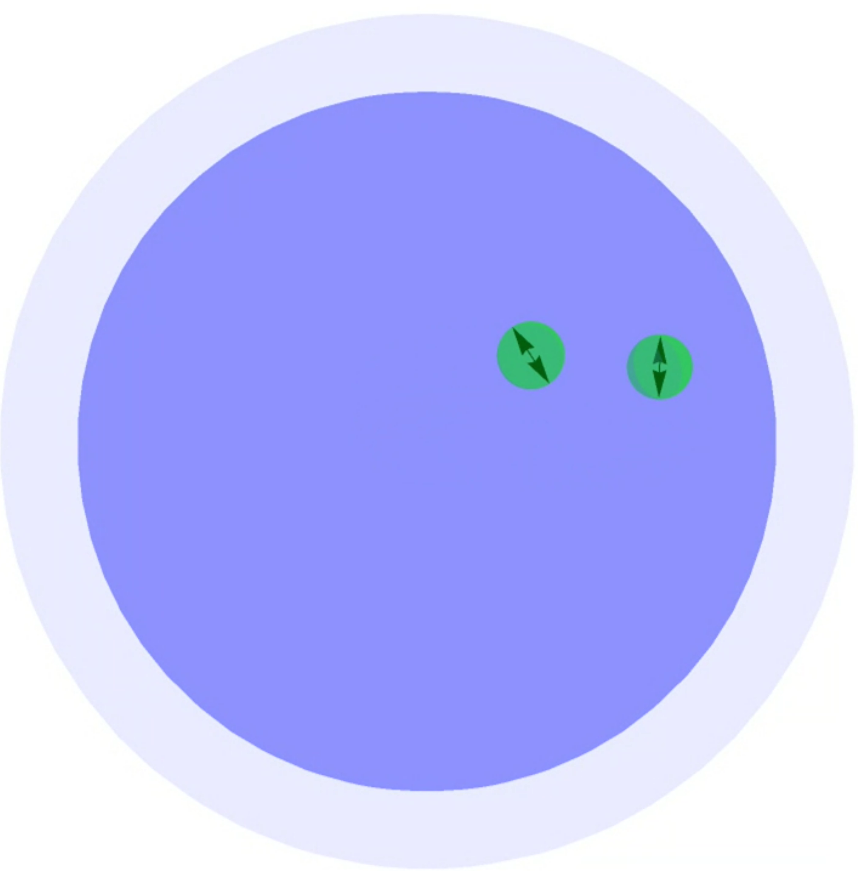}$\pmb{\rightarrow}$&&
\includegraphics[width=3cm]{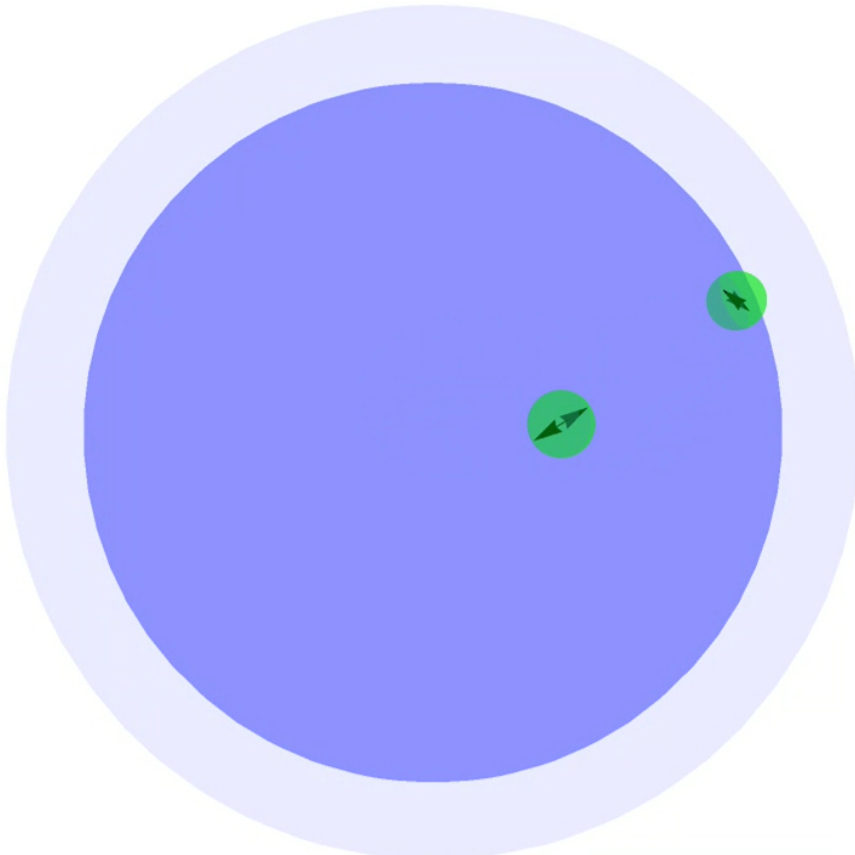}
\end{tabular}
 \caption{(Color online) Pair dynamics at low  curvature $\lambda/R =0.1$ (first and second row), high curvature $\lambda/R =100$ (third and fourth row) and very high curvature $\lambda/R =10^4$ (fifth and sixth row) with strong confinement of the external fluid. In all three regimes, radius R is held fixed and membrane fluid viscosity is tuned to achieve the required  $\lambda/R$.  Time is measured in units of $\frac{\eta R^3}{\kappa}$. Dipole pairs in all regimes start from identical positions and relative orientation as the unconfined membrane Fig.\ref{figpairdyn}. For each regime, in the first row we show the trace of the trajectories and the evolution of position $(\theta,\phi)$ and orientation (measured wrt to $\hat{\phi}$) of the dipole pair wrt time. In the next row we display snapshots of the dynamics at different instants of time, starting with $t=0$ from the left. Particle overlap is prevented by adding a soft repulsion. The dynamics for the first two regimes regimes is non-oscillatory and promotes aggregation of the dipole pair, while in the third regime dipoles fail to aggregate, even under strong confinement.}
 \label{figpairdync} 
\end{figure}

\section{Topological aspects of streamlines}
\label{strm}
Let us now briefly illustrate the topological aspects of the flows sourced by the point force dipole Eq.(\ref{vdp_f}). We plot the flow field in the $(\theta,\phi)$ chart as shown in Fig.\ref{fdp_flow}. It is clear that in both high and low curvature regimes, around the location of the point dipole we have 4 vortical defects, each of which have index $+1$, while at the core of the dipole, we have a saddle of  index $-1$, where the flow is almost radial.  This produces a net index $4-1=3$. The Poincare index theorem on the sphere thus guarantees the existence of a compensating defect of negative index (saddle) in the spherical membrane, such that the total index coincides with the Euler Characteristic of the sphere, which is two. Fig.\ref{fdp_flow} shows this new defect arising from the membrane topology and is absent in the corresponding flows in flat membranes.
\section { Pair dynamics in the spherical membrane}
\label{pair_dyn}
\textbf{Dynamical equations} : We will now subject the central formula for dipole flow Eq.(\ref{vdp_f}) to several tests. For this purpose, we simulate the hydrodynamic interactions of a pair of point force dipoles, each generating the flow described by Eq.(\ref{vdp_f}). For a generic system of N point inclusions, the hydrodynamic interactions are such that each inclusion undergoes a translation and rotation according to the sum of the flow disturbances from remaining inclusions. The rotation rate of the orientation of a given force dipole  is determined by the local vorticity of the flows created by the remaining force dipole inclusions.\footnote { One can also model the rotation rates differently for inclusions of different shapes. Here we choose the simplest model of a probe particle spinning about a vertical axis, whose  angular velocity is proportional to the local vorticity of the fluid velocity field. } In addition,  we focus on the regime where hydrodynamic interactions dominate over thermal fluctuations, hence for simplicity, we ignore the Brownian motion. In such situations, the dynamical equations for an inclusion with coordinates ($\theta_i,\phi_i$)  are given by 
\beqa
&R ~ \dot{\theta}_i=  \sum_{j \neq i}^N  \bm{v}_{\theta}, 
\hspace{1cm}R \sin \theta_i~ \dot{\phi}_i=  \sum_{j \neq i}^N   \bm{v}_{\phi}\nn\\
&\dot{\alpha}_i=  \sum_{j \neq i}^N \underbrace{\frac{1}{2} \left(\nabla^{sp} \times \bm{v}\right).\hat{r}}_{\text{Vorticity induced rotation}} + \underbrace{\frac{1}{R} \cot \theta_i ~ \bm{v}_\phi}_{ \text{Curvature induced rotation}}
\label{dynmeq}
\eeqa

\begin{figure}
\begin{tabular}{lcc}
\includegraphics[width=4cm]{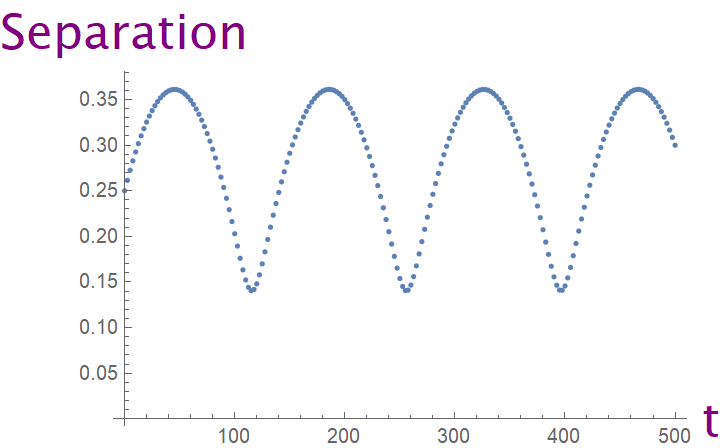}&&
\includegraphics[width=4cm]{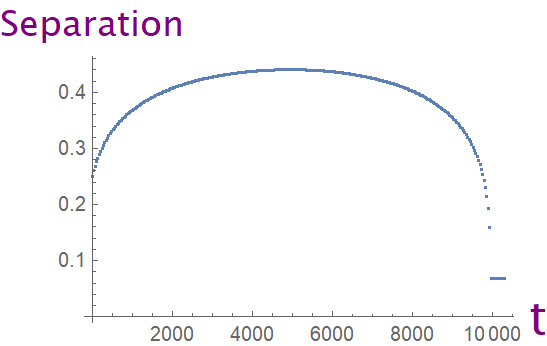}
\end{tabular}
 \caption{(Color online) Inter-particle geodesic separation of the dipole pair wrt time in the high curvature regime ($\lambda/R =100$) for the unconfined membrane (left) and the  strongly confined membrane (right). Time is measured in units of $\frac{\eta R^3}{\kappa}$ and separation is in units of membrane radius R.}
 \label{distcmp_sp} 
\end{figure}

%\beqa
%&R ~ \dot{\theta}_i=  \sum_{j \neq i}^N  \bm{v}_{\theta}, 
%\hspace{1cm}R \sin \theta_i~ \dot{\phi}_i=  \sum_{j \neq i}^N   \bm{v}_{\phi}\nn\\
%&\dot{\alpha}_i=  \sum_{j \neq i}^N \underbrace{\frac{1}{2} \left(\nabla^{sp} \times \bm{v}\right).\hat{r}}_{\text{Vorticity induced rotation}} + \underbrace{\frac{1}{R} \cot \theta_i ~ \bm{v}_\phi}_{ \text{Curvature induced rotation}}
%\label{dynmeq}
%\eeqa
where $\bm{v}_\theta$ and $\bm{v}_\phi$ are the $\theta$ and $\phi$ components of the flow Eq.(\ref{vdp_f}) for the unconfined case and Eq.(\ref{conf_vdp_f}) for the confined situation.  The rate of change of orientation of the i-th dipole, denoted by $\dot{\alpha}_i$, is given by two contributions, as shown in the last line of Eq.(\ref{dynmeq}). The first contribution is from the local vorticity of the flows created by remaining dipoles, denoted by $(\nabla^{sp}\times \bm{v})$ which is the curl of the velocity field in spherical coordinates. The second contribution comes from the curvature\footnote{ The curvature contribution for the i-th dipole can be derived by using a relation similar to Eq.\ref{rotf}: $ \delta \alpha_i = \cot \theta_i \cos \left(\arctan \frac{{\bm{v}_\theta}_i}{{\bm{v}_\phi}_i} \right) \frac{\delta L}{R}$}.\\\\ 
\textbf{Comments on numerics and figures}:  Several remarks about the simulations are in order:  In this paper, we will focus on dipoles with positive strength, in the spirit of ``pushers" studied in the context of microswimming Ref.\cite{lauga2009}. We recall that we have set $\eta_+ =\eta_- \equiv \eta$ ie. the external fluids outside and inside the membrane have same viscosity, leading to a unique Saffman length $\lambda$ ie. $\lambda_+= \lambda_-=\frac{\eta_{2D}}{\eta} \equiv \lambda$. We will be working in units where the dipole strength $\kappa$, the radius R and the 3D viscosity of the external fluids $\eta$ are all set to unity ie. $ \kappa=1, R=1, \eta=1$. Time is measured in units of $\frac{\eta R^3}{\kappa}$.   The low and high curvature regimes arise when $\lambda/R \ll 1 $ and $\lambda/R \gg 1 $ respectively and this is achieved by tuning the viscosity of the membrane fluid $\eta_{2D}$, hence tuning the Saffman length $\lambda$. Fig.\ref{figsym} and Fig.\ref{figpairdyn} illustrate  the interaction of dipole pairs in the unconfined situation where the external fluid outside the sphere extends to infinity. For the low curvature regime, we choose $\lambda =0.1$, denoted by light blue color (represents a less viscous membrane fluid).  For the high curvature regime we choose $\lambda =100$, denoted by deep blue color (represents a more viscous membrane fluid). We will also explore dynamics where the external fluid outside the membrane is confined to a certain radius $(R+H)$, see Fig.\ref{figpairdync}, with the same color codes. In order to probe the effects of strong confinement, we choose $H =10^{-5}$ (in units of R) for both the high and low curvature regimes. Also, we add a soft harmonic repulsion to Eq.(\ref{dynmeq}) to prevent particle overlap in Fig.\ref{figpairdync}. Simulations are performed using Mathematica \cite{wolfram} and Julia \cite{jlsym,mki}.  \\\\
\textbf{Symmetry Check}: As a first check of Eq.({\ref{vdp_f}}), keeping the curvature arbitrary, we consider a pair of mutually perpendicular force dipoles, both situated on the equatorial geodesic. One can check that the rotation rate $\dot{\alpha}$ vanishes for either dipole in this configuration, where $\alpha$ denotes the orientation of the dipoles defined in Eq.(\ref{Tdef}). The dipoles move together along the geodesic with a fixed separation between them. This is illustrated in the first two rows of Fig.\ref{figsym}. The first row shows the trace of the trajectories and the variation of location and orientation of the two dipoles wrt time. The second row shows snapshots of the dynamics at different instants of time starting from $t=0$. Using spherical symmetry, a similar dynamics must emerge when the pair of mutually perpendicular  dipoles move along any arbitrary geodesic. We demonstrate this in third and fourth row of Fig.\ref{figsym} for the case of a longitudinal geodesic. Let us note that without the extra contribution coming from curvature of the membrane, as explained below Eq.(\ref{vdp_f}), this symmetry does not work. Thus, the curvature contribution must be taken into account while  constructing the dipole flows in the spherical membrane and for any curved membrane in general. For example, Fig.\ref{curv} illustrates the focusing action of curvature. Starting from a configuration of two dipole pairs (each pair being mutually perpendicular, see Fig.\ref{curv}), we find that all dipoles propel towards the north pole, with a corresponding decrease in the geodesic separation between the dipole pairs.  In general, curvature effects will become important for the dynamics of larger collection of dipoles distributed across the membrane.  \\\\
\textbf{Dynamics under weak/no confinement}: Let us now proceed to discuss the pair dynamics in the regimes of low and high curvature in the unconfined situation, Fig.\ref{figpairdyn}. In both regimes we have non-linear oscillatory dynamics where the dipoles are attracted and then repelled periodically. This can be understood from the finite rotation rate $\dot{\alpha}$ of the dipole orientations as dictated by Eq.(\ref{dynmeq}). First, we note that in Fig.\ref{fdp_flow}, the streamlines generated by a single dipole oriented along $\hat{\theta}$ (pusher) is repulsive along the orientation of the dipole  (along $\hat{\theta}$) while attractive along the orthogonal directions (around $\hat{\phi}$) . Going to the frame of the first dipole (which is generating the flow as shown in Fig.\ref{fdp_flow}), the second dipole situated in the far-field region initially follows the attractive streamlines around $\hat{\phi}$  and starts approaching the first dipole. However, because of the non-zero rotation rate $\dot{\alpha} \neq 0$ (Eq.(\ref{dynmeq})), the relative orientation of the dipoles changes with time (the dipoles get more aligned as they approach each other) and the second dipole starts facing the repulsive streamlines generated from the first dipole along the other direction (around $\hat{\theta}$ in Fig.\ref{fdp_flow}), thus it is pushed away wrt the first dipole. Once repelled, the relative orientations change ($\dot{\alpha} \neq 0$) and once again favour attraction. This process continues periodically and leads to the oscillatory dynamics in Fig.\ref{figpairdyn} in both low and high curvature regimes, see also the left plot in Fig.\ref{distcmp_sp} where we show how the inter-particle separation oscillates with time in the high curvature case. Let us note that in the low curvature regime, the streamlines have significant azimuthal component (non-radial character) compared to the high curvature regime (Fig.\ref{fdp_flow}).  As a result, the dipole pair undergoes a significant drift in their trajectories during the non-linear oscillations in the low curvature regime. In contrast, in the high curvature regime, since $\lambda \gg R$, the near field radial streamlines (lacking azimuthal character) are more dominant and the dipoles undergo very small drift in their trajectories. Thus, the dipoles explore more surface area of the membrane at low curvatures (first and second rows of Fig.\ref{figpairdyn}) compared to that at high curvatures (third and fourth rows of Fig. \ref{figpairdyn}).  This can be controlled by tuning the fluid viscosities, and hence $\lambda$. Note also that in the high curvature regime the membrane fluid is more viscous, hence from Eq.(\ref{vdp_f}) we see that the dynamical time scales are much larger in the high curvature regime due to reduced mobility. The dynamics we observe in Fig.\ref{figpairdyn} is also consistent with what one expects from the knowledge of the dynamics in flat membranes. For example, the dynamics and trajectories observed at low curvature (small Saffman length $\lambda$) in the curved membrane are consistent with Fig.\ref{azplane} in Appendix \ref{plane_summary} where we have carried out similar simulations in flat membranes at small Saffman length $\lambda$. Meanwhile, the high curvature regime corresponds to larger Saffman length $\lambda$ compared to the sphere radius R, hence the dynamics in this regime closely resembles Fig.\ref{radplane} in Appendix \ref{plane_summary}, carried out at large values of Saffman length $\lambda$.
Let us also mention that at high curvature, the compact geometry of the membrane gives rise to a zero mode that induces a global rotation Ref.\cite{henlev2008,henlev2010} where  the spherical membrane and  internal fluid rotate uniformly as a rigid body. This is clear from the flow sourced by a Stokeslet, please see the second row of Fig.\ref{st} where the dominant flow is a global rotation of the fluid. However, in a force dipole the global rotation generated by the two oppositely oriented forces in a dipole cancel each other and have no effect on the streamlines and dynamics. On a technical note, performing force dipole simulations in closed surfaces without boundaries (such as the sphere) removes possible subtleties associated with PBC (periodic boundary conditions), see for example Ref.\cite{Graaf}. Lastly, it is worth mentioning that one recovers the same dynamics as illustrated in  Fig.\ref{figpairdyn} in weakly confined situations using the confined flow Eq.(\ref{conf_vdp_f}) instead of the unconfined flow Eq.(\ref{vdp_f}) for sufficiently large values of $H$.  \\\\
\textbf{Dynamics under strong confinement} We now consider situations where the external fluid outside the membrane is confined to a thin region $R <r < (R+H)$ such that $H \ll \lambda$. The intuition from flat supported membranes (see Appendix \ref{plane_summary}) and a similar analysis of Eq.(\ref{slc_conf}) suggests that in this regime the additional confinement scale $H$ effectively leads to a screened Saffman length $\lambda_c =\sqrt{\lambda H} \ll \lambda$ for sufficiently small $H$. Beyond the screened Saffman length  $\lambda_c$, the flow is dominated by the traction stress from the confined layer of external fluid, with significant azimuthal component (non-radial character). A unique feature of this flow is that it results in a much slower rotation rate $\dot{\alpha}$ for the hydrodynamic interactions of a dipole pair compared to the unconfined situation (The planar limit of this flow is \textit{vorticity free} in the limit $H \rightarrow 0$, see Eq.(\ref{vdp_conf}) in Appendix \ref{plane_summary}). Thus, in contrast to the unconfined situation, here the second dipole is attracted towards the first dipole but now with much slower change in relative orientation. This results in a remarkably different non-oscillatory dynamics and promotes aggregation of the dipole pair, in both regimes of low and high curvature.  However, even in the confined situation, below the screened Saffman length $\lambda_c$, the near field streamlines are governed by the in-plane membrane stress and generate significant rotation rate for the dipoles, thus re-introducing the oscillatory dynamics that we saw in Fig.\ref{figpairdyn}, hence disfavouring aggregation of the dipoles. Nevertheless, by choosing a sufficiently small value of $H$ (strong confinement), the screened Saffman length $\lambda_c$ can be tuned to be of the order of the inclusion size or smaller, such that the near field radial flows play no role in the dynamics. This also ensures that the effects of curvature on the dynamics essentially vanishes, as long as the curvature is not too high. However, extreme values of curvature will re-introduce the near field radial flows, leading to a faster rotation rate and hence destroying the aggregation of dipoles. The resulting dynamics is presented in Fig.\ref{figpairdync} for 3 regimes of curvature : low ($\lambda/R =0.1$) , high ($\lambda/R =100$) and very high ($\lambda/R =10^4$).  The time evolution is carried out using Eq.(\ref{conf_vdp_f}). For ease of comparison, we use the same initial separation and orientations for the dipole pair in both the unconfined (Fig.\ref{figpairdyn}) and confined (Fig.\ref{figpairdync}) situations. In the regimes of low curvature ($\lambda/R =0.1$) and high curvature ($\lambda/R =100$), we  observe non-oscillatory dynamics favouring aggregation of the dipole pair, while very high curvature ($\lambda/R =10^4$) destroys dipole aggregation .  We compare the time evolution of inter-particle geodesic distance in Fig.\ref{distcmp_sp} and also carry out the aggregation of 3 dipoles in Fig.\ref{three_inc} in the low curvature regime (similar dynamics happens for the high curvature regime as well). Our study thus suggests aggregation effects of dipoles in curved biological membranes in regimes of both low and high curvature by suitably engineering the confinement scale $H$, Fig.\ref{figpairdync}, provided the curvature is not too high. It will be interesting to carry out many-particle simulations including thermal effects along the lines of Ref.\cite{mnk} in curved membrane geometries using the formulas Eq.(\ref{vdp_f}) and Eq.(\ref{conf_vdp_f}).
\section {Summary of simulation results}
\label{sumsim}
We present a summary of the simulations performed in the last section in the following table, focusing on the physically interesting situation of dipole aggregation.
\begin{center}
\begin{tabular}{|l|c|c|c|}
\hline
  \begin{tabular}{l}
  \end{tabular} &
  \begin{tabular}{l}
   $\lambda/R =0.1$  \\
  \end{tabular} &
  \begin{tabular}{l}
   $\lambda/R =100$  \\
  \end{tabular} &
  \begin{tabular}{l}
   $\lambda/R =10^4$  \\
  \end{tabular}
  \\
  \hline
  \begin{tabular}{l}
    Unconfined \\
    \end{tabular} &
  \begin{tabular}{l}
    Oscillatory dynamics \\
      ~~No aggregation \\
  \end{tabular} &
  \begin{tabular}{l}
     Oscillatory dynamics \\
      ~~No aggregation \\
  \end{tabular} &
   \begin{tabular}{l}
     Oscillatory dynamics \\
      ~~No aggregation \\
  \end{tabular} \\
\hline
  \begin{tabular}{l}
    Confined \\
    \end{tabular} &
  \begin{tabular}{l}
    Aggregation observed \\
  \end{tabular} &
  \begin{tabular}{l}
     Aggregation observed \\
  \end{tabular} &
   \begin{tabular}{l}
     Oscillatory dynamics \\
      ~~No aggregation \\
  \end{tabular} \\
  \hline
\end{tabular}
\end{center}
The table indicates that in the unconfined situation, the dynamics of the dipole pair is oscillatory as shown in Fig.\ref{figpairdyn}, in both high and low curvature situations. The situation changes under strong confinement  of the external fluid surrounding the membrane (we choose $H/R =10^{-5}$). Keeping $H$ fixed, we find that the dynamics is non-oscillatory and promotes aggregation of dipole pairs, in low ($\lambda/R =0.1$) as well as high curvature ($\lambda/R =100$ ) situations, see Fig.\ref{figpairdync}. However, at extreme high curvatures ($\lambda/R =10^4$) the membrane stress dominates over the traction from the external confined fluid and we observe that dipoles fail to aggregate, even under strong confinement. Our numerical investigations suggest the threshold curvature  $(\lambda/R)_* \sim 700$, beyond which curvature effects tend to destroy aggregation of the dipole pair. Keeping the confinement depth same and staying within the favourable regime $\lambda/R < 700$, we also find aggregation of three/ more dipoles, see Fig.\ref{three_inc}. 
\clearpage
\section {Conclusion and future extensions}
\label{cncl}
\begin{figure}
\begin{tabular}{lcccccccc}
\includegraphics[width=3cm]{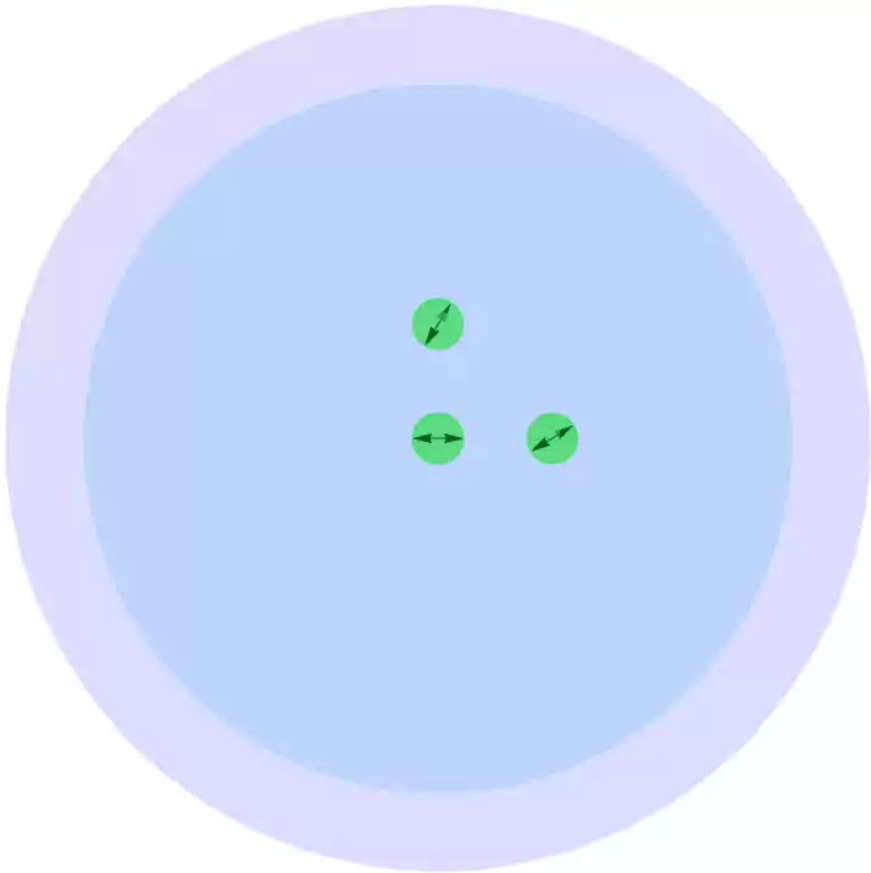}$\pmb{\rightarrow}$&&
\includegraphics[width=3cm]{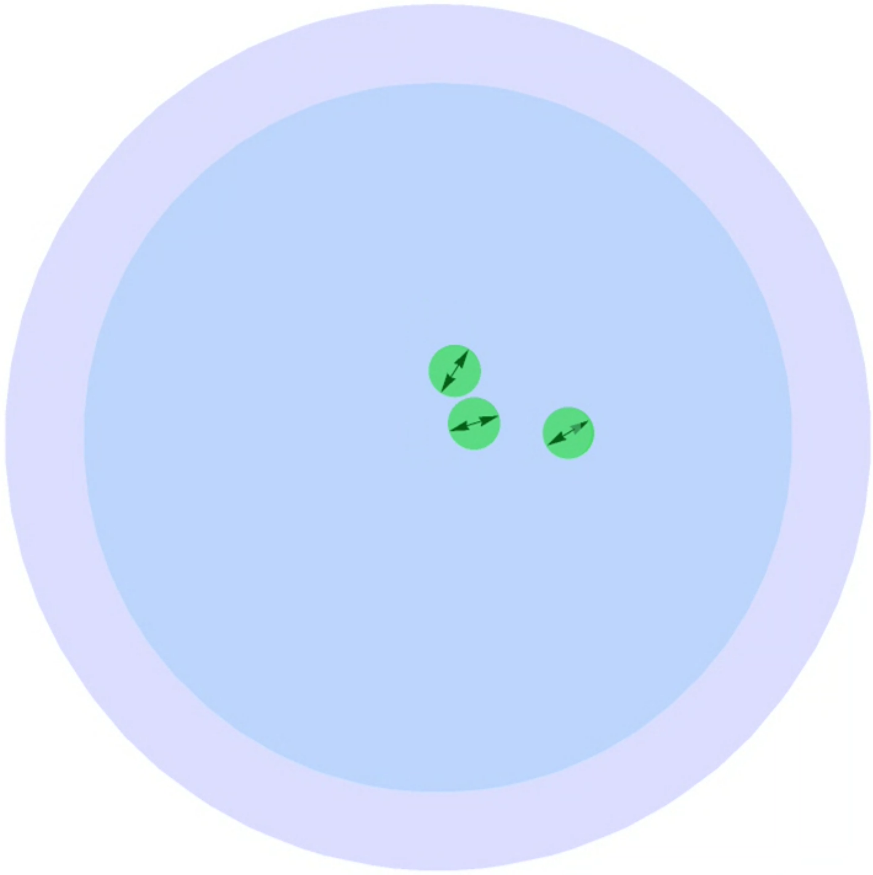}$\pmb{\rightarrow}$&&
\includegraphics[width=3cm]{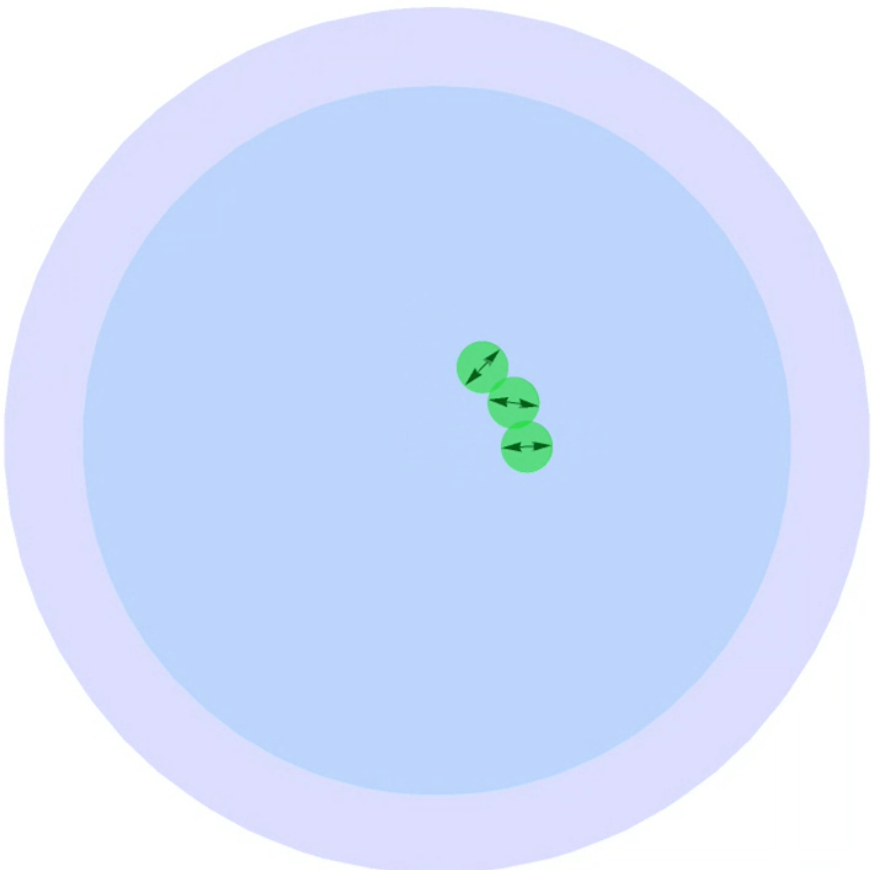}&&
\includegraphics[width=3cm]{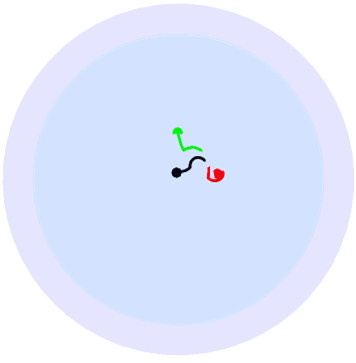}
\end{tabular}
 \caption{(Color online) Aggregation of 3 dipoles with arbitrary initial orientations in the low curvature regime under strong confinement. Starting from left, we show 3 snapshots starting from $t=0$ till aggregation. On the extreme right, the trace of the three trajectories is shown in red, black and green. }
 \label{three_inc} 
\end{figure}
In this paper we present a detailed construction of hydrodynamic flows in curved membranes of fixed geometry, sourced by point force dipoles.  An essential part of the construction of the point dipole limit (Sec.\ref{fdp_cons}) is to keep track of the change in orientation of force vectors due to curvature and this generates an extra term which is crucial for symmetry of dynamical interactions in the curved membrane. We apply this procedure to the case of a spherical membrane and find analytic solution for the flows in the membrane with (Eq.(\ref{conf_vdp_f})) and without confinement (Eq.(\ref{vdp_f})) of the external fluid outside the membrane. The spherical topology of the membrane leads to the creation of a new defect of negative index (Fig.\ref{fdp_flow}) in the flow field  which does not arise in flat membranes. The hydrodynamic interactions of a pair of dipoles is explored next and we subject the central formulas  Eq.(\ref{vdp_f}) and Eq.(\ref{conf_vdp_f}) to several tests for consistency with the geometry of the membrane. Our study suggests aggregation effects of dipoles in biological membranes in regimes of both low and high curvatures, Fig.\ref{figpairdync} and Fig.\ref{three_inc}, under strong confinement. However, very high curvatures tend to destroy dipole aggregation, even under strong confinement. \\\\
One may extend the present work in several directions. We list some of them below:
\begin{itemize}
\item Although the current work is a theoretical investigation, several recent experiments have started exploring the model  we studied here, in particular Ref.\cite{henlev2008,wg2013}. Typically the forces involved are of $\mathcal{O}$(pN) and the membrane radius R is of $\mathcal{O}(\mu m)$. The resulting flows are of $\mathcal{O}(\mu m/s)$.  Accurate particle image velocimetry (PIV) may also be used to study the topological aspects of the flows we constructed here (see Ref.\cite{wg2013}). 
\item Typically for the unconfined membrane the Saffman Length is of $\mathcal{O}(\mu m)$. For nanometer sized inclusions one expects oscillatory dynamics (Fig.\ref{figpairdyn}). However by introducing a confining wall such that $H$ is $\mathcal{O}(nm)$, the screened Saffman length $\lambda_c$ can be made much smaller, leading to dipole aggregation (Fig.\ref{figpairdync}). In a system of many inclusions, one expects formation of dipole clusters, as was recently demonstrated in flat membranes with thermal effects Ref.\cite{mnk}. Although our focus in this paper is the construction of the flow Eq.(\ref{vdp_f}), it would be interesting to perform detailed simulations of  aggregation effects, along the lines of Ref.\cite{mnk}, for a system of inclusions in curved membranes under confinement and explore the effects of membrane geometry and topology on the stability of the aggregated structures. For example, Fig.\ref{curv} shows the focusing action of curvature on dipole motion, while Fig.\ref{figpairdync} shows how extreme curvatures tend to destroy dipole aggregation. In particular, it would be interesting to see how the extra term arising from membrane curvature in Eq.(\ref{vdp_f}) affects the large scale collective dynamics of dipole clusters.  Let us mention in this context that coordinated motion of motor proteins has been reported in recent experiments Ref.\cite{aggexp1,aggexp2}.
\item  It would be worth exploring the dipole dynamics in other situations of interest. For example, one can introduce an asymmetry in the external  fluid viscosities $\eta_+ \neq \eta_-$, as well as consider situations where the internal fluid is also restricted to a certain depth below the membrane surface (the central bulk being a rigid substrate). It would also be interesting to consider hydrodynamic interactions between inclusions of variable  dipole strengths $\kappa$, such as a mixture of ``pusher" and ``puller" type motors.
\item Another direction of investigation is to explore membranes of different geometries, such as cylindrical membranes. Since the cylinder is intrinsically flat, the extra term in Eq.(\ref{vdp_f}) arising from the membrane curvature will be absent. Moreover, unlike the spherical membrane which exhibits global rotation at high curvature Ref.\cite{henlev2010,atz2016}, the cylindrical membrane features an anisotropic mobility tensor and a local  rotation at high curvatures. It would be interesting to see how dipoles interact in such situations.  For more generic geometries, in Sec.\ref{fdp_cons} we commented on how the formula Eq.(\ref{vdp_f}) can be generalized (see Eq.(\ref{rot_gen})), although an explicit solution requires knowledge of the spectrum of the Laplace Beltrami operator on the curved manifold via numerics, along the lines of  Ref.\cite{atz2018}. 
\end{itemize}
We wish to report on some of these exciting topics in future work.
\section{Acknowledgements}
R.S. acknowledges support from DST INSPIRE, India (Grant No. IFA19-PH231). R.S. thanks Ishan Mata and Joe Ninan for stimulating conversations during various stages of the project. 
\appendix
\section {Summary of dynamics of force dipoles in flat membranes}
\label{plane_summary}
In this Appendix, we briefly collect some results from analysis of force dipoles in flat membranes, which may be free (unsupported) or may rest on a substrate. This material is needed to understand some comments and analysis performed in the main text. In preparing this Appendix, we have closely followed Ref.\cite{mnk} and Ref.\cite{evans,lg96,staj,fischer}. For a more detailed and insightful discussion, we refer to Ref.\cite{mnk} and references therein.\\\\
\textbf{Free membrane} : Let us first consider  a free flat membrane situated at $z=0$ and infinitely extended along $x,y$. We denote the membrane pressure by $p$, the  viscosity of the 2D membrane fluid (incompressible) $\eta_{2D}$, $v$ is the in-plane 2D velocity along the membrane,  $\eta_\pm$ denotes the 3D viscosities of external solvents, $v_\pm$ represent the velocities of the 3D external fluids above(+) and below(-) the membrane, respectively. The coupled set of hydrodynamic equations at low Reynolds is given by
\beqa
&\nabla.\vec{v}=0 \nn\\
&f_\alpha  \delta^2(\vec{x}-\vec{x}_0) = \eta_{2D} \nabla^2 v_\alpha -\partial_\alpha p + \left( \sigma_{\alpha z}^+ - \sigma_{\alpha z}^-\right)|_{z\rightarrow 0}\nn\\
& \nabla.\vec{v}_\pm =0, ~~~ \eta_{\pm} \nabla^2 \vec{v}_\pm = -\nabla p_\pm
\label{plane_hydro}
\eeqa
where the bulk fluid stress tensor is given by
\beqa
\sigma_{ij}^\pm = \eta_\pm \left( \partial_i v^\pm_j + \partial_j v^\pm_i\right)
\eeqa
along with the usual no-slip boundary conditions at the membrane surface
\beqa
\vec{v} _\pm |_{z=0}= \vec{v}
\eeqa
Performing a 2D Fourier transform on the incompressibility and momentum balance equation in Eq.(\ref{plane_hydro}), and assuming $\eta_\pm =\eta$, we arrive at
\beqa
&q^\alpha v_\alpha =0, \nn\\
&-\eta_{2D} q^2 v_\alpha(q) - i  q_\alpha p - 2 \eta~ q~ v_\alpha(q) +f_\alpha =0.
\label{mbl}
\eeqa
Contracting the second equation with $q^\alpha$ and using the incompressibility relation,  we get
\beqa
p= \frac{f_\alpha q^\alpha}{i q^2}
\eeqa 
Plugging the membrane pressure back into the momentum balance equation in Eq.(\ref{mbl}), we get 
\beqa
v^{Stokeslet}_i(q) \equiv G^{free}_{i j}(q)~f_j= \frac{\delta_{ij} - \frac{q_i q_j}{q^2} }{\eta_{2D}~ q (q+ \lambda^{-1})}f_j
\label{gfree}
\eeqa
where the length scale $\lambda =\frac{\eta_{2D}}{\eta_+ +\eta_-}$ is the Saffman length and $G^{free}_{ij}(q)$ is the Greens function for the free membrane in Fourier space. Performing the inverse Fourier transform, the expression for the velocity field in real space is given by
\beqa
v^{Stokeslet}_i := G^{free}_{ij} f_j=\frac{1}{4 \pi \eta_{2D}}\left(A\left[\frac{r}{\lambda}\right] \delta_{ij}+B\left[\frac{r}{\lambda}\right]\frac{x_i x_j}{r^2}\right)f_j
\label{plane_pf_main}
\eeqa
where 
\beqa
&A(x) = \pi H_0(x) -\pi \frac{H_1(x)}{x}+\frac{2}{x^2}-\pi \frac{Y_0(x) -Y_2(x)}{2}\nn\\
&B(x) =- \pi H_0(x) + 2\pi \frac{H_1(x)}{x}-\frac{4}{x^2}-\pi Y_2(x)
\eeqa
$H_n$ is the Struve function of order $n$ and $Y_n$ is the Bessel function of second kind of order $n$.\\\\
It is helpful to simplify the above expressions by considering the asymptotic limits :
\beqa
A \rightarrow\begin{cases}
 \log \frac{\lambda}{r},~~~~~~~~~ r\ll \lambda\\
 2 \frac{\lambda^2}{r^2}, ~~~~~~~~~~~ r\gg \lambda
\end{cases}
\eeqa
and
\beqa
B\left(r\right) \rightarrow\begin{cases}
 1~~~~~~~~~~~~~~~ r\ll \lambda\\
 \frac{2 \lambda}{r} -\frac{4 \lambda^2}{r^2} ~~~~~ r\gg \lambda
\end{cases}
\eeqa
From the asymptotic behaviour, we observe the logarithmic Green's function at short distances $r \ll \lambda$, well known to arise in 2D Stokes equations. This  is regulated by the external fluids at large distances, as can be seen from the $r \gg \lambda$ regimes.\\\\

\begin{figure}[h]
\begin{tabular}{lcccccc}
\includegraphics[width=3cm]{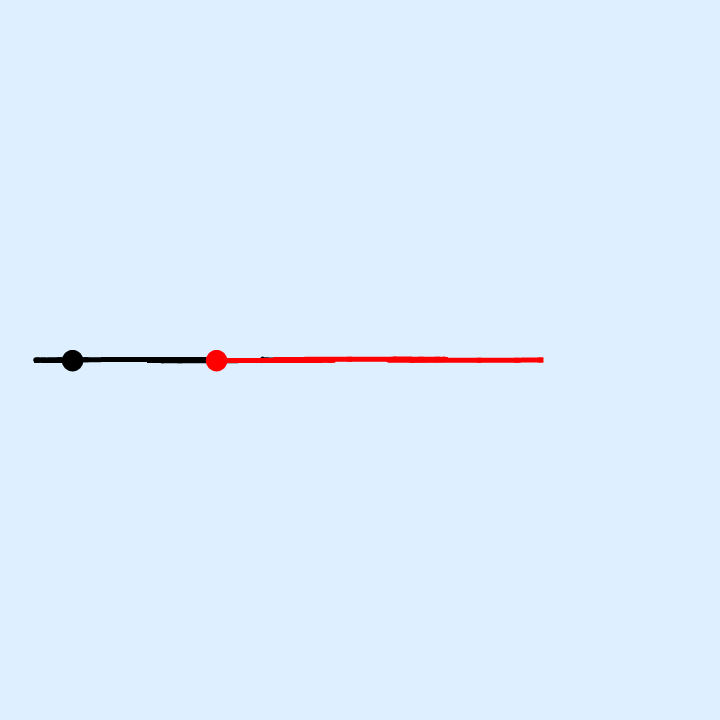}&&
\includegraphics[width=3cm]{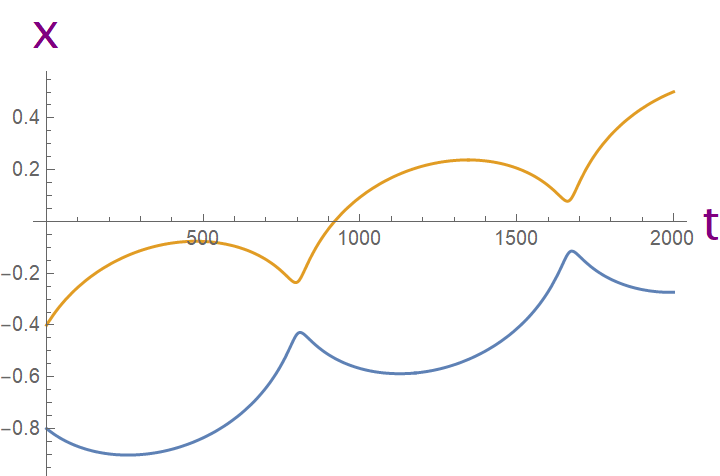}&&
\includegraphics[width=3cm]{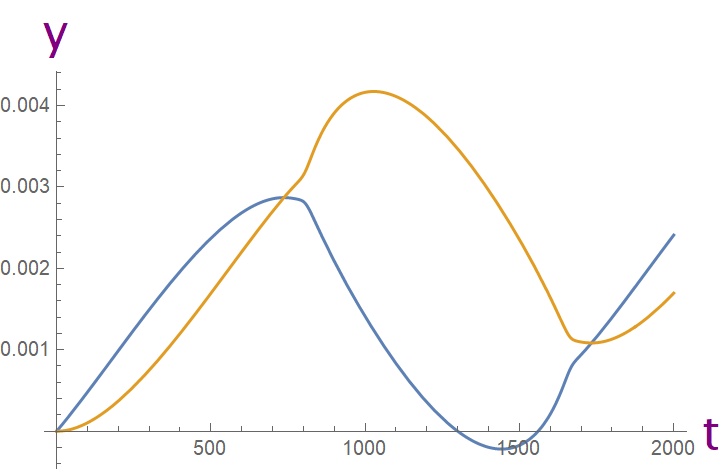}&&
\includegraphics[width=3cm]{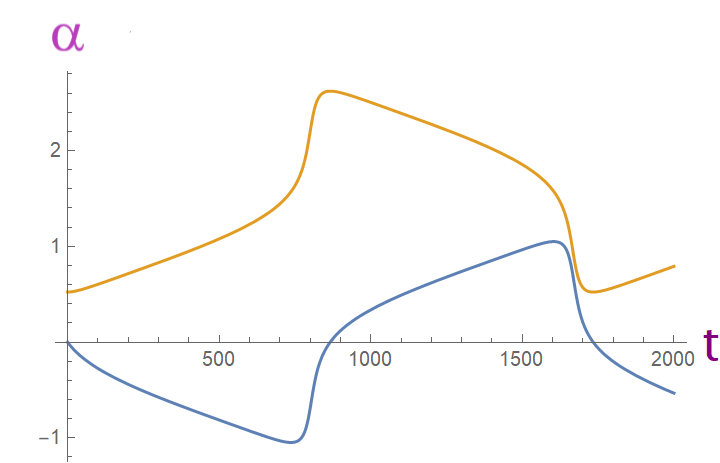}\\
\includegraphics[width=3cm]{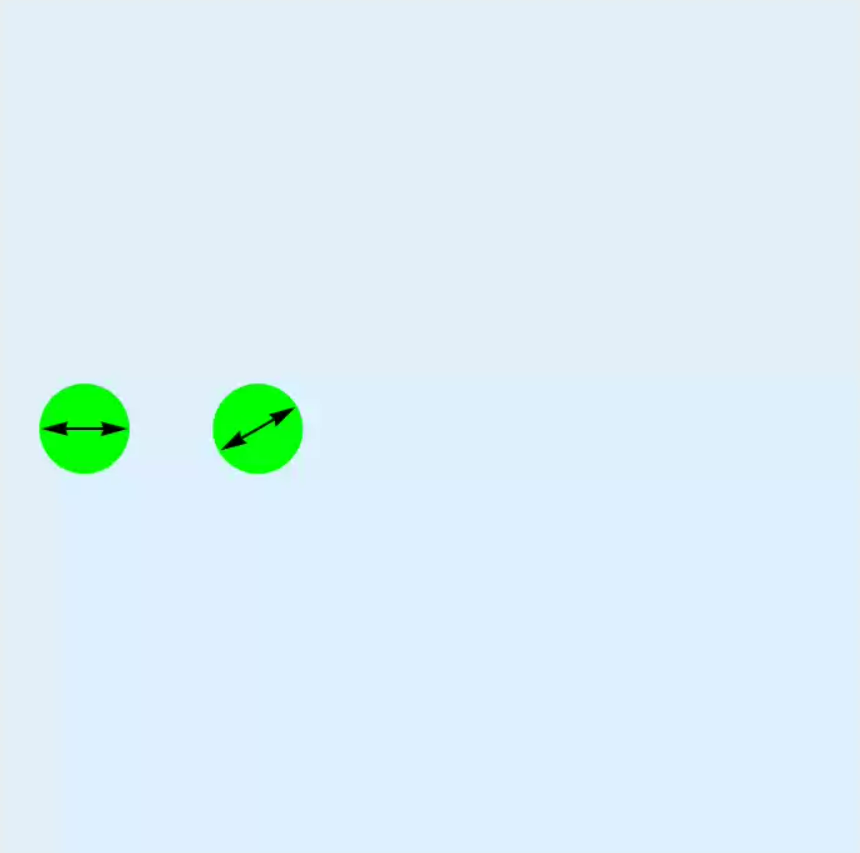}$\pmb{\rightarrow}$&&
\includegraphics[width=3cm]{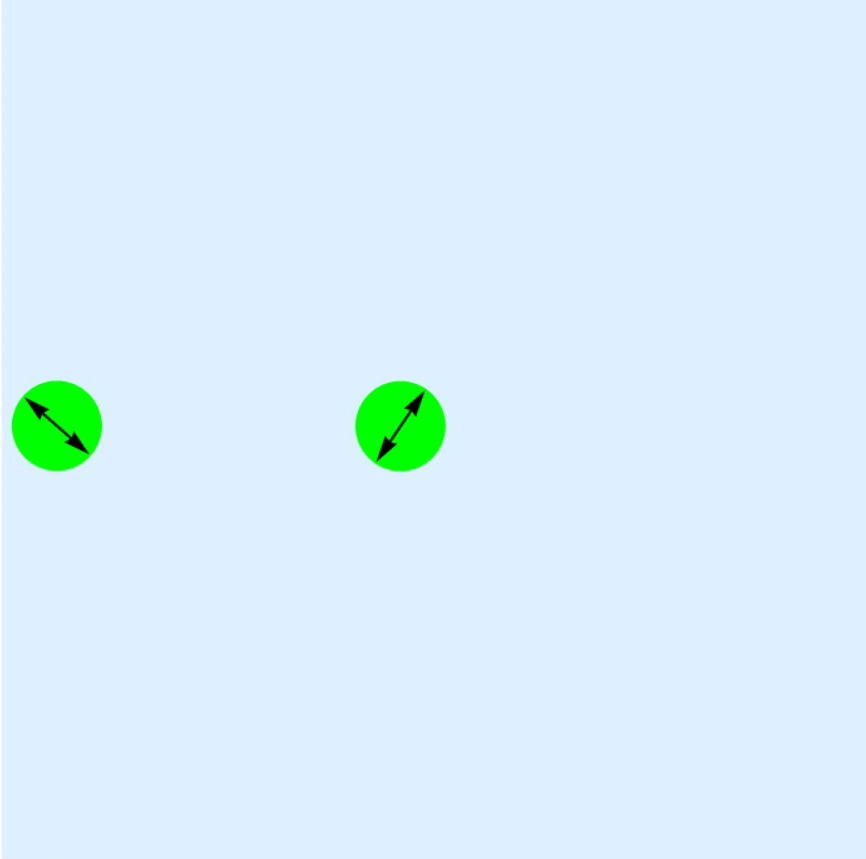}$\pmb{\rightarrow}$&&
\includegraphics[width=3cm]{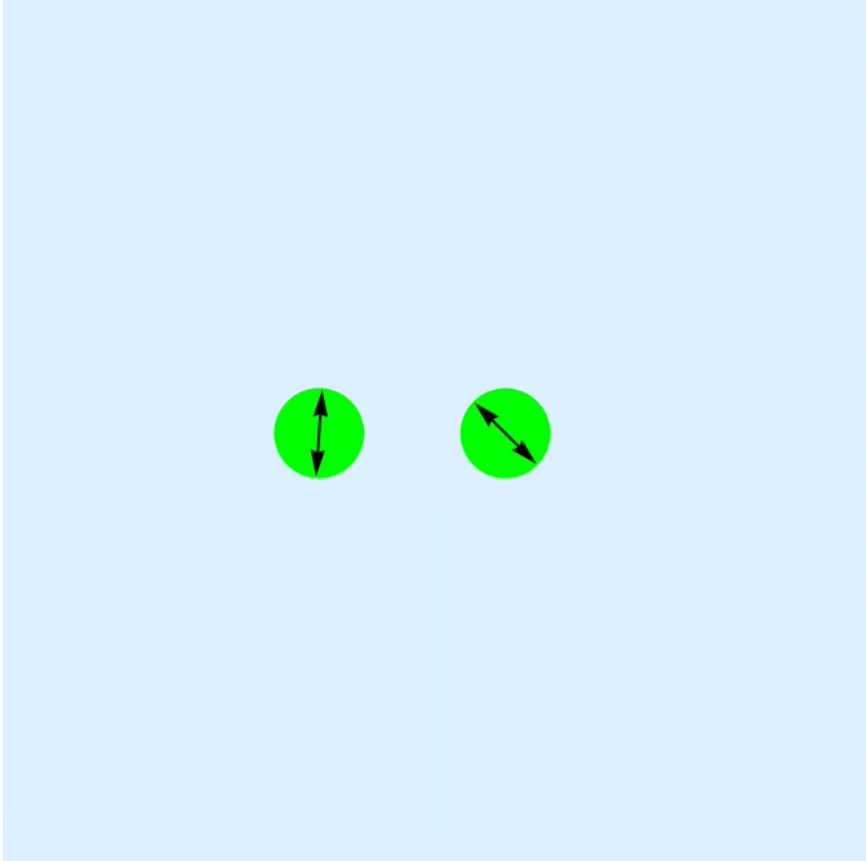}$\pmb{\rightarrow}$&&
\includegraphics[width=3cm]{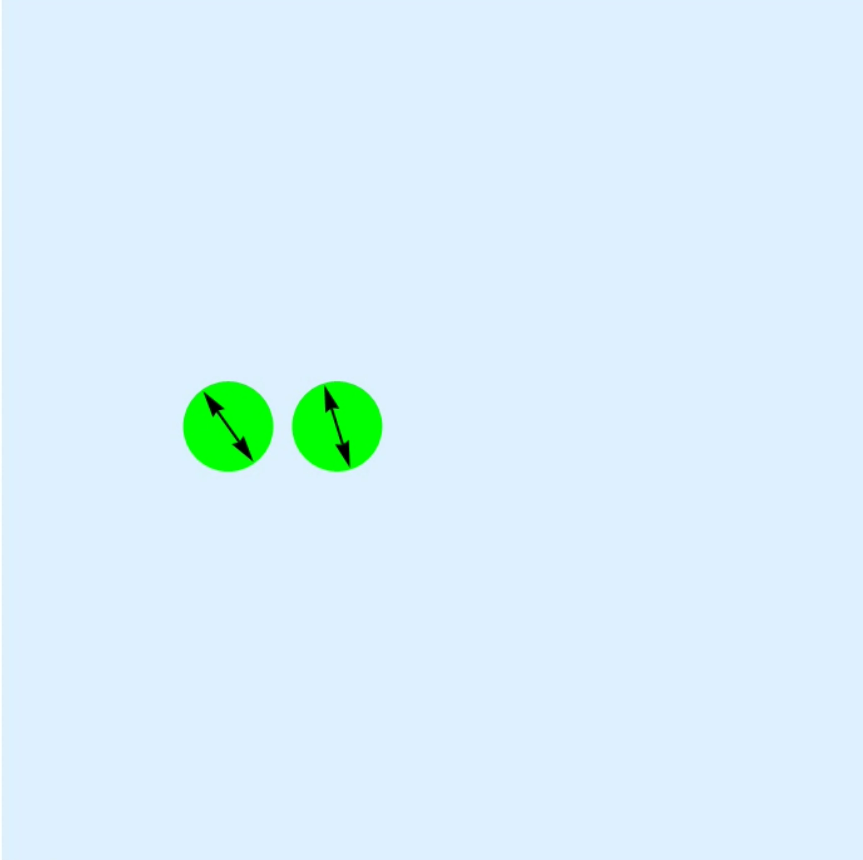}\\
\end{tabular}
\caption{(Color online) \textbf{Interactions at large $\lambda$ in unsupported membrane}: Pair interaction of force dipoles in the unsupported flat membrane for large Saffman length $\lambda/L_0 =125$,  $L_0$ being the initial separation, with initial orientations $\alpha_1=0,~ \alpha_2 =\pi/6$. The time evolution is carried out by Eq.(\ref{vdp_free}), with the Greens function $G^{free}_{ij}$  given by Eq.(\ref{plane_pf_main}) in terms of Struve and Bessel functions. The top row shows the trace of the two trajectories and variation of x, y, orientation $\alpha$. Time is measured in units of $ \frac{125\eta L_0^3}{8 \kappa}$. The second row presents few snapshots of the simulation starting from $t=0$. The dynamics is oscillatory and does not promote aggregation. Compare with high curvature regime of Fig.\ref{figpairdyn} in main text.}
    \label{radplane} 
\end{figure}

\textbf{Supported membrane}: Next we consider a supported flat membrane situated at $z=H$, while the substrate is situated at $z=0$. In contrast to the free membrane, the external fluid below the membrane is now confined within $ 0<z < H$ while  the external fluid in the region $z>H$ above the membrane extends to infinity. For simplicity, let us assume that the fluids above and below the membrane have same viscosity $\eta_+ =\eta_- \equiv \eta$. The confined fluid in the region $ 0<z < H$ thus provides an additional length scale $H$. Allowing both rising and falling modes in the confined region, the Greens function defined in Eq.(\ref{gfree}) is modified to (Ref.\cite{evans,lg96,fischer})
\beqa
v_i^{Stokeslet}(q) \equiv G^{c}_{i j}(q)~f_j= \frac{\delta_{ij} - \frac{q_i q_j}{q^2} }{\eta q (1+ 2\lambda q +\coth{q H})}f_j
\label{gconf}
\eeqa
where $G^{c}_{i j}(q)$ denotes the Greens function in the confined situation. Depending on the length scales $H,\lambda$ and the distance from the source, denoted by $r$, one has different regimes as follows (Ref.\cite{evans,lg96,fischer}):\\\\
\textbf{Strong confinement}~($H \ll \lambda$): Under strong confinement, $\coth(q H) \sim \frac{1}{q H}$ in Eq.(\ref{gconf}) and we get a modified solution where the Saffman length $\lambda$ is reduced to a screened value $\lambda_c = \sqrt{2 \lambda H}$. The Stokeslet solution is the same as Eq.(\ref{plane_pf_main}) with $\lambda$ replaced by $\lambda_c$ and a modified Greens function $G^c$ as follows:
\beqa
&v^{Stokeslet}_i := G^{c}_{ij} f_j=\frac{1}{4 \pi \eta_{2D}}\left(A^c\left[\frac{r}{\lambda_c}\right] \delta_{ij}+B^c\left[\frac{r}{\lambda_c}\right]\frac{x_i x_j}{r^2}\right)f_j\nn\\
&A^c(x) =  -\frac{2}{x^2} +2 K_0(x) +\frac{2 K_1(x)}{x}\nn\\
&B^c(x) =\frac{4}{x^2}-2 K_0(x) -4 \frac{K_1(x)}{x}
\label{gconf_realspace}
\eeqa
where $K_n$ is the modified Bessel function of order n. The near field region $r \ll \lambda_c$ in this situation is dominated by the membrane stress and scales as $\log(\frac{r}{\lambda_c})$ while the far field region  $r \gg \lambda_c$ is dominated by the stress from the confined fluid, giving rise to a power law scaling $ v \sim \frac{1}{r^2}$. Thus to leading order, the Greens function  $G^c_{ij}(r)$  is
\beqa
G^c_{ij}(r) \rightarrow\begin{cases}
\frac{1}{4 \pi \eta_{2D}} \left(\delta_{ij} \left(\log \frac{2 \lambda_c}{r} -\gamma -\frac{1}{2} \right) + \frac{r_i r_j}{r^2} \right) +\mathcal{O}(r/\lambda_c),~~~~~~~~~ r\ll \lambda_c\\\\
 \frac{H}{2 \pi \eta r^2}\left( -\delta_{ij} + 2 \frac{r_i r_j}{r^2}\right) + \mathcal{O}(r^{-\frac{1}{2}}e^{-r/\lambda_c}), \hspace{2.5cm} r\gg \lambda_c
\end{cases}
\label{strongconf}
\eeqa
\textbf{Weak confinement}~($H \gg \lambda$): Here the Greens function  $G^c_{ij}(r)$  reduces to (Ref.\cite{evans,lg96,fischer})
\beqa
G^c_{ij}(r) \rightarrow\begin{cases}
\frac{1}{4 \pi \eta_{2D}} \left(\delta_{ij} \left(\log \frac{2 \lambda }{r} -\gamma -\frac{1}{2} \right) + \frac{r_i r_j}{r^2} \right) +\mathcal{O}(r/\lambda),\hspace{2.5cm} r\ll \lambda \ll H\\\\
 \frac{1}{4 \pi \eta~ r}\left( \frac{r_i r_j}{r^2}\right) + \mathcal{O}(\frac{\lambda^2}{r^2}), \hspace{6.1cm} \lambda \ll r\ll H\\\\
  \frac{H}{2 \pi \eta r^2}\left( -\delta_{ij} + 2 \frac{r_i r_j}{r^2}\right) + \mathcal{O}(r^{-3}), \hspace{4.75cm}  \lambda \ll H \ll r 
\end{cases}
\label{weakconf}
\eeqa
\textbf{Flows sourced by force dipoles}: The velocity field sourced by a dipole can be easily related to the directional derivative of the stokeslet flows along the orientation of the force (as explained in the main text). Thus for the free membrane, the flow sourced by a force dipole is given by (Ref.\cite{mnk})
\beqa
[v_{i}]^{dipole} = -\kappa \hat{f}_k ~ \partial_{x_k} \left[ G^{free}_{ij} \hat{f}_j \right]
\label{vdp_free}
\eeqa
where $G^{free}_{ij}$ is the Greens function for the free membrane given by Eq.(\ref{plane_pf_main}) and $\kappa = |\bm{f}| ~ dL $ is a constant in the limit of a point dipole $dL \rightarrow 0 ,~ |\bm{f}| \rightarrow \infty$. One gets two regimes for the resulting velocity field of a dipole, placed at the origin and directed along $\hat{f}$:
\beqa
\bm{v}^{dipole}=\begin{cases}
-\frac{\kappa}{2 \pi \eta r^2} \left( (1-3 \cos^2 \theta) \hat{r} +\cos \theta ~\hat{f}\right) ~~~~~ r \gg \lambda\nn\\
-\frac{\kappa}{4 \pi \eta_{2D} r}  (1-2 \cos^2 \theta) \hat{r}  ~~~~~~~~~~~~~~~~~~~~~ r \ll \lambda
\end{cases}
\label{vdp_rad_az}
\eeqa
where $ \cos \theta = \hat{r} . \hat{f}$.\\
Similarly for the confined case, one has
\beqa
[v_{i}]^{dipole}= -\kappa \hat{f}_k ~ \partial_{x_k} \left[ G^{c}_{ij} \hat{f}_j \right]
\label{vdpc}
\eeqa
with the Greens function ~$G^{c}_{ij} $ given by Eq.(\ref{gconf}) and its asymptotic limits Eq.(\ref{strongconf}) and  Eq.(\ref{weakconf}) for strong and weak confinements respectively.\\\\ In particular, plugging the Greens function $G^{c}_{ij} $ that arises in the far field region under strong confinement (second row of Eq.(\ref{strongconf})) into Eq.(\ref{vdpc}) yields a flow of the following form :
\beqa
\vec{v}^{dipole}=
-\frac{\kappa H}{ \pi \eta r^3} \left( (1-4 \cos^2 \theta) \hat{r} +2 \cos \theta ~\hat{f}\right) ~~~ r \gg \lambda_c
\label{vdp_conf}
\eeqa
where $ \cos \theta = \hat{r} . \hat{f}$ and $\hat{r}$ is the position vector of the response location. A distinct feature of Eq.(\ref{vdp_conf}) is that it is \textit{vorticity free} and thus in this regime a system of dipoles will not change their orientations. Following the arguments presented in the main text, this flow is suitable for non-oscillatory dynamics and is favourable for aggregation of dipoles. However, the near field flow below the screened Saffman length $\lambda_c$ has significant vorticity and leads to oscillatory dynamics similar to Fig.\ref{radplane}. But one can make $H$ sufficiently small such that the screened Saffman length $\lambda_c$ is of the order of the inclusion size or smaller, thus effectively eliminating the near field radial flows arising from membrane hydrodynamics.
\begin{figure}[h]
\begin{tabular}{lcccccc}
\includegraphics[width=3cm]{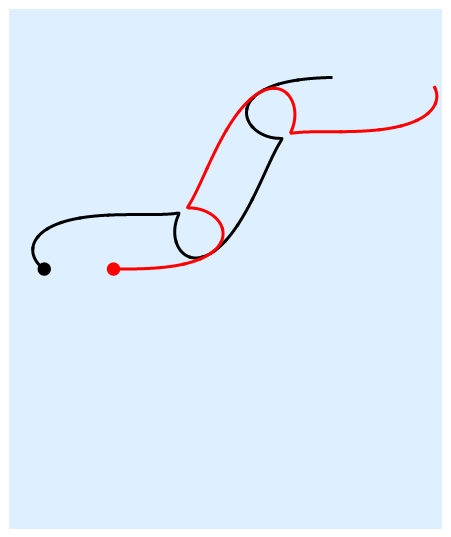}&&
\includegraphics[width=3cm]{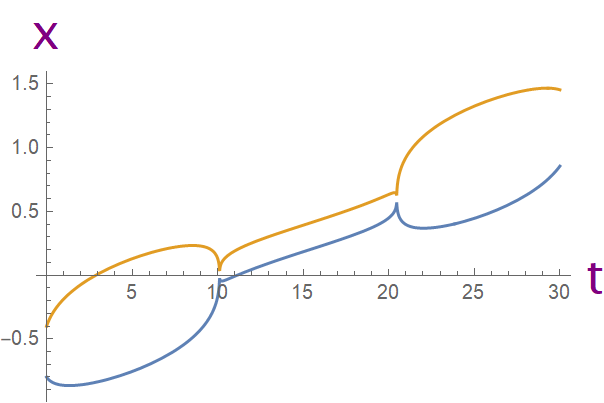}&&
\includegraphics[width=3cm]{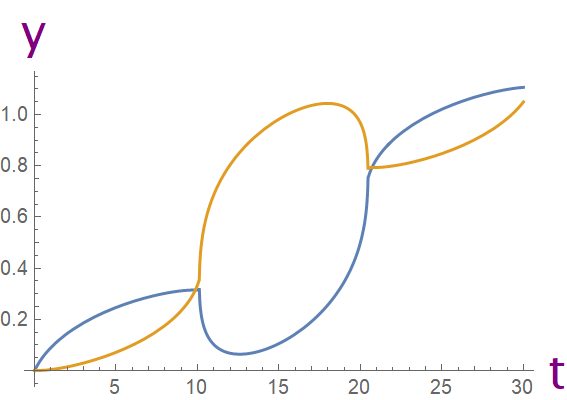}&&
\includegraphics[width=3cm]{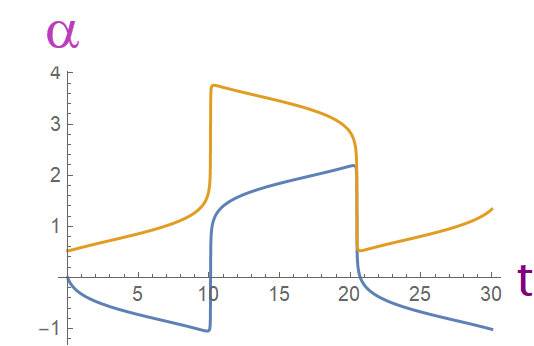}\\
\includegraphics[width=3cm]{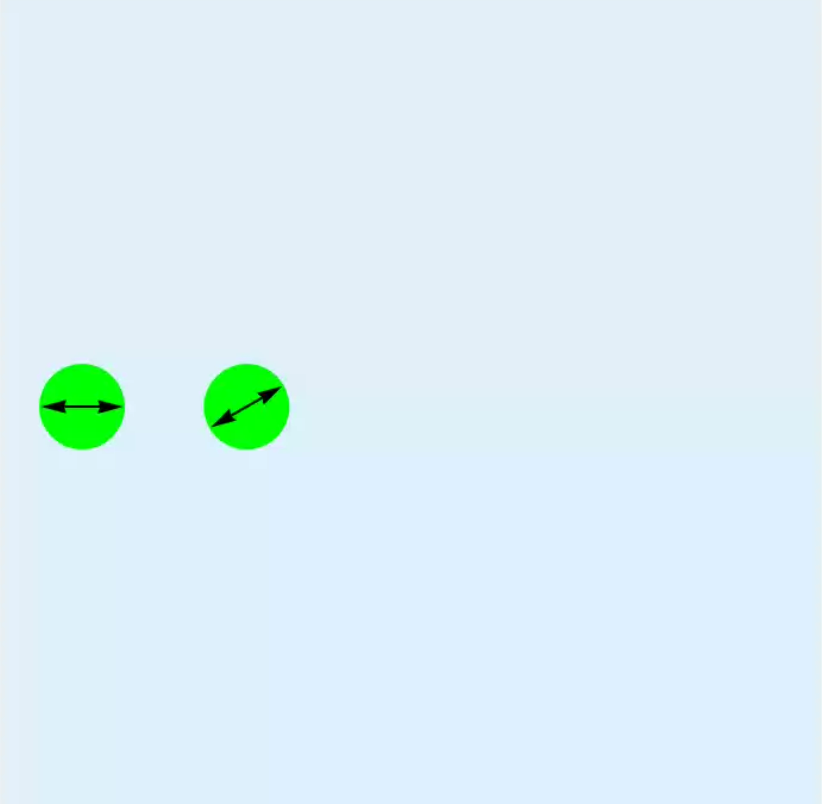}$\pmb{\rightarrow}$&&
\includegraphics[width=3cm]{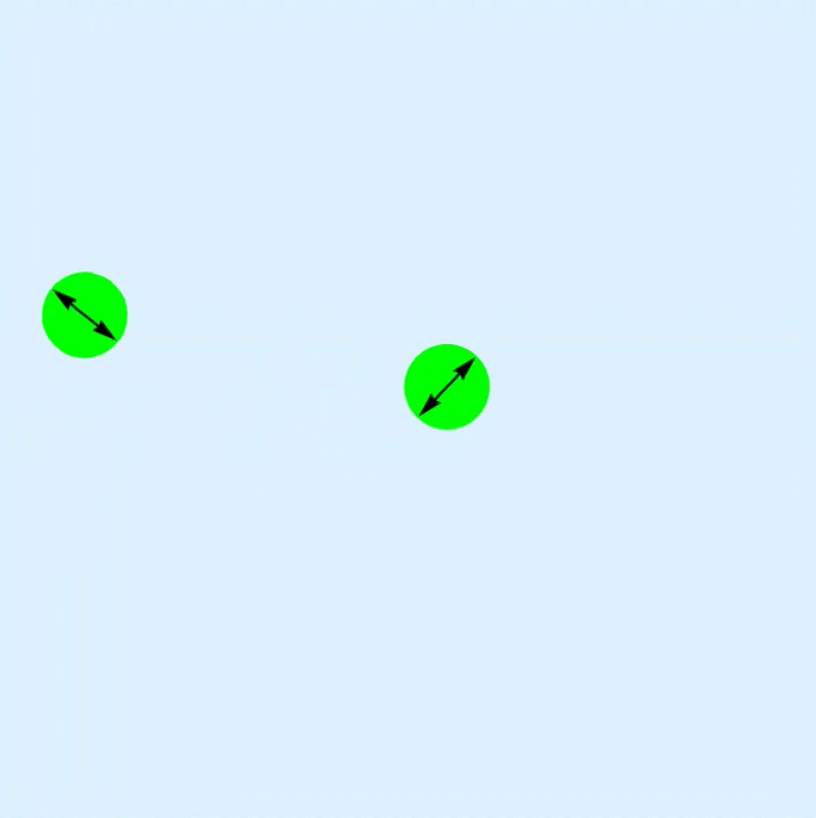}$\pmb{\rightarrow}$&&
\includegraphics[width=3cm]{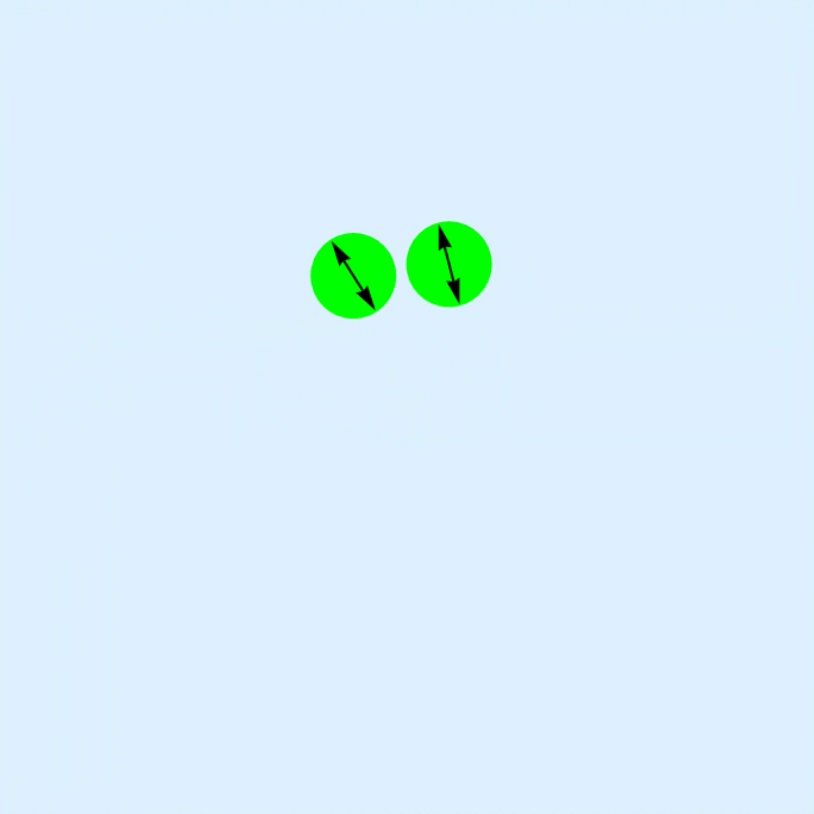}$\pmb{\rightarrow}$&&
\includegraphics[width=3cm]{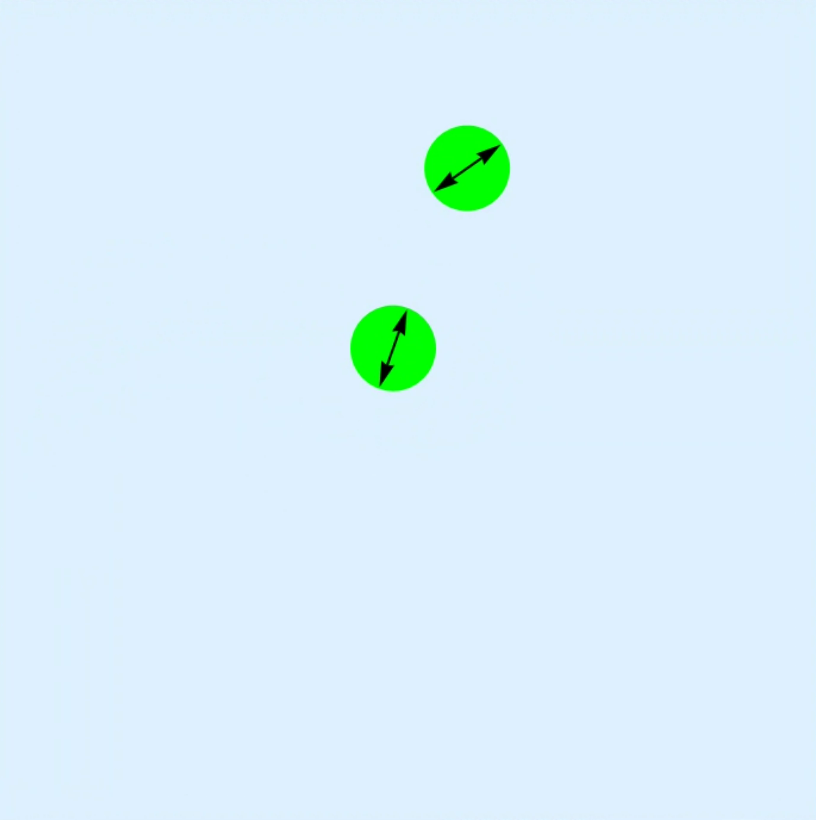}
\end{tabular}
\caption{(Color online)  \textbf{Interactions at small $\lambda$ in unsupported membrane}:  Pair interaction of force dipoles in the unsupported flat membrane for small Saffman length $\lambda/L_0 =0.125$, $L_0$ being the initial separation, with initial orientations $\alpha_1=0,~ \alpha_2 =\pi/6$. The time evolution is carried out by Eq.(\ref{vdp_free}), with the Greens function $G^{free}_{ij}$  given by Eq.(\ref{plane_pf_main}) in terms of Struve and Bessel functions.  The top row shows a trace of the two trajectories and the variation of x, y, orientation $\alpha$  wrt time. Time is measured in units of $ \frac{125\eta L_0^3}{8 \kappa}$. The second row presents snapshots of the simulation starting from $t=0$ for one oscillation cycle.  The dynamics is oscillatory and does not promote aggregation. Compare with low curvature regime of Fig.\ref{figpairdyn} in main text.}
    \label{azplane} 
\end{figure}
\begin{figure}[h]
\begin{tabular}{lcccccc}
\includegraphics[width=3cm]{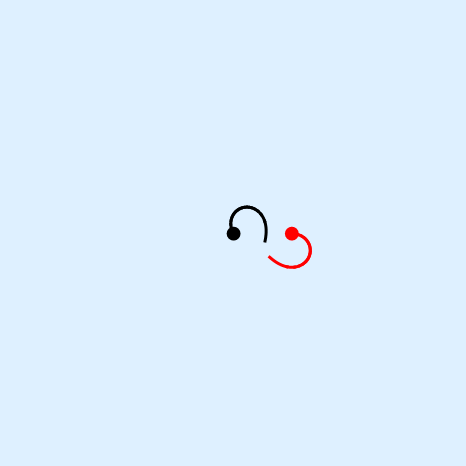}&&
\includegraphics[width=3cm]{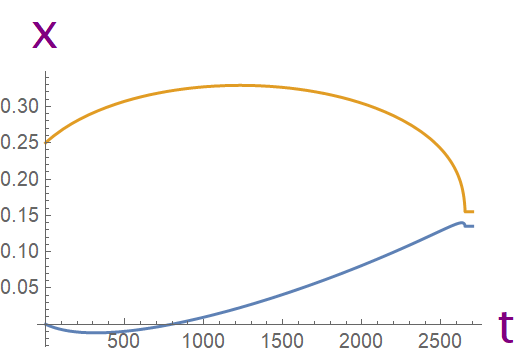}&&
\includegraphics[width=3cm]{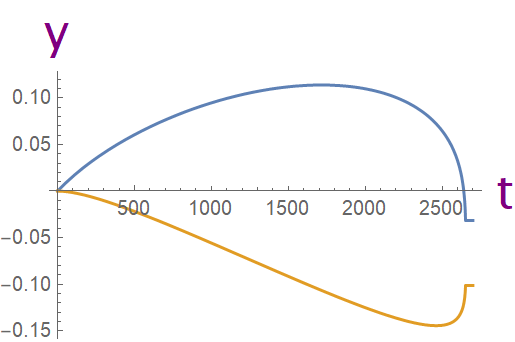}&&
\includegraphics[width=3cm]{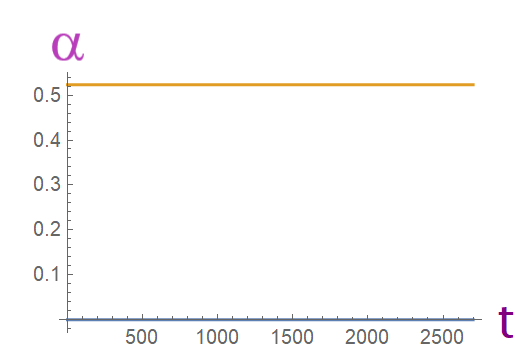}\\
\includegraphics[width=3cm]{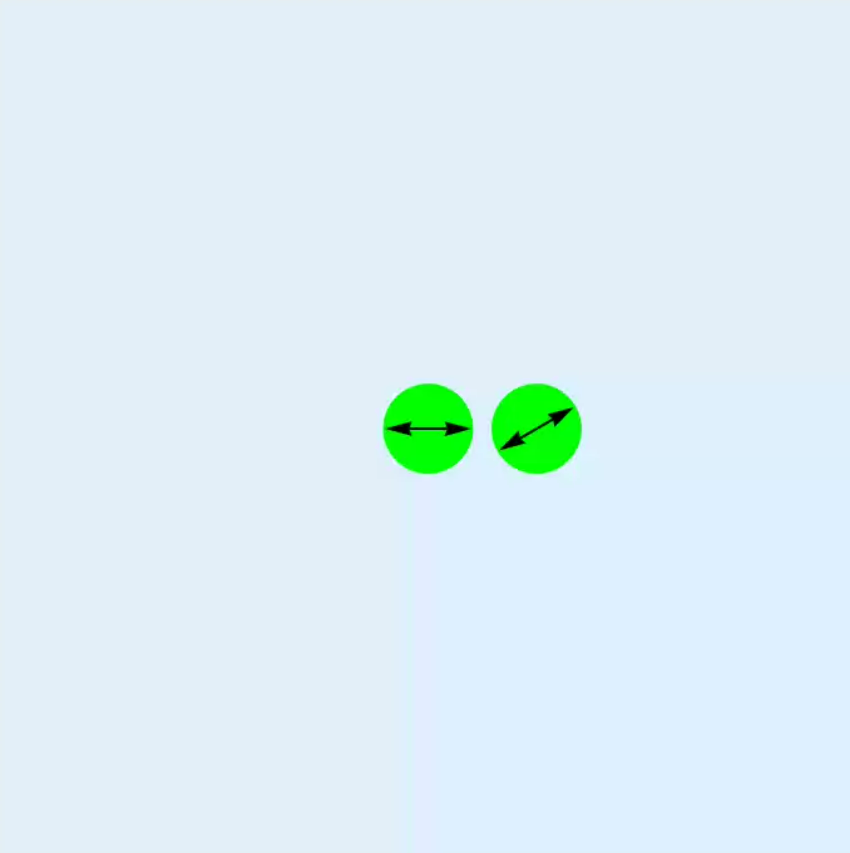}$\pmb{\rightarrow}$&&
\includegraphics[width=3cm]{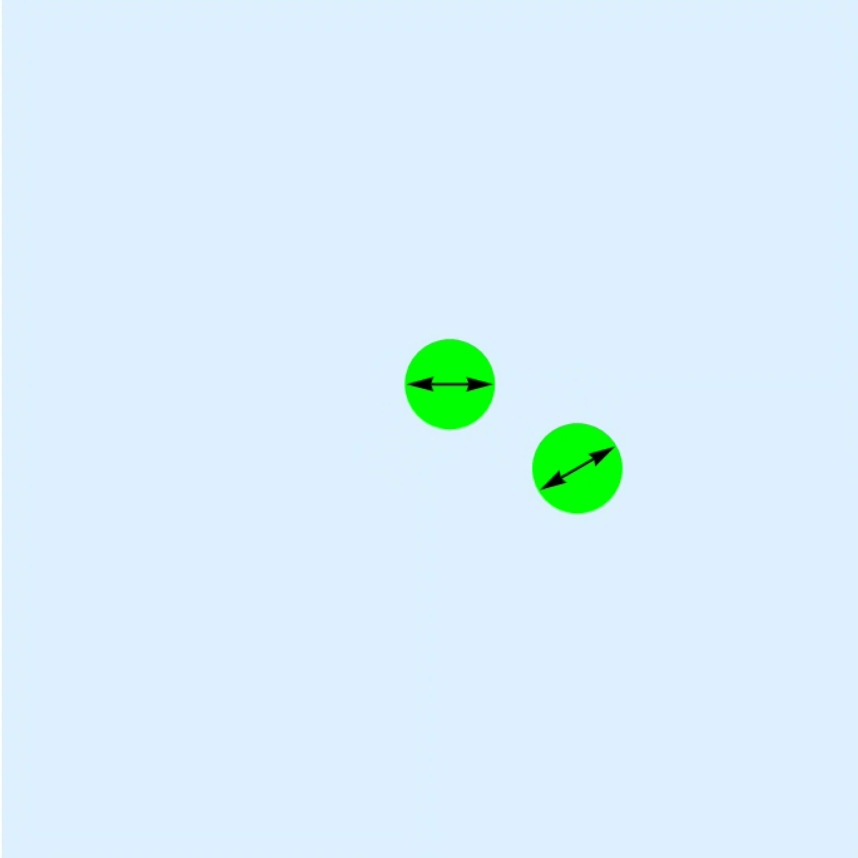}$\pmb{\rightarrow}$&&
\includegraphics[width=3cm]{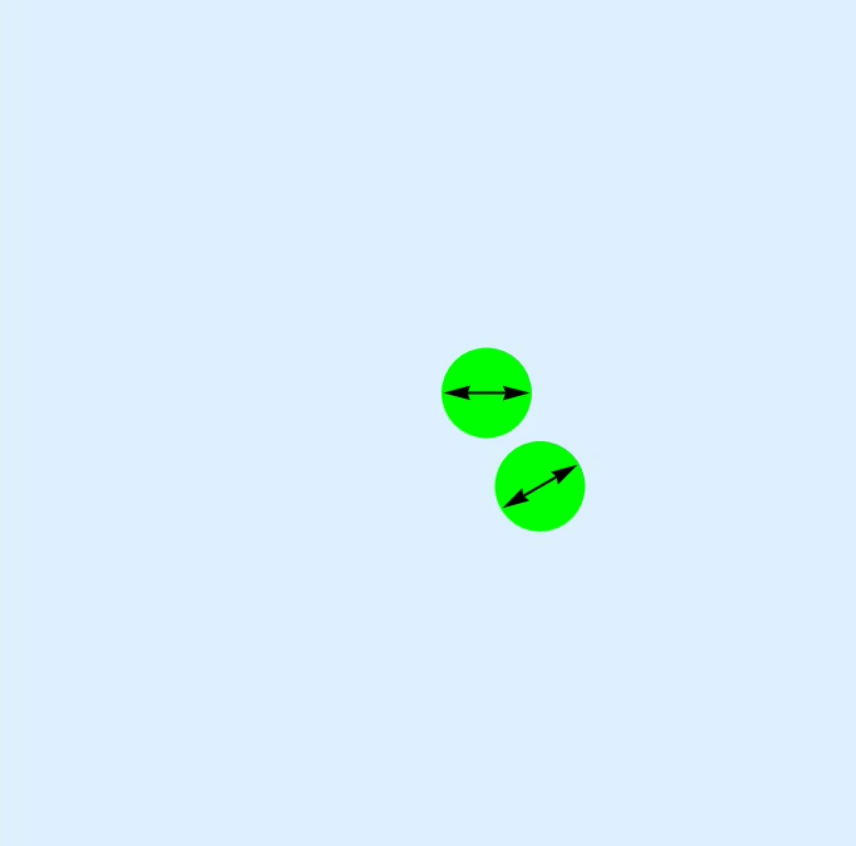}$\pmb{\rightarrow}$&&
\includegraphics[width=3cm]{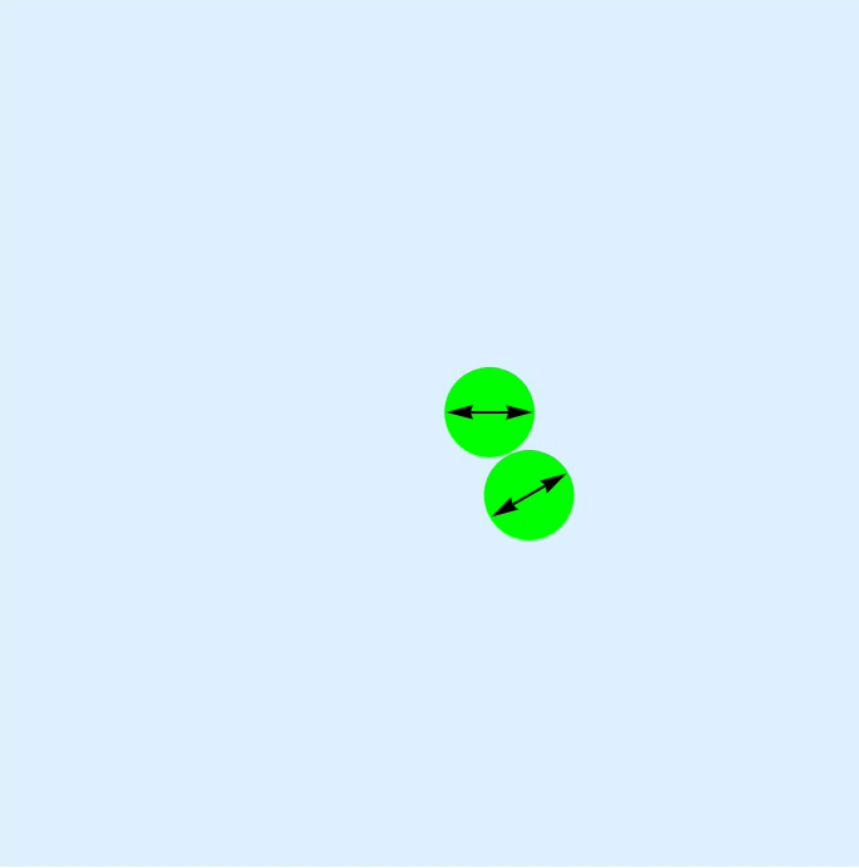}\\
\end{tabular}
\caption{(Color online)  \textbf{Strong confinement and small $\lambda$} : Pair interaction of force dipoles in the flat membrane under strong confinement ($H/\lambda \ll 1$). The parameters are chosen such that Saffman  length $\lambda$ is small, $\lambda/L_0 =0.2$, where $L_0$ is the initial separation, and the screened Saffman length $\lambda_c$ is smaller than particle size. The initial orientations (measured wrt to $\hat{x}$) are $\alpha_1=0,~ \alpha_2 =\pi/6$. The time evolution is carried out by Eq.(\ref{vdpc}) and the Greens function in the strongly confined situation $G^c$ written in terms of modified Bessel functions, see Eq.(\ref{gconf_realspace}). The top row shows a trace of the two trajectories and the variation of x, y and orientation $\alpha$  wrt time. Time is measured in units of $ \frac{64\eta L_0^3}{ \kappa}$. The dynamics is non-oscillatory and promotes aggregation. The second row presents snapshots of the simulation till aggregation. Soft repulsion is added to prevent particle overlap. Compare with low curvature regime of Fig.\ref{figpairdync} in main text.}
    \label{conf1} 
\end{figure}
\begin{figure}[h]
\begin{tabular}{lcccccc}
\includegraphics[width=3cm]{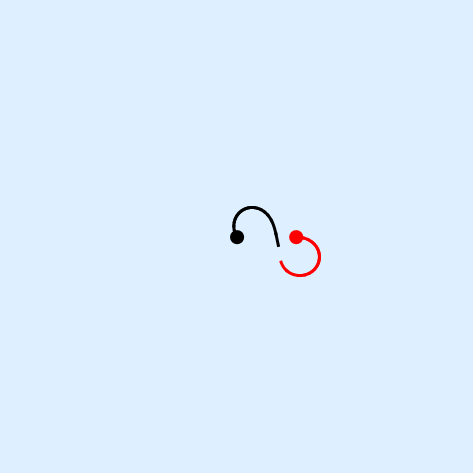}&&
\includegraphics[width=3cm]{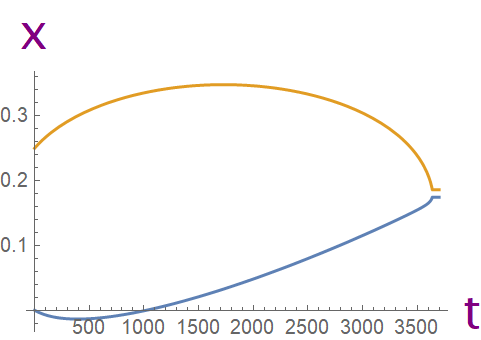}&&
\includegraphics[width=3cm]{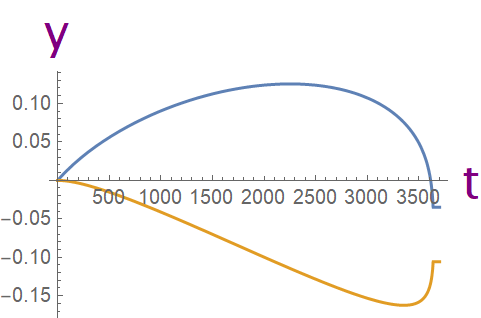}&&
\includegraphics[width=3cm]{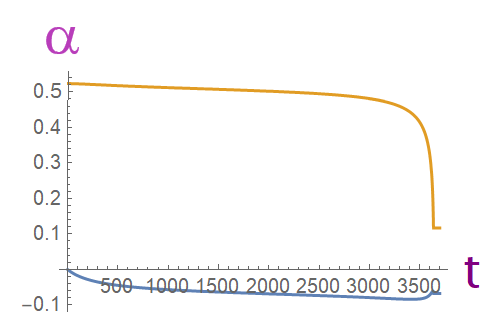}\\
\includegraphics[width=3cm]{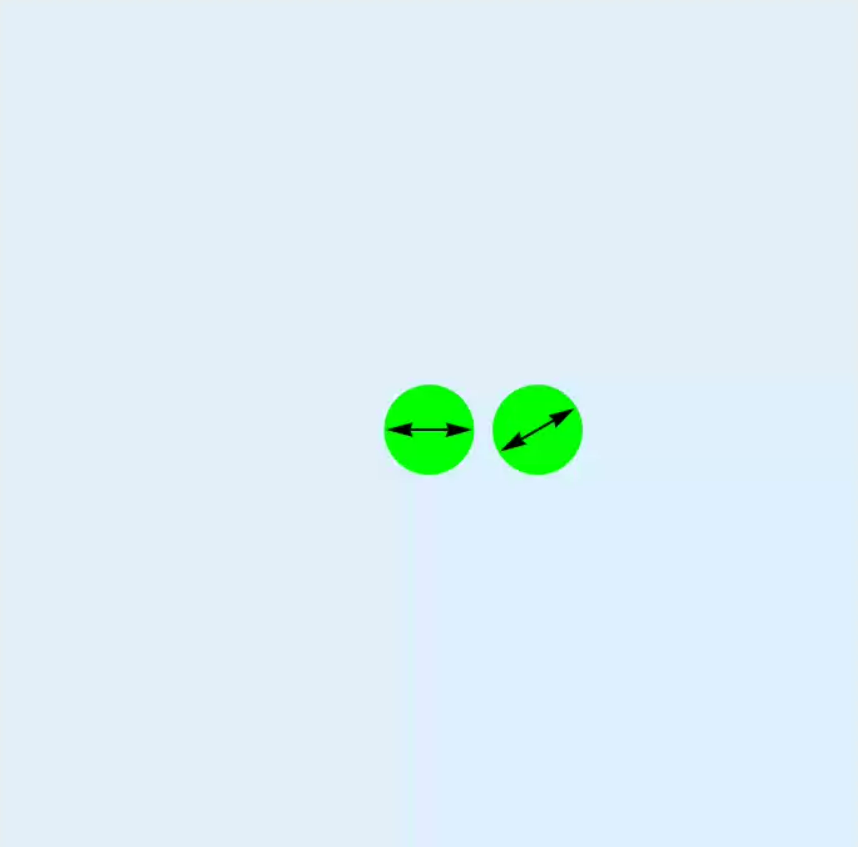}$\pmb{\rightarrow}$&&
\includegraphics[width=3cm]{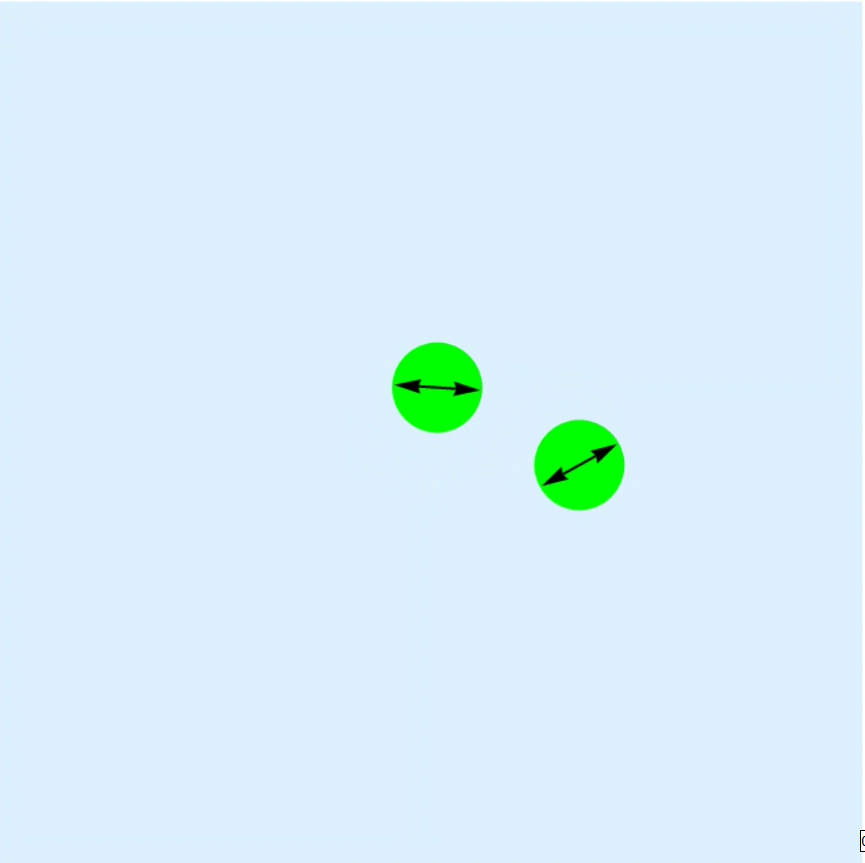}$\pmb{\rightarrow}$&&
\includegraphics[width=3cm]{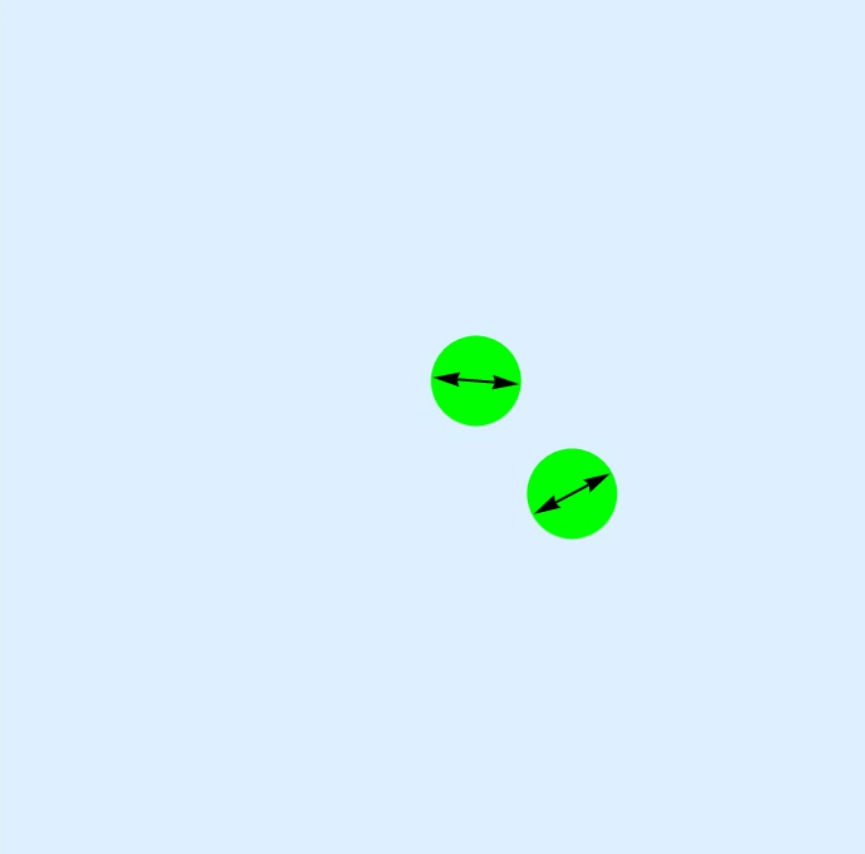}$\pmb{\rightarrow}$&&
\includegraphics[width=3cm]{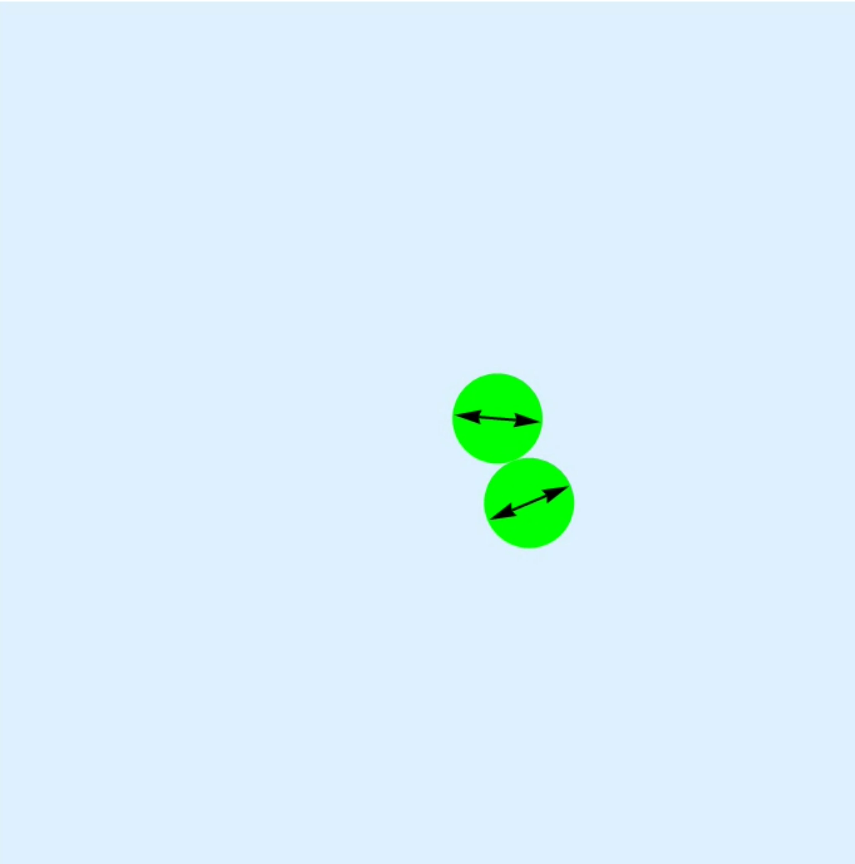}\\
\end{tabular}
\caption{(Color online)  \textbf{Strong confinement and large $\lambda$}: Pair interaction of force dipoles in the flat membrane under strong confinement ($H/\lambda \ll 1$). The parameters are chosen such that Saffman length $\lambda$ is large $\lambda/L_0 =400$ but the screened Saffman length $\lambda_c$ is still of the order of particle size. The initial orientations (measured wrt to $\hat{x}$) being $\alpha_1=0,~ \alpha_2 =\pi/6$. The time evolution is carried out by Eq.(\ref{vdpc}) and the Greens function in the strongly confined situation $G^c$ written in terms of modified Bessel functions, see Eq.(\ref{gconf_realspace}). The top row shows a trace of the two trajectories and the variation of x, y and orientation $\alpha$  wrt time. Time is measured in units of $ \frac{\eta L_0^3}{64 \kappa}$. The dynamics is non-oscillatory and promotes aggregation. The second row presents snapshots of the simulation till aggregation. Soft repulsion is added to prevent particle overlap. Compare with high curvature regime of Fig.\ref{figpairdync} in  main text.}
    \label{conf0} 
\end{figure}
\begin{figure}[h]
\begin{tabular}{lcccccc}
\includegraphics[width=3cm]{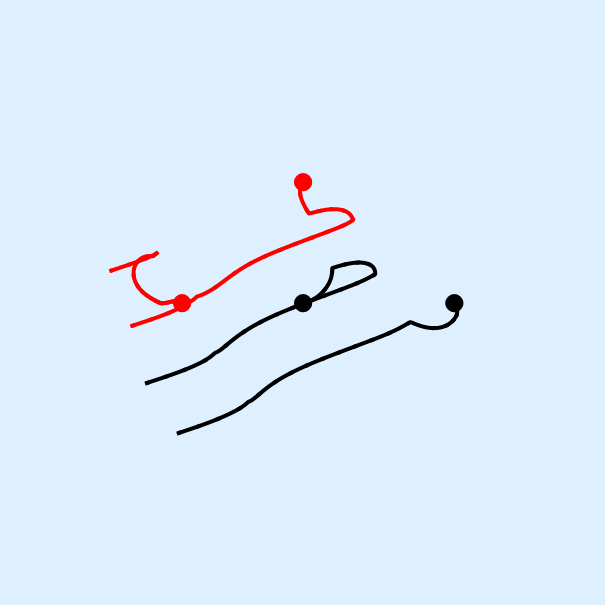}&&
\includegraphics[width=3cm]{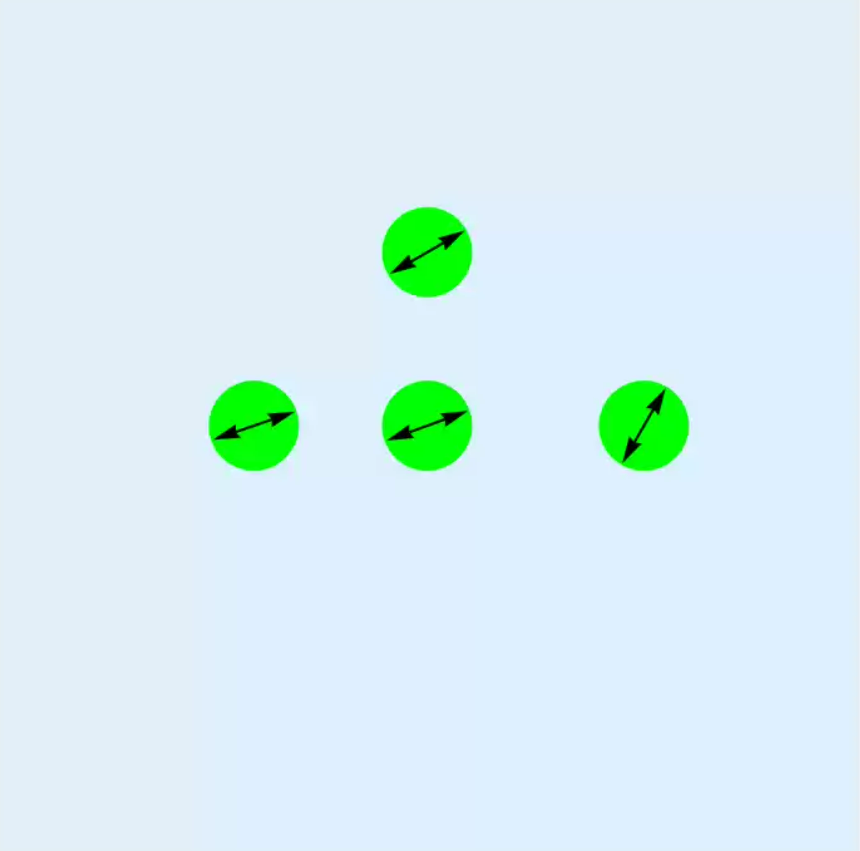}$\pmb{\rightarrow}$&&
\includegraphics[width=3cm]{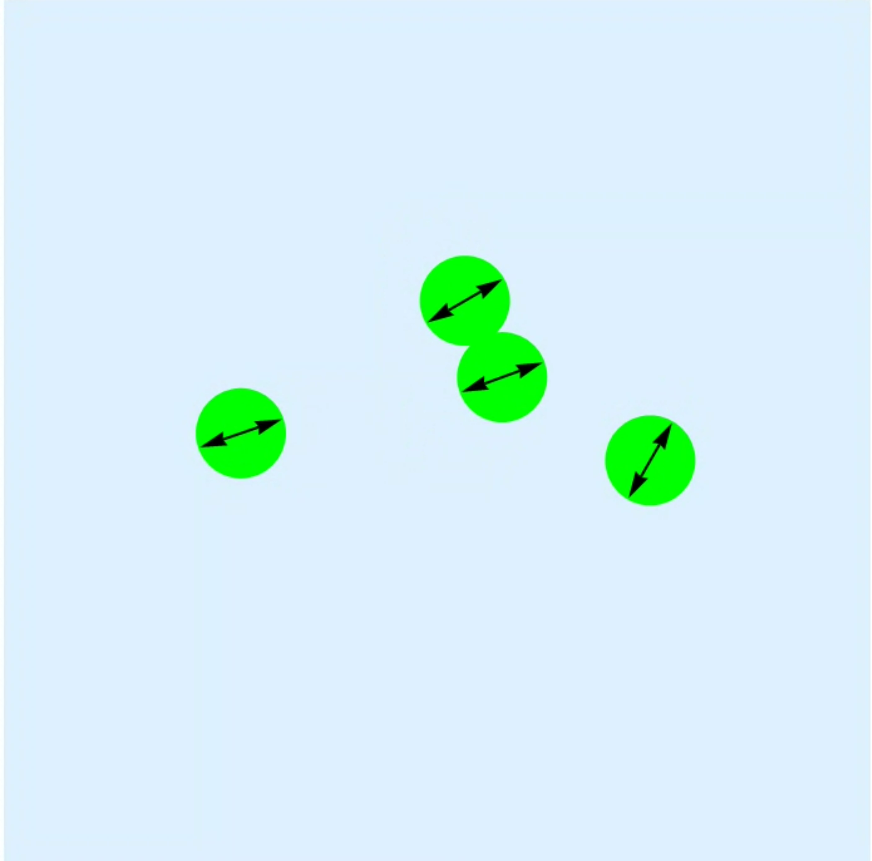}$\pmb{\rightarrow}$&&
\includegraphics[width=3cm]{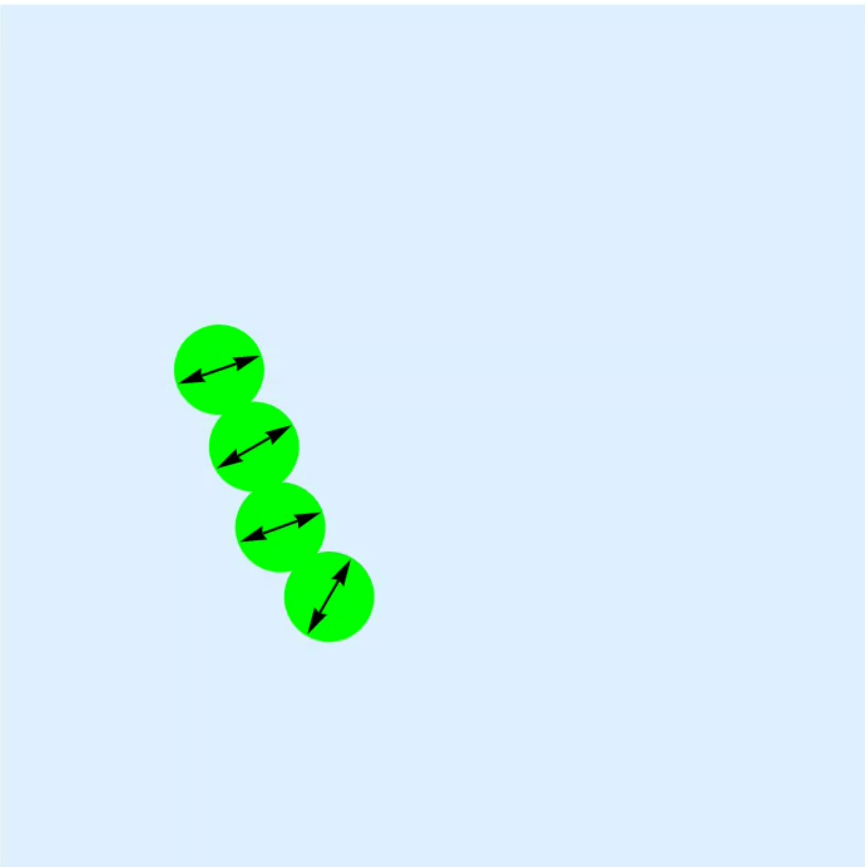}
\end{tabular}
\caption{(Color online)  Hydrodynamic aggregation of a system of four force dipoles under strong confinement. Simulations performed using the vorticity free flow under strong confinement Eq.(\ref{vdp_conf}). Trace of the trajectories is shown on the left, followed by three snapshots of the motion starting from $t=0$ till aggregation.}
    \label{conf2} 
\end{figure}
\begin{figure}[h]
\begin{tabular}{lcccc}
\includegraphics[width=4cm]{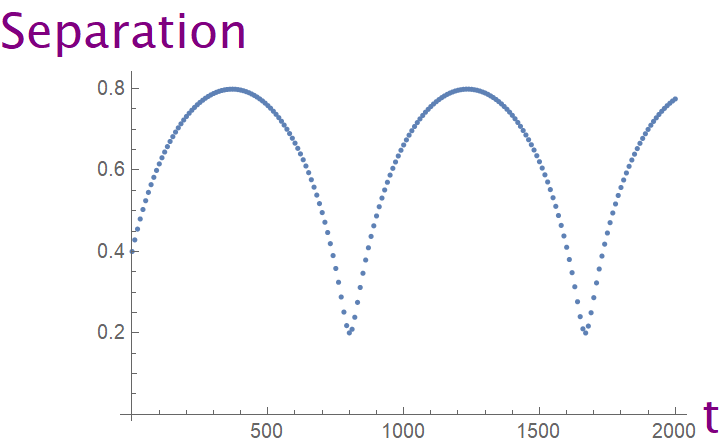}&&
\includegraphics[width=4cm]{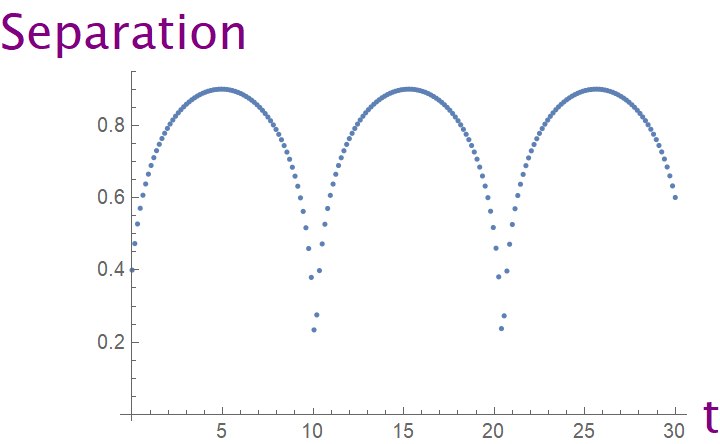}&&
\includegraphics[width=4cm]{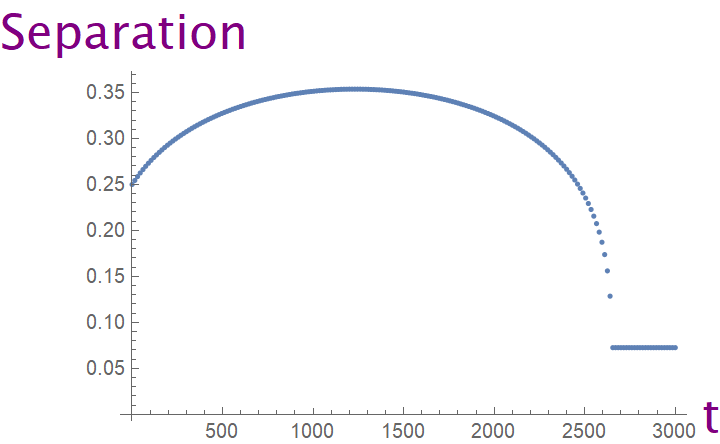}
\end{tabular}
\caption{(Color online) Comparison of evolution of inter-particle separation with time. From left to right, the separation is plotted for large  Saffman length  (Fig.\ref{radplane}), small Saffman length (Fig.\ref{azplane}), pair dynamics in strong confinement (Fig.\ref{conf1}), with the appropriate time units mentioned in earlier figures.}
    \label{trj_plane} 
\end{figure}
\\\\
\textbf{Study of hydrodynamic interactions of force dipoles}: The hydrodynamic interactions of a pair of force dipoles is illustrated in Fig.\ref{radplane} and Fig.\ref{azplane} for large and small values of the Saffman length $\lambda$ in the unconfined situation. In each of these simulations, we used the same initial condition for the dipole pair  orientations. The dynamical evolution is constructed in the same way as explained in Eq.(\ref{dynmeq}) of the main text, adapted to planar (x,y) coordinates. The time evolution is carried out using the velocity field in the unconfined situation,  Eq.(\ref{vdp_free}), with the Greens function $G^{free}_{ij}$  given by Eq.(\ref{plane_pf_main}) written in terms of Struve and Bessel functions. We find that both regimes of small and large $\lambda$ display similar non-linear oscillatory dynamics, not suitable for aggregation. However, as explained in main text, in the far field regime, the dipole pair dynamics happens over an an extended surface area of the membrane compared to the near field regime because of the dominant azimuthal (non-radial) nature of the streamlines at small $\lambda$. We next explore the pair dynamics under strong confinement  using Eq.(\ref{vdpc}) and the Greens function Eq.(\ref{gconf_realspace}) written in terms of modified Bessel functions. The dynamics is illustrated in Fig.\ref{conf1} and Fig.\ref{conf0} for small and large Saffman length $\lambda$. In both situations, the confinement $H$ is sufficiently small such that the screened Saffman length $\lambda_c$ is the smallest length scale in the problem. The rate of change of orientations of the dipoles is much slower (exactly vanishes in the limiting case, see Eq.(\ref{vdp_conf})) and this promotes non-oscillatory dynamics and favours aggregation, as was illustrated in Ref.\cite{mnk}. We also demonstrate this in a four particle simulation carried out in Fig.\ref{conf2} and compare the time evolution of dipole pairs in several scenarios in Fig.\ref{trj_plane}. In all figures, $L_0$ denotes the initial separation between dipole pairs.
\section { Details of viscous hydrodynamics in curved membranes}
\label{chyd_summary}
In this appendix section, we briefly review the formulation of hydrodynamics in curved membranes of fixed geometry following  Ref.\cite{henlev2008,henlev2010} and then specialize to Stokeslet flows in spherical memrbane.
Following the same notations as the main text, the incompressibility and momentum balance equations in the Low Reynolds regime in a generic curved membrane is given by
\beqa
D^\alpha v_\alpha =0, ~~~ D^\beta \Pi_{\alpha \beta}=0
\label{curvedstokes}
\eeqa
where 
\beqa
 \Pi_{\alpha \beta}=p g_{\alpha \beta} - \eta _{\alpha \beta \mu \gamma} D^{\mu} v^{\gamma}
 \eeqa
 and 
 \beqa
 \eta_{\alpha \beta \mu \gamma } = \eta_{2D} \left( g_{\alpha \mu} g_{\beta \gamma } + g_{\alpha \gamma } g _{\mu \beta }\right) + ( \xi - \eta_{2d} ) g_{\alpha \beta} g_{\mu \gamma}
 \eeqa
where v represents the 2D fluid velocity, $D$ is the 2D covariant Levi Civita Connection appropriate for the curved surface, p is the local membrane pressure and $g_{\mu \nu}$ is the surface metric. $\eta_{2D}$ and  $\xi$ are the shear and bulk viscosities respectively.\\\\
Using the incompressible nature of the membrane fluid $D^\alpha v_\alpha =0$ and the properties of the Levi Civita connection $D_{\alpha} g_{\mu \nu}=0$, we can simplify the momentum balance equation of  Eq.(\ref{curvedstokes})
\beqa
&D^{\beta} \Pi_{\alpha \beta} =0\nn\\
&\Rightarrow D^{\beta}\left( p g_{\alpha \beta} - \left( \eta_{2D} (g _{\alpha \mu} g_{\beta \gamma} + g_{\alpha \gamma} g _{\mu \beta}) +(\xi -\eta_{2D}) g_{\alpha \beta} g_{\mu \gamma}\right) D^{\mu} v^{\gamma}\right)=0\nn\\
&\Rightarrow D_{\alpha} p - \eta_{2D} \left(D_{\gamma} D_{\alpha} v^{\gamma} +D_\mu D^\mu v_\alpha\right)+(\xi - \eta_{2D}) D_{\alpha} \cancelto{0}{(D_\gamma v^\gamma)} =0\nn\\
&\Rightarrow D_{\alpha} p - \eta_{2D} \left(\underbrace{D_{\gamma} D_{\alpha} v^{\gamma}} +D_\mu D^\mu v_\alpha\right)=0\nn\\
\label{stk1}
\eeqa
Considering the term under braces in the above equation, we note that the derivatives $D_\gamma$ and $D_{\alpha}$  do not commute in curved surfaces, hence while flipping the their order one picks up a commutator term proportional to the local Gaussian curvature $K(x)$ ie.
\beqa
D_{\gamma} D_{\alpha} v^{\gamma}= D^{\gamma} D_{\alpha} v_{\gamma}=\commutator{ D^{\gamma}}{ D_{\alpha}} v_\gamma + D_\alpha \cancelto{0}{D^\gamma v_\gamma }=K(x)~ v_\alpha.
\eeqa
Using this in Eq.(\ref{stk1}) we get
\beqa
& D_{\alpha} p - \eta_{2D} \left(K(x) ~ v_\alpha +D_\mu D^\mu v_\alpha\right)=0\nn\\
\label{stk2}
\eeqa
In order to extract the flows due to point sources in the curved membrane, it is important to understand the spectrum of the operator $\eta_{2D} \left( K(x)  + \Delta\right)$,
\beqa
\eta_{2D} \left( K(x)  + \Delta\right) v_\alpha (\vec{x},s) = \lambda_s v_\alpha(\vec{x},s)
\label{egveq}
\eeqa
where $s$ labels the eigenvalues $\lambda_s$ and corresponding eigenfunctions $v_\alpha(\vec{x},s)$.
A generic velocity field in the curved geometry can thus be written as a superposition of these eigenmodes  $v_\alpha(\vec{x},s)$ as follows:\\
\beqa
 v_\alpha(x) =\sum_s A_s v_\alpha(\vec{x},s)
 \label{vdcmp}
\eeqa
In view of the incompressibility of the velocity eigenmodes, we introduce a stream function  $\phi(\vec{x},s)$:
\beqa
v_\alpha(\vec{x},s) =\epsilon_{\alpha \gamma} D^{\gamma} \phi(\vec{x},s)
\label{vstrm}
\eeqa
In terms of this stream function, Eq.(\ref{egveq}) takes the form

 \beqa
 \eta_{2D} \left( 2 K(x) D^{\gamma} \phi(\vec{x},s)  + D^{\gamma} \Delta \phi(\vec{x},s)\right)  = \lambda_s  D^{\gamma} \phi(\vec{x},s)\nn\\
\label{finalegvl}
 \eeqa
For generic curved membranes, one can construct numerical solutions. But for surfaces of constant curvature, analytic progress can be made since the equation has the simple structure of a Laplace eigenvalue equation :
\beqa
  \Delta~ \phi(\vec{x},s)  =  \frac{\lambda_s  - 2 K \eta_{2D}}{\eta_{2D}} ~~ \phi(\vec{x},s)\nn\\
\label{constk}
 \eeqa
Having discussed the hydrodynamic equations for a generic curved membrane, we now briefly discuss the external viscous fluids, described by the usual 3D Stokes equations. Hydrodynamic quantities like velocity, pressure, viscosity for $r>R$ carries a suffix $+$ while those for $r<R$ carries a $-$.
 \beqa
 \vec{\nabla} \cdot \vec{v}_\pm=0, ~~ \eta_\pm \nabla^2 v_\pm = \vec{\nabla} p_\pm
 \eeqa
The membrane fluid equations together with the 3D fluid equations need to be solved with the boundary conditions
 \beqa
v_\pm | _{r=R}=v
 \label{stick}
 \eeqa
and stress balance condition
 \beqa
 \sigma^{ext}_\alpha= D_\alpha p -\eta_{2D} \left(K(x) +\Delta\right) v_\alpha + T_\alpha.
 \label{stressbalance}
 \eeqa
Here $\sigma_\alpha^{ext}$ is the stress from the external point source generating the flow  while $T_\alpha$ represents the traction stress due to coupling of the membrane fluid with the external fluids.
 \beqa
 T_\alpha= \sigma_{\alpha r}^{-} -\sigma_{\alpha r}^+ |_ {r=R}, ~~ \sigma_{ij}^\pm =\eta_\pm \left(D_i v_j^\pm +D_j v_i^\pm\right)- g_{ij} p_{\pm}
 \label{trc}
 \eeqa
In terms of stream function defined in Eq.(\ref{vstrm}), one can recast the stress balance equation Eq.(\ref{stressbalance}) (after taking an antisymmetric derivative to remove the membrane pressure term)
  \beqa
\epsilon^{\alpha \beta} D_{\beta} \sigma^{ext}_\alpha=  - \sum_s A_s \lambda_s  \Delta \phi_s +  \epsilon^{\alpha \beta} D_{\beta} T_\alpha\nn\\
 \label{stressbalancemain}
 \eeqa
The discussions above hold for generic curved membranes, however from now we specialize to spherical membrane and construct the Stokeslet flows.\\\\
\textbf{Spherical membrane}: The spectrum of the Laplace Beltrami operator on a sphere  obeys
\beqa
\Delta \phi_{lm} = - \frac{l(l+1)}{R^2} \phi_{lm}.
\label{knownsp}
\eeqa
Comparing Eq.(\ref{constk}) and Eq.(\ref{knownsp})  we can read off the eigenvalues $\lambda$, 
\beqa
\lambda_l = \frac{2 -l(l+1)}{R^2} ~ \eta_{2D}
\label{sphereev}
\eeqa
and the eigenfunctions are the usual spherical harmonics
\beqa
\phi_{lm} = Y_{lm}(\theta, \phi)
\label{sphereef}
\eeqa
The velocity field on the spherical membrane can thus be written as
\beqa
v_\alpha =\sum_{lm} A_{lm} \epsilon_{\alpha \gamma} D^{\gamma} Y_{lm}
\label{vdcmps2}
\eeqa  
One solves the unknown coefficients $A_{lm}$ from the stress balance condition Eq.(\ref{stressbalance}). Using the well known Lamb's solution  Ref.\cite{lamb} for the external solvents, one can compute the traction term in Eq.(\ref{stressbalancemain}),
\beqa
T_\alpha =\sum_{lm} \left( \frac{\eta_-}{R} (l-1) + \frac{\eta_+}{R}(l+2)\right) A_{lm} \epsilon_{\alpha \beta} D^{\beta} Y_{lm}(\theta, \phi) .
\label{trcfinal}
\eeqa
Since we are interested in Stokeslet flows, we insert a point force localized at $(\theta_0,\phi_0)$ and decompose it via
\beqa
\sigma^{ext}_\alpha=\frac{F_{0_\alpha}}{R^2}\underbrace{\sum_{l=0}^{\infty} \sum_{m=-l}^{l} Y_{lm}(\theta, \phi) Y_{lm}^*( \theta_0,\phi_0)}_{\frac{1}{\sin \theta_0} \delta(\theta -\theta_0) \delta (\phi -\phi_0)}.
\eeqa
One can now solve $A_{lm}$ in terms of force components $ F_{0_\alpha}$ from Eq.(\ref{stressbalancemain})

\beqa
A_{lm} 
=\frac{\csc \theta_0}{\eta_{2D} s_l l (l+1)} 
\left(F_{\theta_0} \partial_{\phi_0} Y_{lm}^* (\theta_0 , \phi_0) - F_{\phi_0} \partial_{\theta_0} Y_{lm}^* (\theta_0 , \phi_0)\right)
\label{almsol}
\eeqa
where $s_l =l(l+1) -2 +\frac{R}{\lambda_-}(l-1)+\frac{R}{\lambda_+}(l+2)$ and $\lambda_{\pm}= \frac{\eta_{2D}}{\eta_{\pm}}$.\\\\
Plugging Eq.(\ref{almsol}) into Eq.(\ref{vdcmps2}) leads us to the final form of the Stokeslet flow Eq.(\ref{ptf_flow}) of the main text, which we repeat here for convenience : 
\beqa
\bm{v}_{Stokeslet} =  \frac{1}{4 \pi \eta_{2D}} \tilde{\bm{\nabla}}_{\theta,\phi}\left( \bm{f} \cdot \tilde{\bm{\nabla}}_{\theta_0, \phi_0}~ S[\gamma] \right)
\eeqa
where 
$\tilde{\nabla}_{\theta,\phi}$ is the twisted gradient operator $ \left(\csc \theta ~ \partial_\phi, -\partial_ \theta\right)$ at the response location $(\theta,\phi)$ and $\tilde{\nabla}_{\theta_0,\phi_0}$ is a similar gradient at the source location and
\beqa
S[\gamma]:=\sum_l \frac{2l+1}{s_l l (l+1)} P_l[\cos \gamma]
\label{appsdef} 
\eeqa
In our numerical simulations, we used analytic versions of Eq.(\ref{appsdef}) in real space, performing the sum over Legendre modes with some care.  The roots of $s_l=0$ appearing in the denominator of  Eq.(\ref{appsdef}) are given by
\beqa
l_p= \frac{ -(\eta_{2d} + R \eta_- + R \eta_+) + \sqrt{9 \eta_{2d}^2 + 6R \eta_{2d} (\eta_- -\eta_+)+ R^2 (\eta_- + \eta_+)^2}}{2 \eta_{2d}}\nn\\
l_m= \frac{ -(\eta_{2d} + R \eta_- + R \eta_+) - \sqrt{9 \eta_{2d}^2 + 6R \eta_{2d} (\eta_- -\eta_+)+ R^2 (\eta_- + \eta_+)^2}}{2 \eta_{2d}}\nn\\
\label{roots}
\eeqa
The model parameters thus control the location of the roots in the real line.  While $l_p$ lies in the range $-2 < l_p \leq 1$, the other root $l_m$ is always negative. Using this knowledge about the roots, one can now perform a partial fraction decomposition of Eq.(\ref{appsdef}) and sum the parts separately, leading to different real space representations of S depending on the sign of the root $l_p$, listed in Eq.(\ref{repf1}) and Eq.(\ref{repf2}) below.\\\\
\textbf{Case 1 } : $-2 < l_p < 0$  ( Low curvature regime)
\beqa
&S_{l_p<0}= \frac{1}{l_m l_p} \left(\log[2] -\log(-\cos \gamma + \sqrt{2- 2 \cos \gamma}+1)\right)+ \frac{1}{(1+l_m)(1+l_p)} \log[\frac{\cos \gamma -  \sqrt{2- 2 \cos \gamma}-1}{\cos \gamma-1}]+  \nn\\
&\frac{1+2 l_m}{l_m (1+l_m)(l_m-l_p)} A[l_m] +
\frac{1+2 l_p}{l_p (1+l_p)(l_p-l_m)} A[l_p]
\label{repf1}
\eeqa
where the function $A[l_m]$ is expressed in terms of Appell Hypergeometric function.
\beqa
&A[l_m]=\nn\\
&\frac{(-1+l_m) l_m ~\mathcal{A}[2-l_m,\frac{1}{2},\frac{1}{2}, 3-l_m, e^{i \gamma}, e^{-i \gamma}] -(-2+l_m)\left((-1+l_m) ~\mathcal{A}[-l_m,-\frac{1}{2},-\frac{1}{2}, 1-l_m, e^{i \gamma} , e^{-i \gamma}]+ 2 l_m~ \mathcal{A}[1-l_m, \frac{1}{2},\frac{1}{2}, 2 -l_m, e^{i \gamma}, e^{- i \gamma}]~ Cos ~\gamma \right)}{(-2+l_m)(-1+l_m) l_m}\nn\\
\label{Adef}
\eeqa
and a similar relation for  $A[l_p]$.\\\\
\textbf{Case 2} : $0< l_p <1$ (High curvature regime).\\
In this situation,
\beqa
&S_{l_p>0}= \frac{1}{l_m l_p} \left(\log[2] -\log(-\cos \gamma + \sqrt{2- 2 \cos \gamma}+1)\right)+ \frac{1}{(1+l_m)(1+l_p)} \log[\frac{\cos \gamma -  \sqrt{2- 2 \cos \gamma}-1}{\cos \gamma-1}]+  \nn\\
&\frac{1+2 l_m}{l_m (1+l_m)(l_m-l_p)} A[l_m] +
\frac{1+2 l_p}{l_p (1+l_p)(l_p-l_m)} B[l_p]
\label{repf2}
\eeqa 
where
\beqa
B[l_p]= -\frac{1}{l_p}+ \frac{1- \mathcal{A}~[-l_p,\frac{1}{2},\frac{1}{2},1-l_p,e^{i \gamma}, e^{-i \gamma}]}{l_p}
\label{Bdef}
\eeqa
Some special cases not captured by the above representations are listed below for completeness:\\\\
\textbf{Case 3} : one of the roots is zero \\
In terms of the other root, denoted by $\tilde{l} =\frac{\eta_-+4 \eta_+}{\eta_--2 \eta_+}$ we get
\beqa
S=  \sum_l \frac{2l+1}{(l-\tilde{l}) l^2 (l+1)} P_l[\cos \gamma]
=\frac{-1- \tilde{l}}{\tilde{l}^2} \left(\log[2] -\log(-\cos \gamma + \sqrt{2- 2 \cos \gamma}+1)\right) \nn\\
-\frac{1}{\tilde{l}} S_0 + \frac{1}{1+\tilde{l}} A[-1] +\frac{1+2 \tilde{l}}{\tilde{l}^2 (1+\tilde{l})} A[\tilde{l}]\nn\\
\eeqa
where
\beqa
S_0=\sum_{l=1}^\infty \frac{P_l[\cos \gamma]}{l^2}
\eeqa
 is convergent and needs numerical evaluation.\\
\textbf{Case 4} : one of the roots is -1\\
In this situation the other root is $\tilde{l} =\frac{2( \eta_-+ \eta_+)}{2\eta_-- \eta_+}$ and we have
\beqa
S=  \sum_l \frac{2l+1}{(l-l_p) l (l+1)^2} P_l[\cos \gamma]
=\frac{-1}{\tilde{l}} \left(\log[2] -\log(-\cos \gamma + \sqrt{2- 2 \cos \gamma}+1)\right) + \frac{1}{-\tilde{l}-1} \tilde{S}_0 \nn\\
+  \frac{\tilde{l}}{(\tilde{l}+1)^2} A[-1] +\frac{1+2 \tilde{l}}{\tilde{l} (1+\tilde{l})^2} A[\tilde{l}] \nn
\eeqa
where
\beqa
\tilde{S}_0=\sum_{l=1}^\infty \frac{P_l[\cos \gamma]}{(l+1)^2}\nn
\eeqa
 is convergent and needs numerical evaluation.\\

\begin{footnotesize}

\end{footnotesize}

\end{document}